\documentclass[final,3p,times]{elsarticle}

\usepackage[utf8]{inputenc}
\usepackage[T1]{fontenc}
\usepackage{graphicx}
\usepackage{amsmath}
\usepackage{booktabs}
\usepackage{subfig}
\usepackage{geometry}
\usepackage{multirow}

\usepackage[unicode]{hyperref}


\journal{(Insert Journal Name)}

\begin{document}

\begin{frontmatter}

\title{Enhancing Phase Clustering in Nanomechanical Property Maps of Multiphase Materials Using Kernel-Averaged Mechanical Mismatch}


\author[1]{D. Mercier\corref{cor1}}
\ead{david.mercier@synopsys.com}

\author[1,2]{Y. El Gharoussi}

\cortext[cor1]{Corresponding author}

\affiliation[1]{organization={Ansys, part of Synopsys, ICME Team at Synopsys Innovation Group},
addressline={}, 
city={France}}

\affiliation[2]{organization={Polytech Clermont},
addressline={}, 
city={France}}

\begin{abstract}
This work presents a novel approach for improving phase identification in nanomechanical property maps of multiphase materials, such as those obtained by nanoindentation or atomic force microscopy (AFM). A major difficulty in validating clustering strategies for such data lies in the absence of ground-truth phase labels in experimental measurements and in the tendency of overly simplistic synthetic datasets to artificially inflate algorithmic performance. To address this gap, we construct controlled yet non-trivial synthetic benchmarks featuring tunable mechanical contrast, graded interfaces, curved boundaries, and diffuse morphologies, enabling rigorous and realistic evaluation of clustering robustness.

Conventional clustering methods based solely on elastic modulus ($E$) and hardness ($H$) often struggle to resolve distinct phases when mechanical contrast is low or when diffuse interphase regions are present. We introduce the \textit{Kernel-Averaged Mechanical Mismatch} (KAMM), a neighborhood-informed feature that quantifies local mechanical heterogeneity through neighboring comparisons in the $(E,H)$ space. When incorporated into a three-dimensional clustering space $(E, H, \mathrm{KAMM})$, this framework improves phase separability, enhances interphase detection, and increases robustness to noise. By enabling more reliable segmentation of mechanical domains under realistic contrast conditions, the proposed method facilitates the generation of representative volume elements (RVEs) and supports more accurate extraction of phase-specific properties in heterogeneous microstructures.
\end{abstract}

\section*{Graphical Abstract}

\begin{center}
\includegraphics[width=\linewidth]{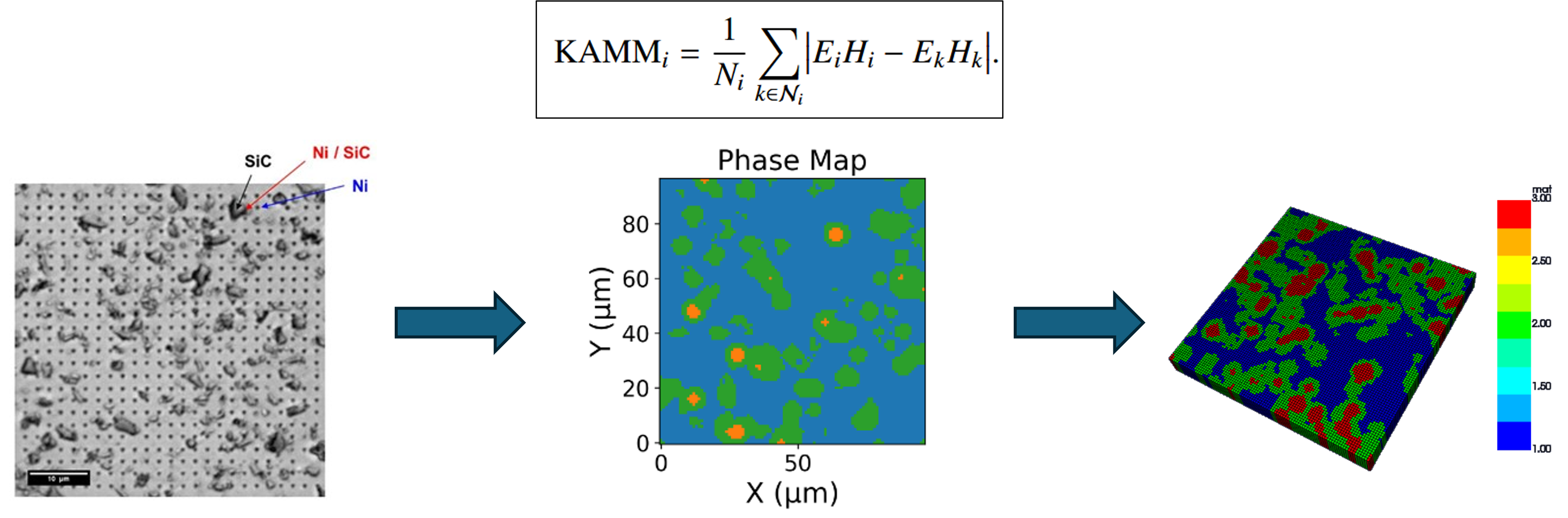}
\end{center}

\begin{keyword}
Nanomechanical Property Maps \sep Clustering \sep Multiphase materials \sep Nanoindentation \sep Kernel-Averaged Mechanical Mismatch (KAMM) \sep Neighborhood-based descriptors \sep Phase map \sep Modelling workflow
\end{keyword}

\end{frontmatter}

\section{Introduction}

High-speed nanoindentation mapping has become a widely adopted technique for characterizing the local mechanical behavior of heterogeneous materials. By enabling the acquisition of neighborhood-resolved maps of hardness and elastic modulus, it provides a powerful means to investigate complex microstructures in metals and alloys, metallic glasses, composites, concretes, stones, and other multiphase systems~\cite{de_vasconcelos_grid_2016, kossman_mechanical_2019, nohazic_coating_2021, liang_spatial_2022, shi_determining_2023, dhal_mapping_2023, guo_machine_2025, chen_curve_2024, rossi_revealing_2025}. In parallel, atomic force microscopy (AFM)--based nanomechanical mapping has emerged as a complementary approach, offering higher spatial resolution and the ability to probe compliant or delicate surfaces with limited permanent deformation~\cite{randall_nanoindentation_2009, coq_germanicus_quantitative_2020, he_atomic_2018, garcia_advances_2025}. Together, these techniques enable high-throughput mechanical characterization at the submicron scale, bridging local mechanical responses with mesoscale microstructural features ~\cite{Miracle2021, Gianola2023}.

The growing availability of large-scale nanomechanical maps has driven the adoption of data-driven analysis methods. In particular, unsupervised machine learning approaches-most notably clustering algorithms-have been increasingly employed to automatically segment material domains exhibiting distinct mechanical behavior~\cite{mercier_tridimap_2018, mercier_microstructural_2019, bernachy-barbe_data_2019, chen_clustering_2021, giolando_ai-dente_2023, jentner_unsupervised_2023, zhang_unsupervised_2024}. When constituent phases display well-separated hardness (H) and elastic modulus (E) values, clustering based on these two features is generally effective. However, many heterogeneous materials exhibit overlapping mechanical responses due to similar phase stiffness, compositional gradients, or microstructural interactions. In such cases, clustering becomes ambiguous and may fail to reliably separate phases \cite{Constantinides2006}.

These ambiguities are often associated with the presence of interfaces and interphases, which play a central role in governing the macroscopic behavior of composites and multiphase materials. Interfaces-sharp boundaries between discrete phases-can act as mechanical discontinuities or stress concentrators, potentially introducing measurement artefacts. Interphases-graded or diffuse transition regions-are frequently responsible for adhesion, damage tolerance, and transport properties~\cite{hausild_statistical_2019, hodzic_nano-indentation_2000, hausild_determination_2016}. Despite their importance, interfaces and interphases remain difficult to resolve experimentally. Their spatial extent may lie below the resolution limits of conventional microscopy, while their mechanical signatures can be masked by measurement noise, sampling artefacts, or insufficient contrast in H and E. As a result, interphase data points may be misclassified by clustering algorithms as belonging to adjacent phases or grouped into ambiguous intermediate clusters, thereby biasing phase identification and phase-specific property extraction~\cite{mercier_microstructural_2019, konstantopoulos_classification_2020, alizade_comparative_2024, ortizmembrado_2025}

To address these challenges, several strategies have been proposed. On the algorithmic side, alternative clustering methods and comparative benchmarking of clustering techniques have been explored to improve robustness in the presence of overlapping phase properties~\cite{alizade_comparative_2024}. On the feature level, additional descriptors beyond H and E have been introduced, including the H/E ratio, energy dissipation, indentation curve shape parameters, and machine-learning–derived features~\cite{jakes_nanoindentation_2009, lu_extraction_2020, kossman_pop-identification_2021, besharatloo_influence_2021, park_deep_2023, trost_explainable_2025}. Improved discrimination has also been achieved by correlating nanoindentation data with complementary characterization techniques, such as EBSD in metallic systems~\cite{bruno_advanced_2024}, or by performing indentation at multiple depths to probe subsurface heterogeneity~\cite{konstantopoulos_classification_2020}. More recently, physics-informed synthetic nanoindentation datasets have been proposed to systematically reproduce measurement variability, uncertainty, and artefacts, providing a traceable framework for controlled investigations of phase discrimination limits and the validation of data analysis and classification methodologies~\cite{maculotti_synthetic_2025}.

Despite these advances, many existing approaches remain constrained by the absence of explicit consideration of the spatial relationships between neighboring measurement points, which becomes essential when phase boundaries are diffuse or mechanically subtle. Without accounting for how adjacent data points are arranged and interact locally, clustering algorithms may fail to distinguish gradual transitions from true phase regions. This shortcoming hampers the robust identification of interfaces and interphases in complex material systems, thereby motivating the development of methodologies that integrate neighborhood-informed information, uncertainty quantification, and physics-informed data representations~\cite{puchi-cabrera_machine_2023}. In heterogeneous materials, neighborhood-based descriptors—including phase topology, morphology, size distributions, interfacial characteristics, spatial correlations, and orientation distributions—play a decisive role in determining the effective macroscopic properties. Whether the system consists of fibers, particles, grains, pores, or multiphase domains, neglecting the spatial arrangement of mechanical properties can lead to significant errors in property prediction and model calibration. Such frameworks enable the generation of realistic microstructural, RVE-based simulations that faithfully capture both morphological and mechanical heterogeneity~\cite{breuer_rve_2020, pardoen_nanomechanics_2021}.

In this work, we introduce a new neighborhood-informed feature to support clustering-based phase identification and the generation of realistic representative volume elements (RVEs): the \textit{Kernel-Averaged Mechanical Mismatch} (KAMM). Inspired by orientation-based neighborhood metrics used in EBSD analysis~\cite{parish_cluster_2022, mtex_kam}, KAMM quantifies local mechanical heterogeneity by measuring the deviation of a given local experimental measurement point from its surrounding neighborhood in the $(E,H)$ space. High KAMM values are typically associated with proximity to interfaces or interphases, whereas low values indicate locally homogeneous mechanical behavior. By embedding KAMM into a three-dimensional clustering space $(E, H, \mathrm{KAMM})$, we introduce neighborhood-based awareness into data-driven phase segmentation.

The performance of the proposed approach is assessed using synthetic nanoindentation datasets designed to replicate representative multiphase microstructures, including systems with sharp phase boundaries, diffuse interphases, and graded mechanical transitions. These controlled datasets provide access to ground truth phase labels, enabling a rigorous quantitative evaluation of clustering accuracy. The results demonstrate that KAMM significantly improves classification robustness, enhances the separability of mechanically overlapping phases, and enables the identification and exclusion of ambiguous interphase regions, thereby improving the reliability of phase-wise property extraction.

The structure of the paper is as follows. Section~\ref{sec:methods} describes the synthetic data generation procedure, details the clustering methodology, and introduces the different KAMM-based features, illustrated using a dedicated five-region benchmark specimen. Section~\ref{sec:results} presents quantitative clustering performance metrics with and without KAMM across multiple two-phase reference microstructures, both with and without interphases. This section also includes applications to experimental nanoindentation datasets from the literature, demonstrating the practical benefits of incorporating kernel-based mechanical mismatch features. Section~\ref{sec:discussion} discusses the implications and limitations of the proposed framework, including its integration into multiscale materials modeling workflows. Finally, Section~\ref{sec:conclusion} summarizes the main findings and outlines directions for future research.

\section{Materials and Methods}
\label{sec:methods}

\subsection{Synthetic Specimen Generation}
\label{sec:synthetic_specimens}

\paragraph{Rationale and benchmarking challenge}

A central difficulty in evaluating clustering algorithms for nanomechanical phase identification lies in the absence of ground truth for experimental datasets. Real materials exhibit unknown phase labels, uncontrolled noise sources, spatial correlations, and scale-dependent interactions, making quantitative benchmarking inherently ambiguous. Conversely, overly simplistic synthetic datasets (e.g., perfectly separable Gaussian clusters without spatial structure) lead to artificially optimistic clustering performance and fail to represent the complexity of multiphase microstructures.

Constructing a meaningful synthetic benchmark therefore requires balancing three competing constraints:
(i) exact knowledge of phase labels,
(ii) tunable mechanical contrast and noise levels,
and (iii) realistic spatial organization, including sharp interfaces, diffuse transitions, curvature effects, and non-compact morphologies.

To address this, we developed a fully parametric Python-based generation framework that programmatically constructs synthetic nanomechanical maps under controlled and reproducible conditions. The script defines the spatial grid, assigns phase geometries, imposes user-defined mechanical contrast via $(E_\mathrm{ratio}, H_\mathrm{ratio})$, and injects configurable intra-phase Gaussian noise and optional property gradients. All geometric parameters (e.g., inclusion size, interphase width, particle distribution, curvature), mechanical ratios, and noise levels are explicitly tunable, enabling systematic exploration of clustering performance across a wide range of controlled scenarios.

Using this framework, we generated four families of synthetic nanomechanical datasets that emulate representative microstructural patterns encountered in multiphase materials. Each dataset provides exact phase labels, user-defined mechanical contrast, controlled intra-phase variability, and optional neighborhood-level gradients. For every configuration, the generator automatically exports:
\begin{itemize}
    \item a synthetic indentation dataset ($H$, $E$ per pixel),
    \item the corresponding ground-truth phase map (image and CSV),
    \item configuration metadata (all input parameters),
    \item and phase-level summary statistics.
\end{itemize}

It is assumed that the spatial resolution is sufficient for each indentation test (i.e., each synthetic pixel) to probe a single phase. This requires appropriate scale separation: the maximum indentation depth is kept smaller than the characteristic microstructural length scale (e.g., $h_\text{max} \lesssim 0.1D$, with $D$ the mean particle diameter or interphase width), while remaining larger than the surface roughness (e.g., $h_\text{max} > 3R_q$). In addition, the indentation spacing $d$ is chosen larger than approximately twice the plastic zone radius to prevent interaction between adjacent indents. Under these conditions, each measurement reflects intrinsic phase properties rather than composite or roughness-affected responses~\cite{randall_nanoindentation_2009}.

Rather than creating trivially separable clusters, the objective of these datasets is to systematically probe clustering robustness under:
\begin{itemize}
    \item low to high mechanical contrast,
    \item overlapping property distributions,
    \item sharp versus graded interfaces,
    \item curved versus planar boundaries,
    \item and diffuse, spatially ambiguous morphologies.
\end{itemize}

\paragraph{Five-region quadrant + center system (feature exploration)}

\begin{figure}[htbp]
\centering
\includegraphics[width=0.65\columnwidth]{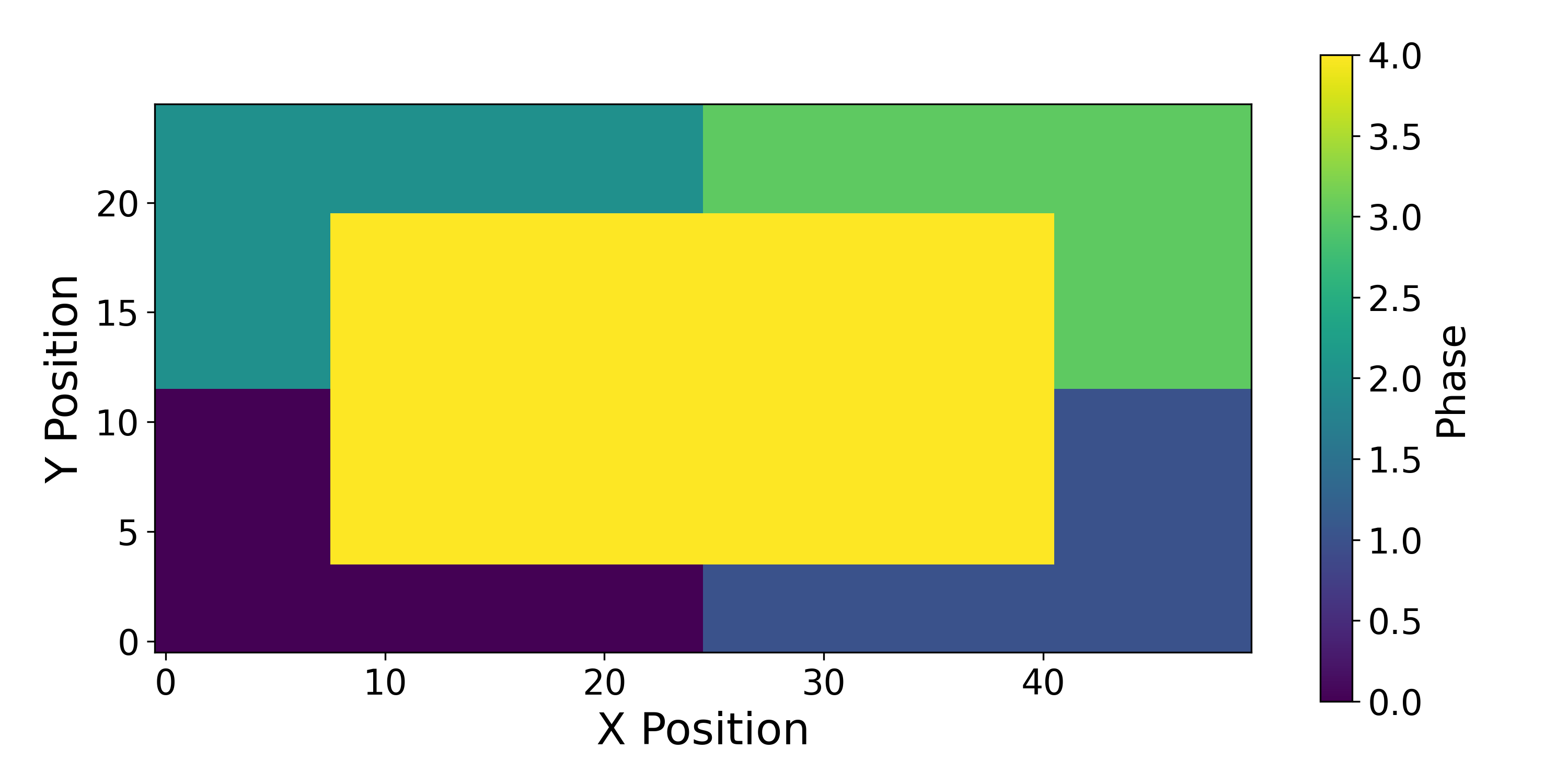}
\caption{Five-region specimen used to illustrate sensitivity of scalar and neighborhood-based descriptors.}
\label{fig:five_region_example}
\end{figure}

This benchmark divides the grid into four quadrants with distinct mechanical properties and a central compliant inclusion (Fig.~\ref{fig:five_region_example}). The assigned phase properties span a broad range (from $H\!\approx\!0.2$~GPa up to $H\!\approx\!5$~GPa and from $E\!\approx\!10$--$50$~GPa), enabling exploration of low-, moderate-, and high-contrast configurations through the implied ratios $(E_\mathrm{ratio},H_\mathrm{ratio})$.

Gaussian noise is added within each region (typically $\sigma_H=0.1$~GPa and $\sigma_E=0.2$~GPa), introducing intra-phase dispersion and partial overlap between clusters. Because geometry is fixed and certain phase pairs may exhibit subtle contrast, this dataset is primarily intended for feature-behavior analysis rather than pure clustering validation. It enables controlled comparison of scalar descriptors ($H$, $E$, $H/E$, $H^3/E^2$) and neighborhood-based descriptors (e.g., KAMM-based features), particularly near boundaries and in regions of moderate contrast.

\paragraph{Rectangular composite with sharp or graded interface}

\begin{figure*}[htbp]
\centering

\subfloat[Rectangular composite, sharp interface]{
    \includegraphics[width=0.32\textwidth]{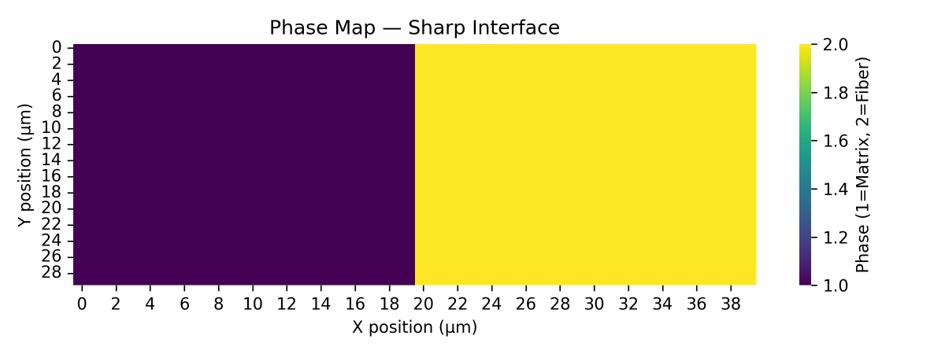}
}
\hfill
\subfloat[Rectangular composite, small interphase]{
    \includegraphics[width=0.32\textwidth]{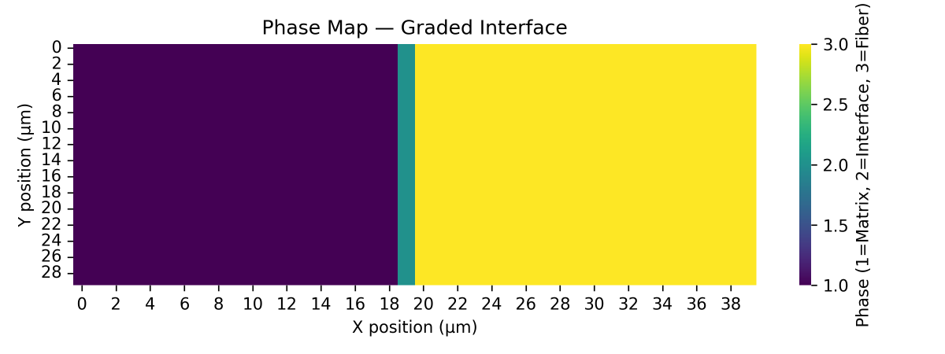}
}
\hfill
\subfloat[Rectangular composite, large interphase]{
    \includegraphics[width=0.32\textwidth]{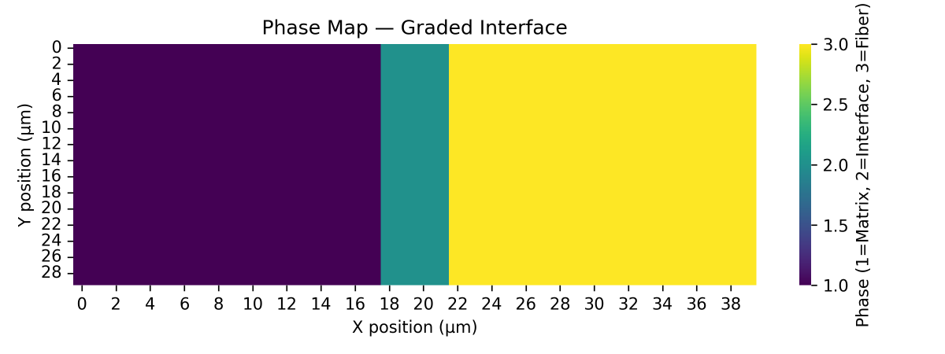}
}

\vspace{0.6em}

\subfloat[Matrix--fiber composite (no interphase)]{
    \includegraphics[width=0.32\textwidth]{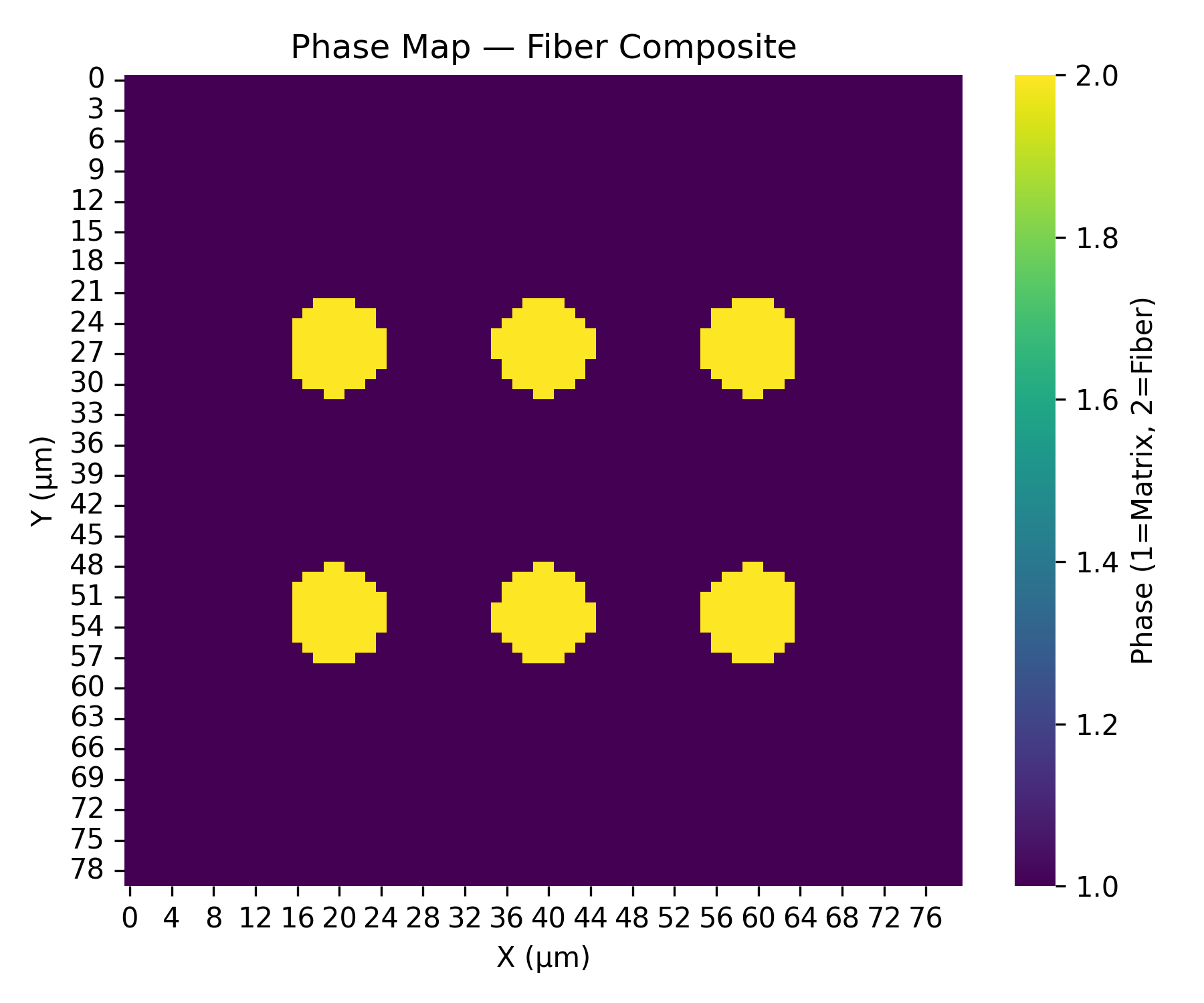}
}
\hfill
\subfloat[Matrix--fiber composite (with annular interphase)]{
    \includegraphics[width=0.32\textwidth]{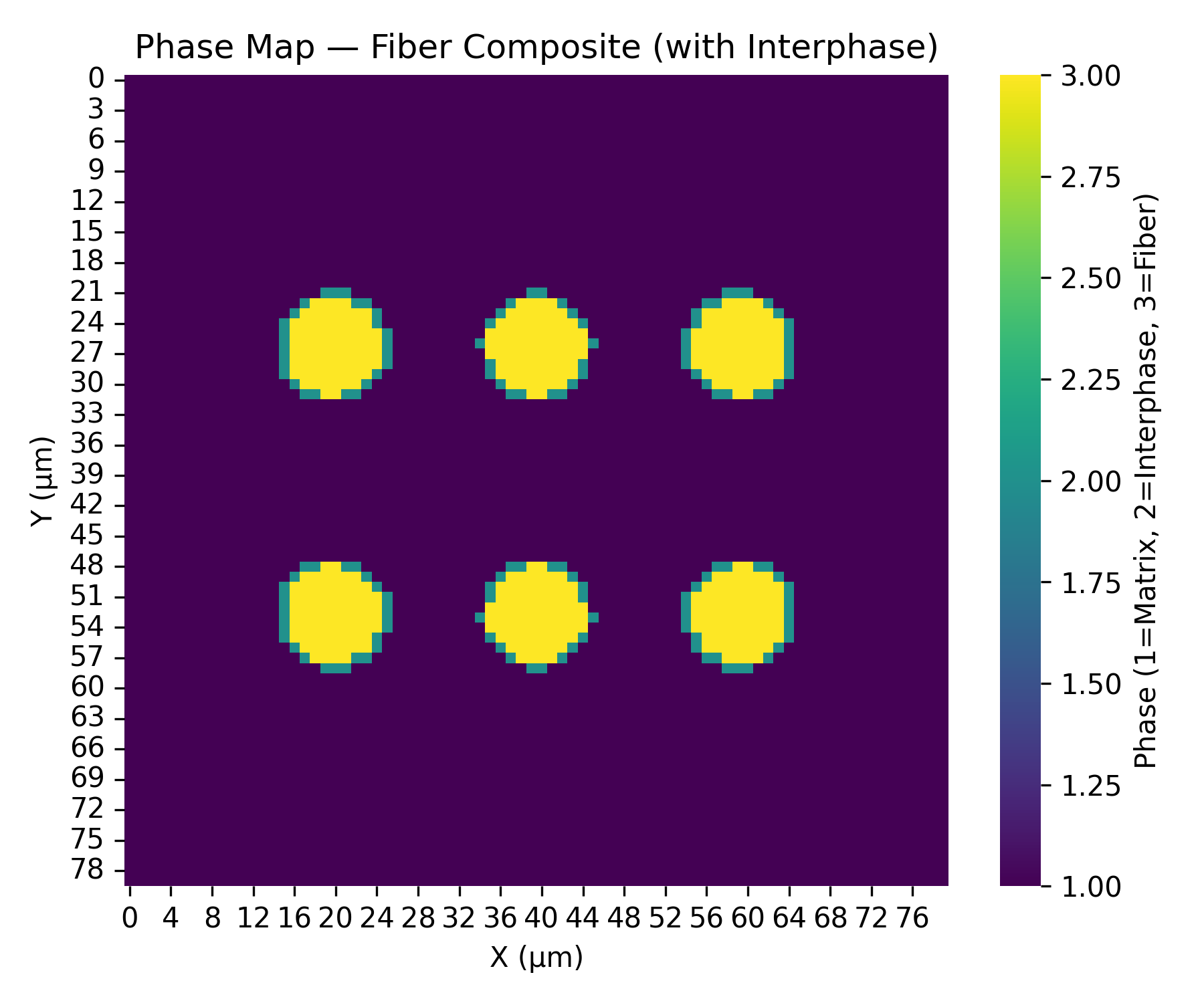}
}
\hfill
\subfloat[Matrix--particle composite with diffuse inclusions]{
    \includegraphics[width=0.32\textwidth]{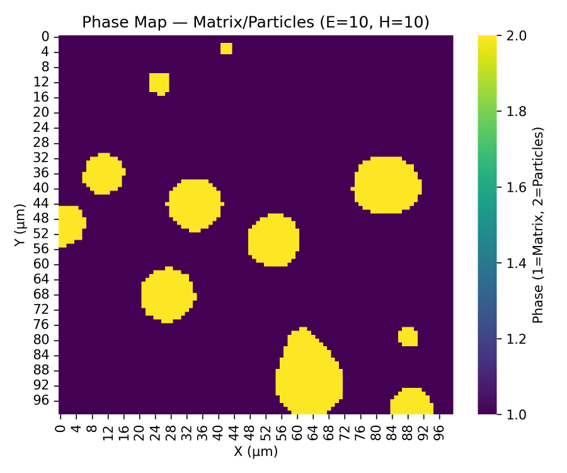}
}

\caption{Synthetic specimen families used for benchmarking clustering performance. 
The configurations span sharp and graded planar interfaces (a--c), periodic circular inclusions with and without interphase layers (d--e), and diffuse, randomly distributed particles (f). 
Together, these geometries provide controlled yet structurally diverse scenarios for evaluating clustering robustness under varying mechanical contrast and morphological complexity.}
\label{fig:synthetic_examples}
\end{figure*}

This one-dimensional configuration consists of a matrix (left), a fiber phase (right), and an optional interphase region (Fig.~\ref{fig:synthetic_examples}(a--c)). When the interphase width is zero, the system reduces to a sharp two-phase composite (Fig.~\ref{fig:synthetic_examples}(a)). Nonzero widths introduce a graded transition in which $H$ and $E$ vary smoothly across the interface (Fig.~\ref{fig:synthetic_examples}(b,c)). Relative interphase widths of 0.03 and 0.05 were used to simulate realistic gradients. Mechanical contrast is controlled via $(E_\mathrm{ratio}, H_\mathrm{ratio})$, while low-amplitude Gaussian noise is added within each region (typically $\sigma_H = 0.1$~GPa and $\sigma_E = 0.1$~GPa) to emulate experimental variability. Despite its simple geometry, this configuration is non-trivial from a clustering standpoint: graded interfaces blur phase boundaries, reduce cluster separability in feature space, and introduce spatial correlation. It therefore provides a controlled framework for evaluating sensitivity to diffuse transitions and boundary localization accuracy.

\paragraph{Matrix--fiber composite with optional annular interphase}

This dataset simulates a periodic fiber-reinforced composite with circular inclusions arranged on a regular grid (Fig.~\ref{fig:synthetic_examples}(d,e)). An interphase layer of user-defined thickness may surround each fiber, resulting in either a two-phase (matrix/fiber) configuration (Fig.~\ref{fig:synthetic_examples}(d)) or a three-phase (matrix/interphase/fiber) system (Fig.~\ref{fig:synthetic_examples}(e)).

Mechanical properties are sampled from Gaussian clusters whose means are defined by $(E_\mathrm{ratio}, H_\mathrm{ratio})$. Curved boundaries and concentric gradients introduce classification ambiguity near interfaces, particularly when contrast is low. This configuration challenges clustering methods in the presence of non-planar geometry and radially varying properties.

\paragraph{Matrix--particle composite with diffuse random particles}

This two-phase system consists of a homogeneous matrix embedding soft, irregular inclusions generated by superposition of Gaussian peaks at random positions (Fig.~\ref{fig:synthetic_examples}(f)). Unlike the fiber case, morphology is neither periodic nor sharply defined.

The resulting diffuse transitions mimic poorly resolved or chemically graded particle boundaries. Since geometry is not explicitly prescribed, particle regions are inferred by thresholding the hardness field. This dataset is intentionally challenging: clusters may be non-compact in feature space, spatially irregular, and partially overlapping. It probes algorithm robustness under low contrast, random morphology, and spatial ambiguity.

\paragraph{Mechanical parameterization and contrast scaling}

The mechanical properties assigned to the synthetic geometries described above are defined per pixel (except for the five-region feature-exploration case, which uses fixed values). For the composite configurations (rectangular, matrix--fiber, and matrix--particle), mechanical contrast is introduced systematically via scaling ratios:

\[
E_\mathrm{ratio},\, H_\mathrm{ratio} \in [1.0,\,3.0],
\]

with selected high-contrast cases extending to 5 or 10 for the matrix--particle configuration (see Fig.~\ref{fig:synthetic_examples}(f)).

The lower bound ($\approx 1$) intentionally produces nearly indistinguishable phases, where clustering becomes ill-posed. Moderate ratios (2--3) correspond to common multiphase systems, whereas higher ratios represent extreme ceramic/metal or glass/polymer mismatches. This controlled variation enables systematic evaluation of clustering stability as a function of separability in mechanical feature space.

Figure~\ref{fig:synthetic_examples} summarizes the synthetic specimen families, highlighting the diversity of spatial structures---sharp, graded, circular, and diffuse---used to construct a benchmark that is controlled yet sufficiently complex to avoid trivial clustering scenarios.

\paragraph{Experimental validation}

To assess practical relevance beyond controlled synthetic conditions, the clustering framework was also applied to experimental nanomechanical data from a nickel--silicon carbide (Ni--SiC) composite coating reported in~\cite{mercier_microstructural_2019}. This system features hard ceramic inclusions dispersed within a metallic matrix and provides a realistic test case where phase labels are not explicitly known, allowing qualitative comparison with microstructural observations.

\subsection{KAMM features and clustering pipeline}
\label{sec:kamm_method}

This subsection introduces the neighborhood-based mechanical descriptors and the clustering framework used for phase identification. In particular, we define the \textbf{Kernel-Averaged Mechanical Mismatch (KAMM)}, inspired by the \textit{Kernel Average Misorientation (KAM)} metric used in EBSD. By quantifying local mechanical heterogeneity through comparisons of each indentation point’s $(E,H)$ values with those of its neighbors, KAMM enhances boundary detection and improves phase separability under limited mechanical contrast.

\begin{figure}[htbp]
\centering
\includegraphics[width=0.5\columnwidth]{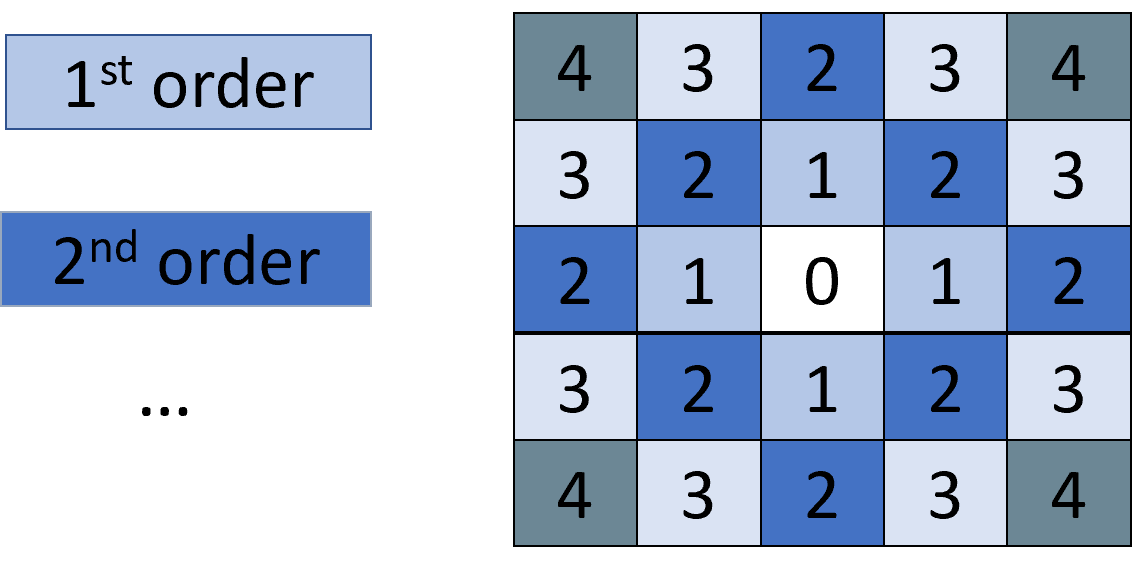}
\caption{Pixel connectivity schemes used for KAMM, with first-order neighborhood (4 nearest neighbors in light blue) and second-order neighborhood (includes diagonals and two-pixel offsets in dark blue).}
\label{fig:kam_neighbors}
\end{figure}

\paragraph{Standard KAMM}
For a point \(i\) with a neighbor set \(\mathcal{N}_i\) of size \(N_i\), the default KAMM metric is defined as the average absolute mismatch of the modulus--hardness product:
\begin{equation}
\label{eq:kamm}
\text{KAMM}_i
=
\frac{1}{N_i}
\sum_{k \in \mathcal{N}_i}
\bigl| E_i H_i - E_k H_k \bigr|.
\end{equation}

\paragraph{Component-wise mismatch metrics}
Two scalar projections isolate the mismatch in each property:
\begin{align}
\label{eq:kaem}
\text{KAEM}_i &=
\frac{1}{N_i}
\sum_{k \in \mathcal{N}_i}
\lvert E_i - E_k \rvert,
\\[4pt]
\label{eq:kapm}
\text{KAPM}_i &=
\frac{1}{N_i}
\sum_{k \in \mathcal{N}_i}
\lvert H_i - H_k \rvert.
\end{align}

\paragraph{Alternative KAMM variants}
Two additional formulations capture relative or coupled variations between modulus and hardness:
\begin{itemize}

    \item \textbf{KAMM\textsubscript{RATIO}:}
    \begin{equation}
    \text{KAMM}_{\mathrm{ratio},\,i}
    =
    \frac{1}{N_i}
    \sum_{k \in \mathcal{N}_i}
    \left\lvert
    \frac{E_i - E_k}{\,H_i - H_k + \varepsilon\,}
    \right\rvert,
    \end{equation}
    where \(\varepsilon\) is a small positive constant to avoid division by zero.

    \item \textbf{KAMM\textsubscript{NORMPROD}:}
    \begin{equation}
    \text{KAMM}_{\mathrm{normprod},\,i}
    =
    \frac{1}{N_i}
    \sum_{k \in \mathcal{N}_i}
    \frac{
        (E_i - E_k)(H_i - H_k)
    }{
        \sigma_E \, \sigma_H
    },
    \end{equation}
    where \(\sigma_E\) and \(\sigma_H\) are the (global) standard deviations of \(E\) and \(H\).
\end{itemize}

These variants emphasize absolute, relative, or coupled mismatch, and can highlight diffuse gradients or subtle transitions in the modulus–hardness fields.

\paragraph{Neighborhood connectivity}
KAMM is computed over spatial neighbors defined by the connectivity schemes illustrated in Fig.~\ref{fig:kam_neighbors}:
\begin{itemize}
    \item \textbf{First-order} ($\mathcal{O}_1$): 4-connectivity (up, down, left, right)
    \item \textbf{Second-order} ($\mathcal{O}_2$): includes diagonals and two-pixel offsets (up to 8 neighbors)
\end{itemize}

\begin{figure}[htbp]
\centering

\begin{minipage}{0.45\linewidth}
    \subfloat[Hardness ($H$), units: GPa]{
        \includegraphics[width=0.9\linewidth]{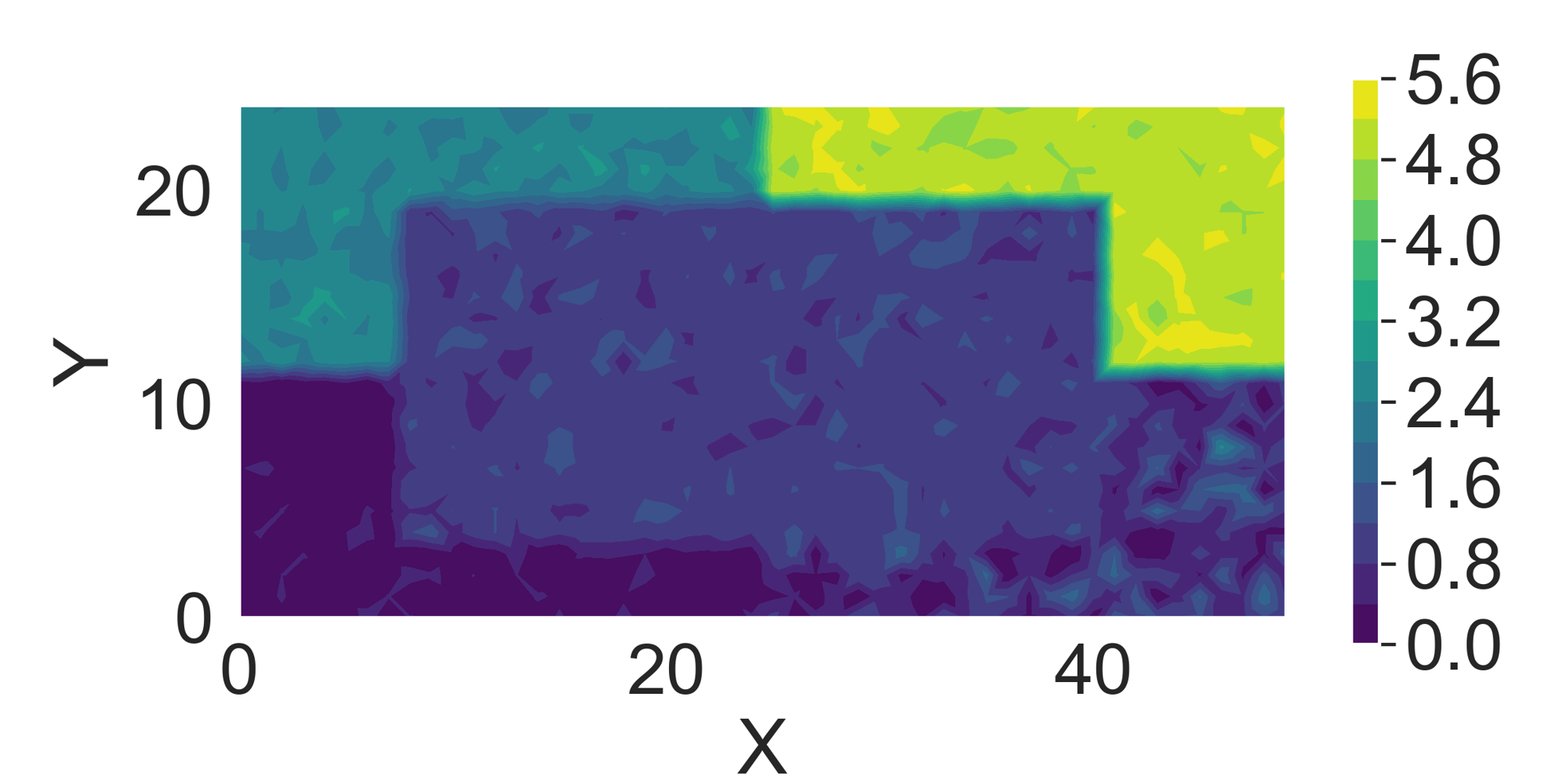}
    }
\end{minipage}
\hfill
\begin{minipage}{0.45\linewidth}
    \subfloat[Modulus ($E$), units: GPa]{
        \includegraphics[width=0.9\linewidth]{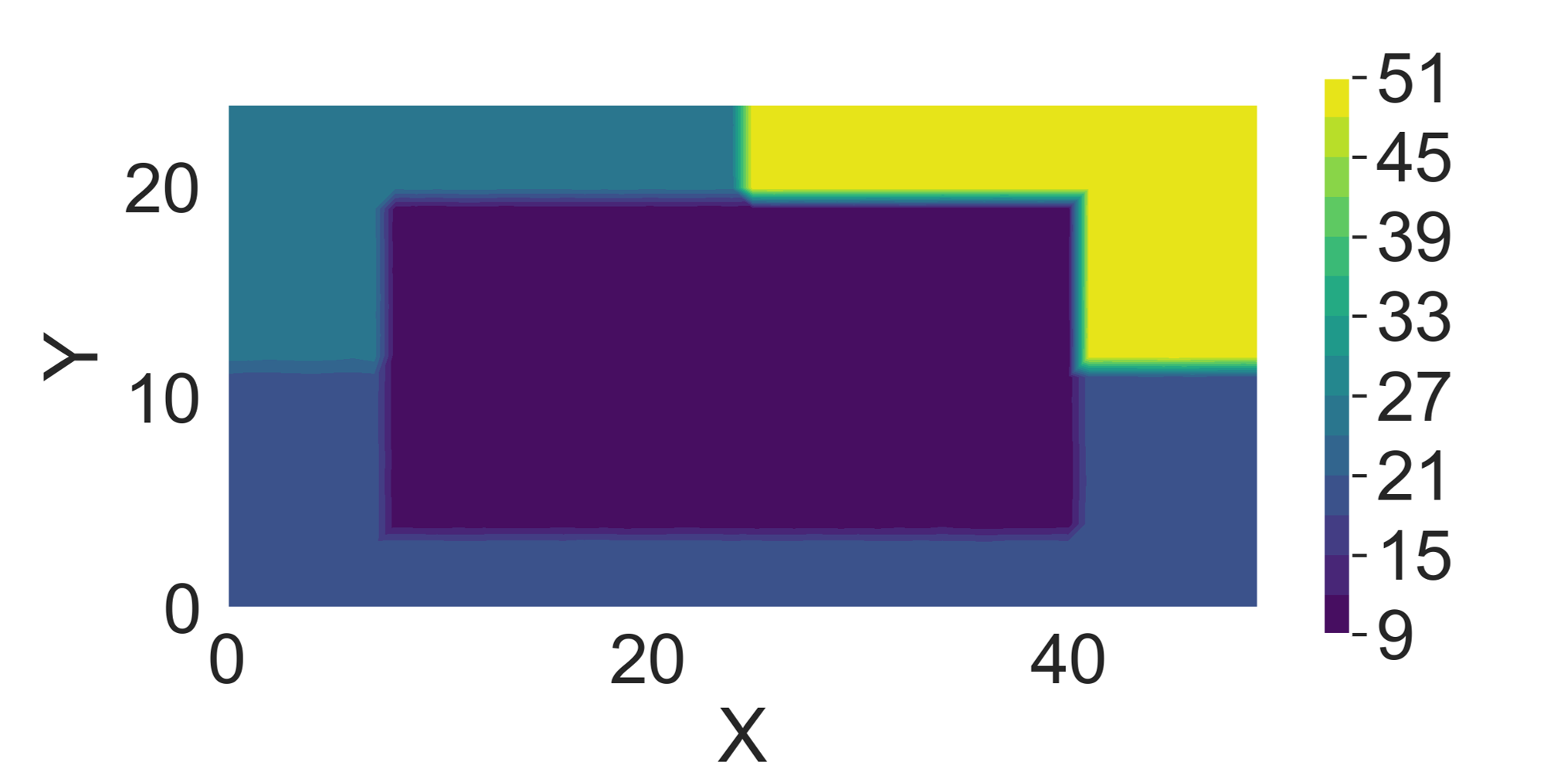}
    }
\end{minipage}

\vspace{0.2em}

\begin{minipage}{0.45\linewidth}
    \subfloat[$H/E$ ratio, dimensionless]{
        \includegraphics[width=0.9\linewidth]{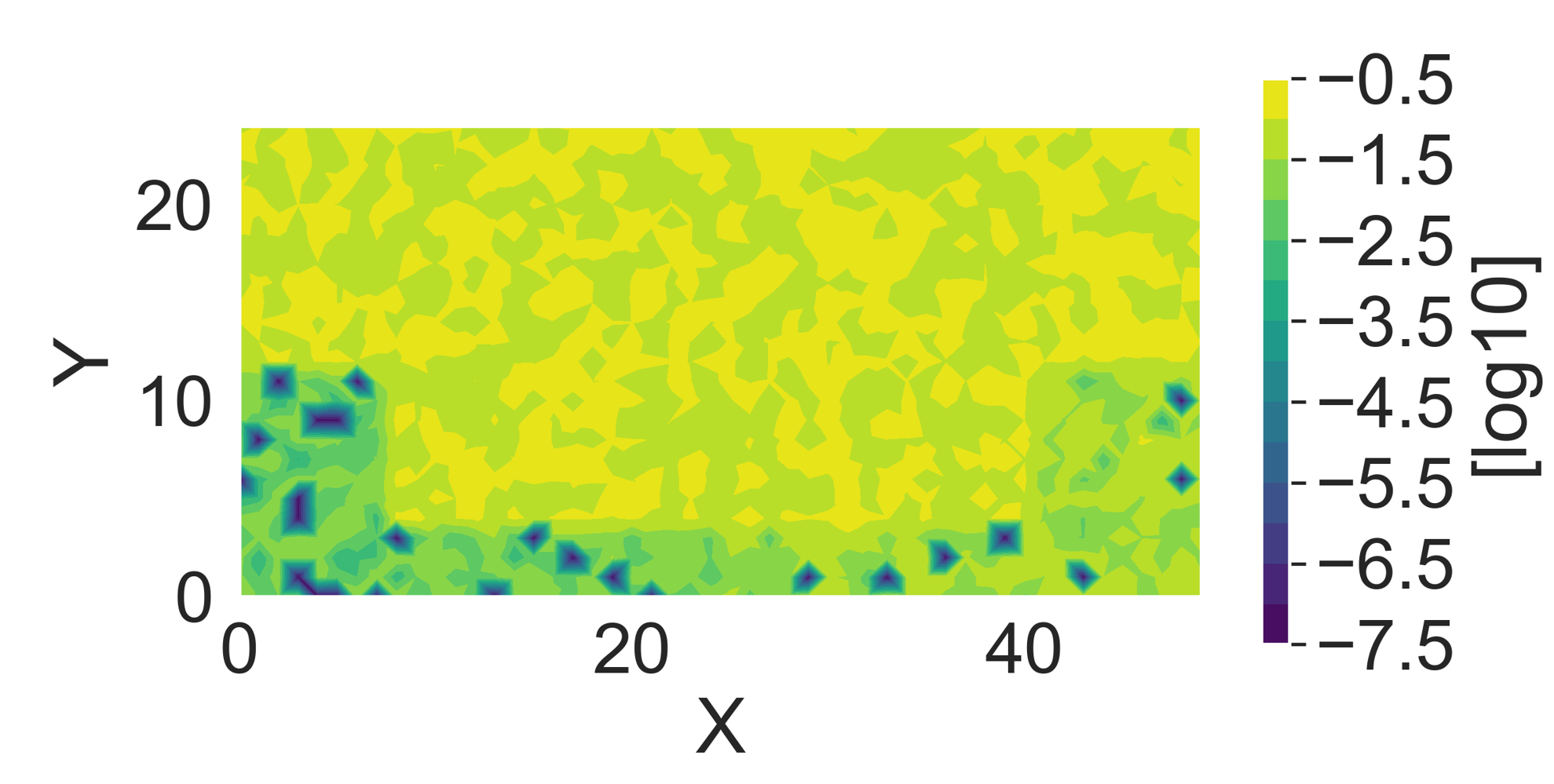}
    }
\end{minipage}
\hfill
\begin{minipage}{0.45\linewidth}
    \subfloat[$H^3/E^2$ ratio, units: GPa]{
        \includegraphics[width=0.9\linewidth]{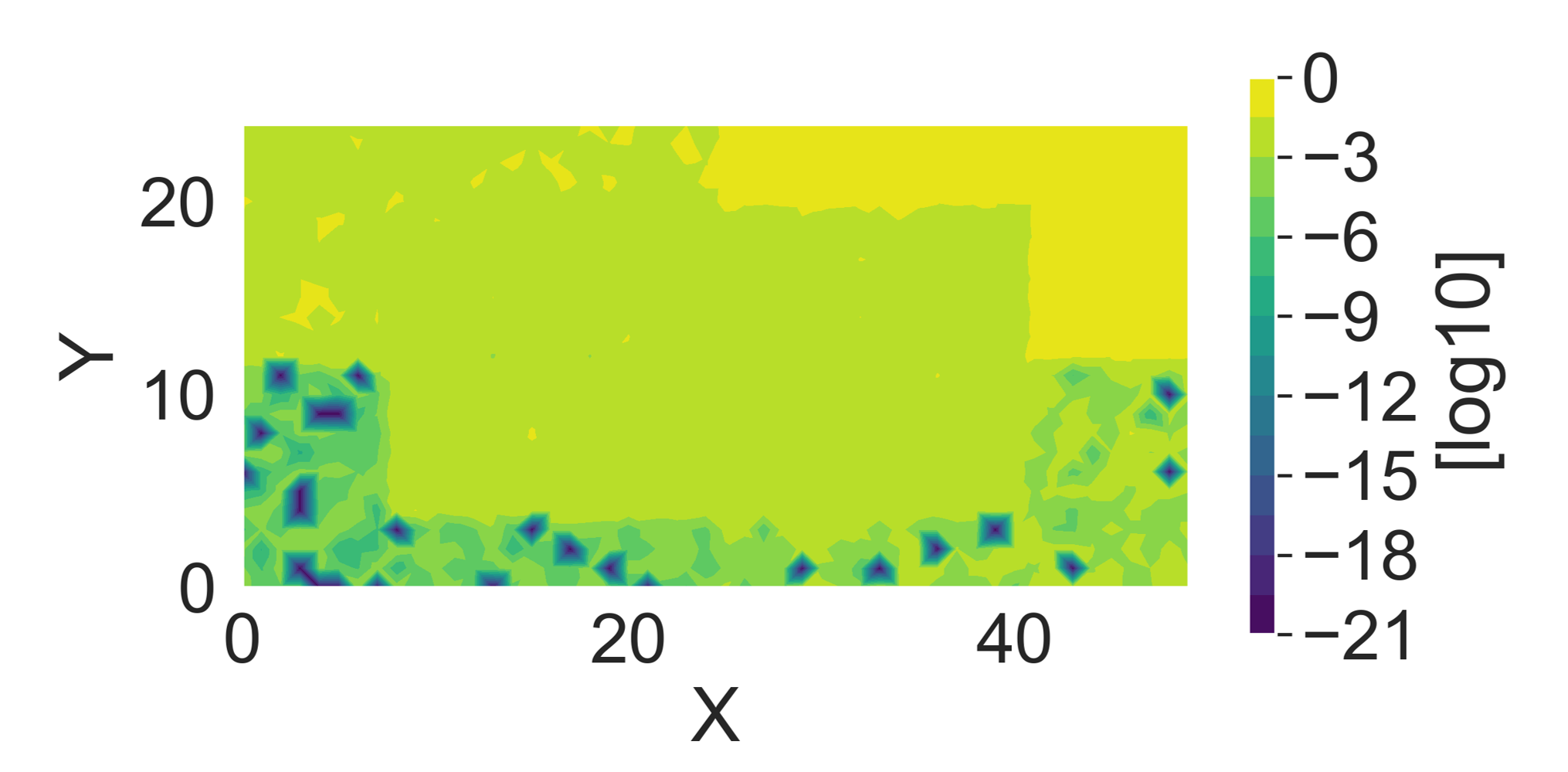}
    }
\end{minipage}

\vspace{0.2em}

\begin{minipage}{0.45\linewidth}
    \subfloat[KAMM (1st order), units: GPa$^2$]{
        \includegraphics[width=0.9\linewidth]{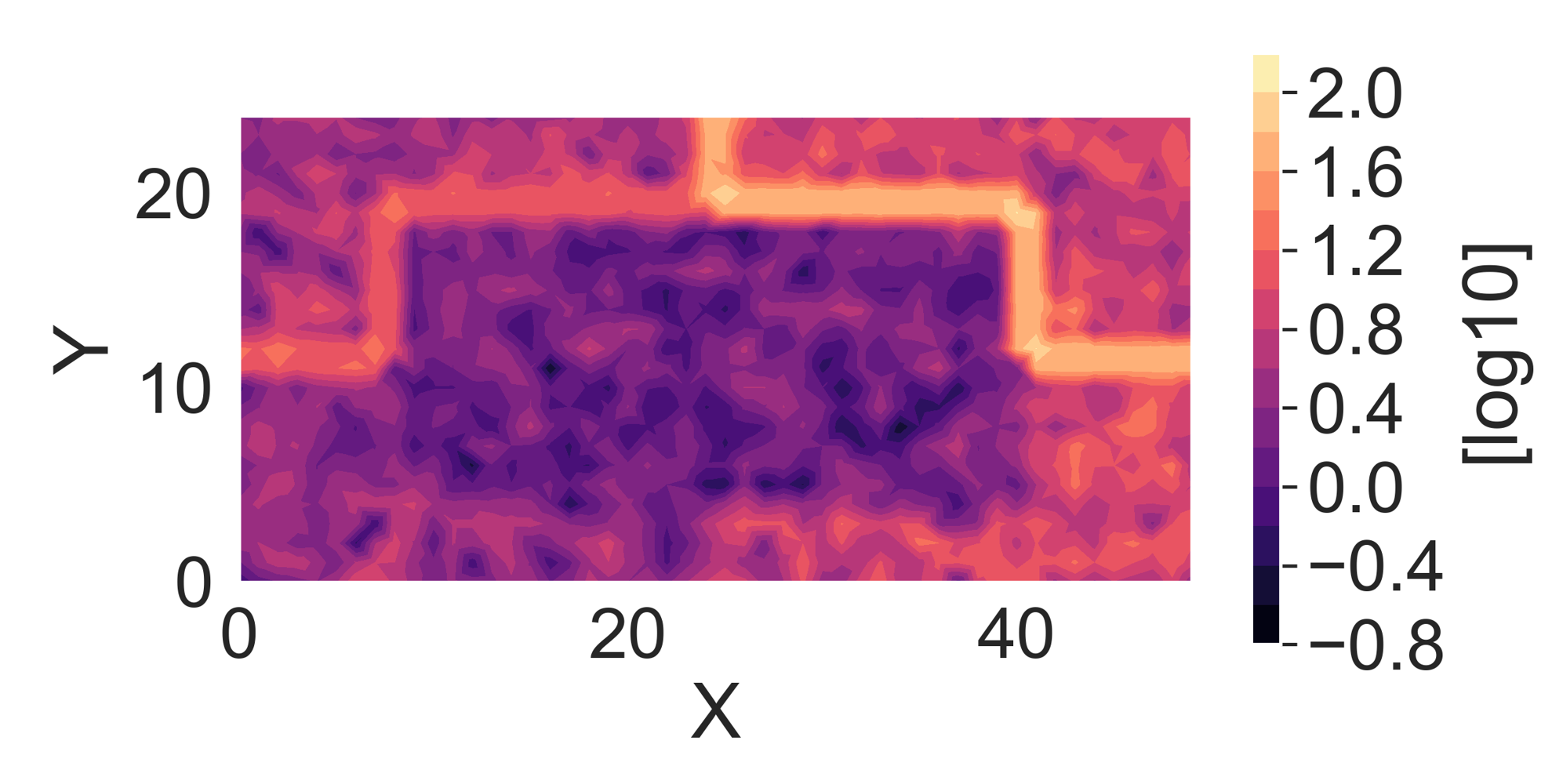}
    }
\end{minipage}
\hfill
\begin{minipage}{0.45\linewidth}
    \subfloat[KAMM (2nd order), units: GPa$^2$]{
        \includegraphics[width=0.9\linewidth]{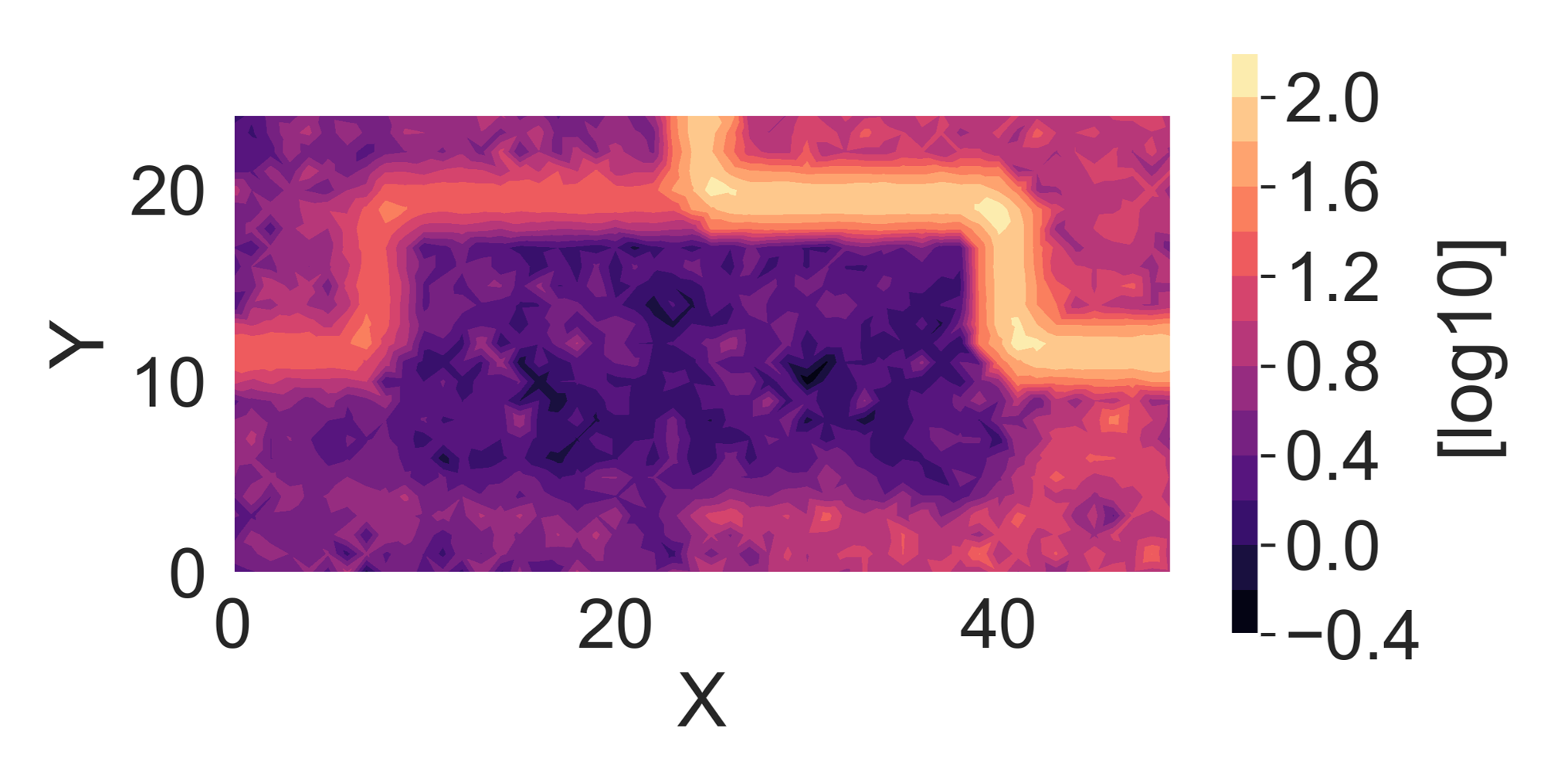}
    }
\end{minipage}

\vspace{0.2em}

\begin{minipage}{0.45\linewidth}
    \subfloat[KAPM (hardness mismatch), units: GPa]{
        \includegraphics[width=0.9\linewidth]{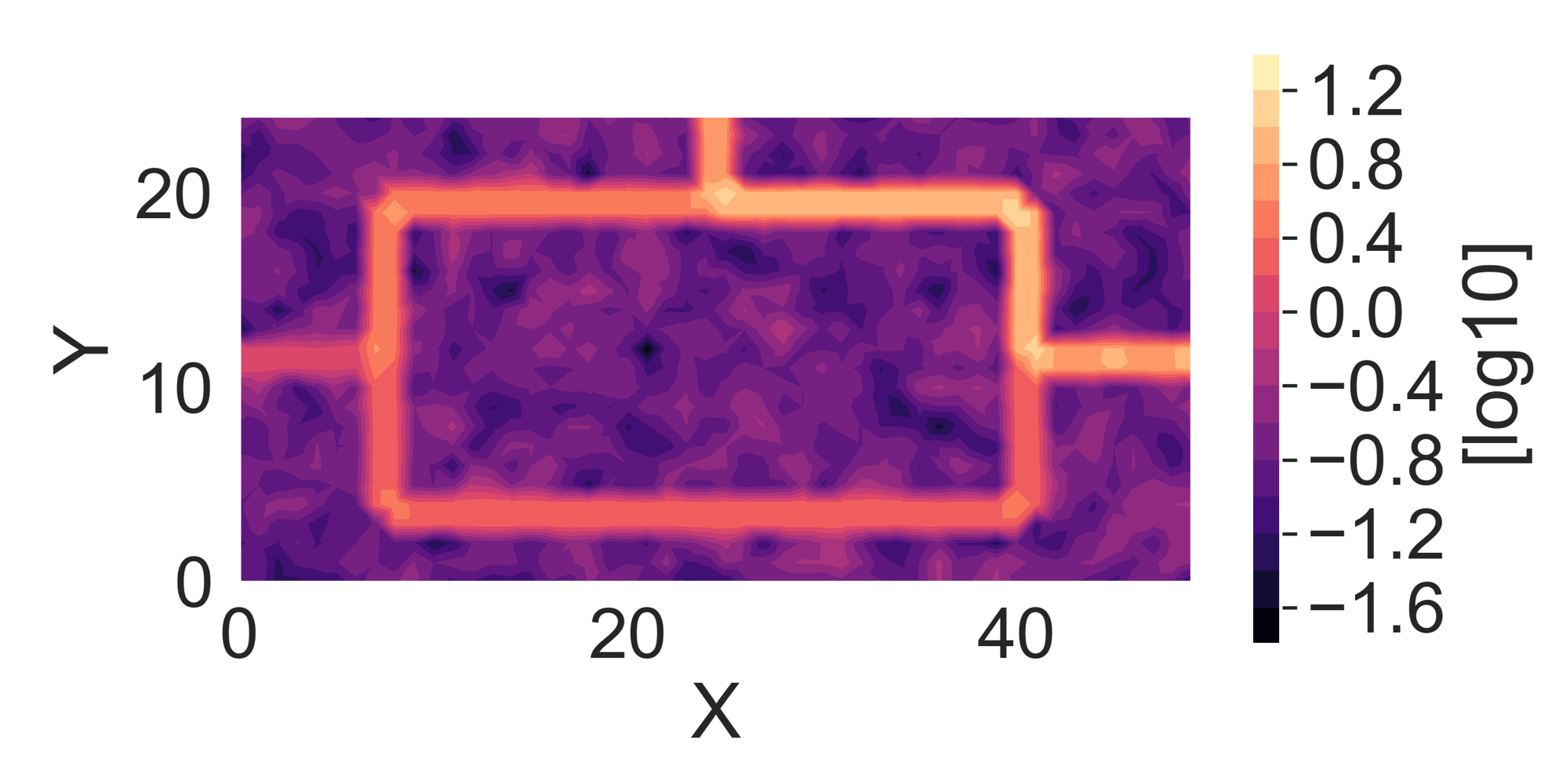}
    }
\end{minipage}
\hfill
\begin{minipage}{0.45\linewidth}
    \subfloat[KAEM (modulus mismatch), units: GPa]{
        \includegraphics[width=0.9\linewidth]{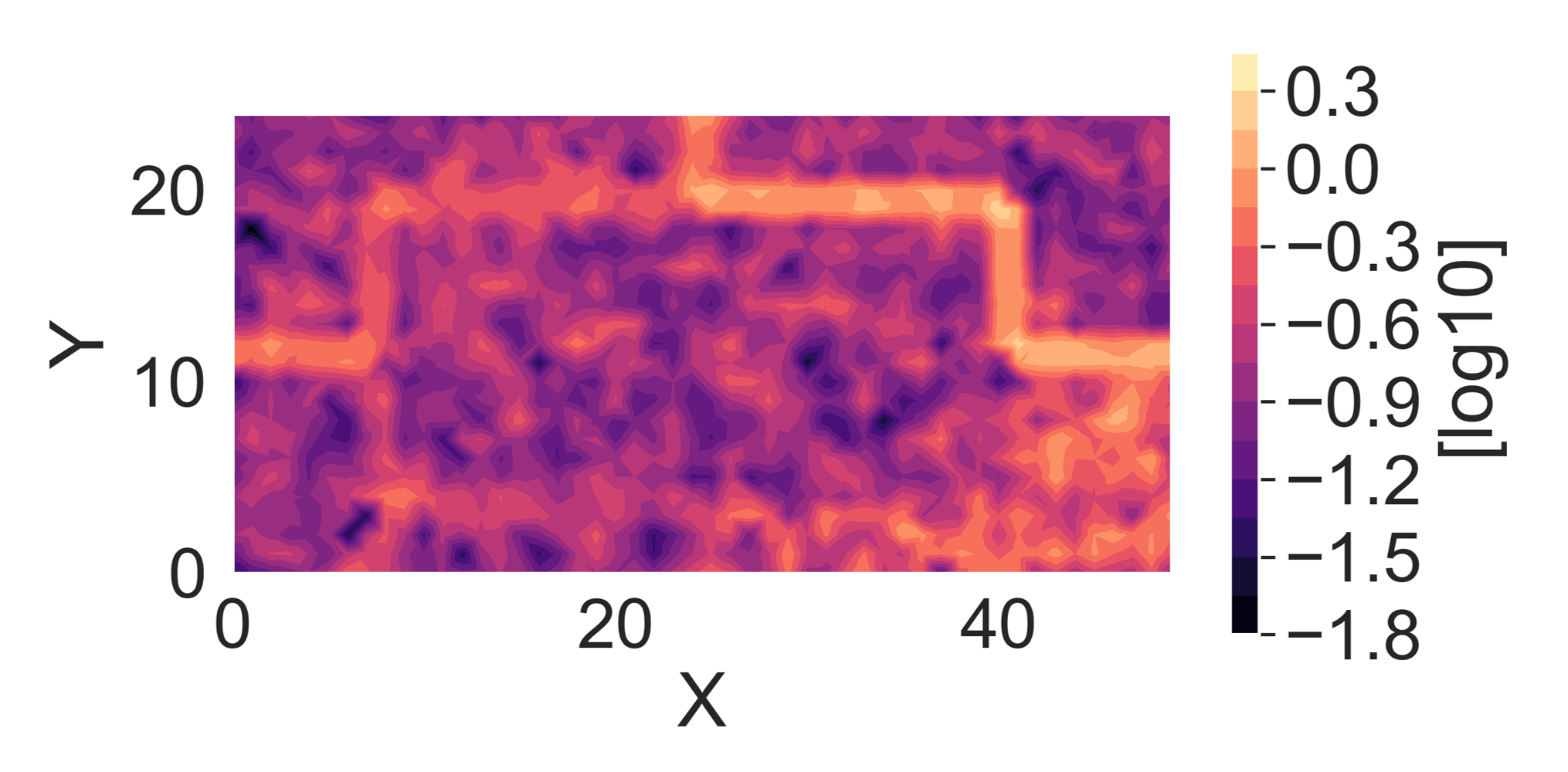}
    }
\end{minipage}

\vspace{0.2em}

\begin{minipage}{0.49\linewidth}
    \subfloat[KAMM\textsubscript{RATIO}, dimensionless]{
        \includegraphics[width=0.9\linewidth]{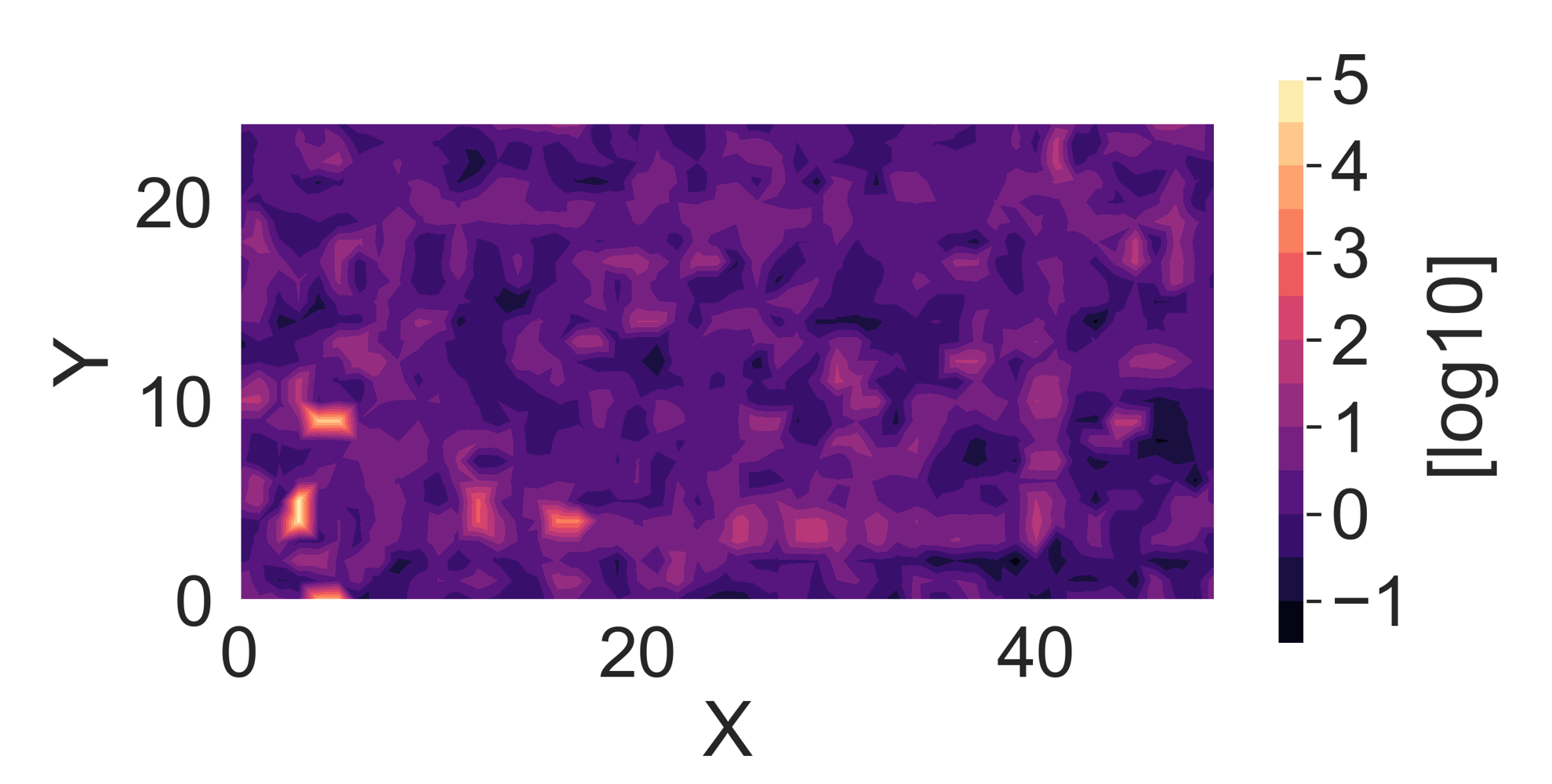}
    }
\end{minipage}
\hfill
\begin{minipage}{0.49\linewidth}
    \subfloat[KAMM\textsubscript{NORMPROD}, units: GPa]{
        \includegraphics[width=0.9\linewidth]{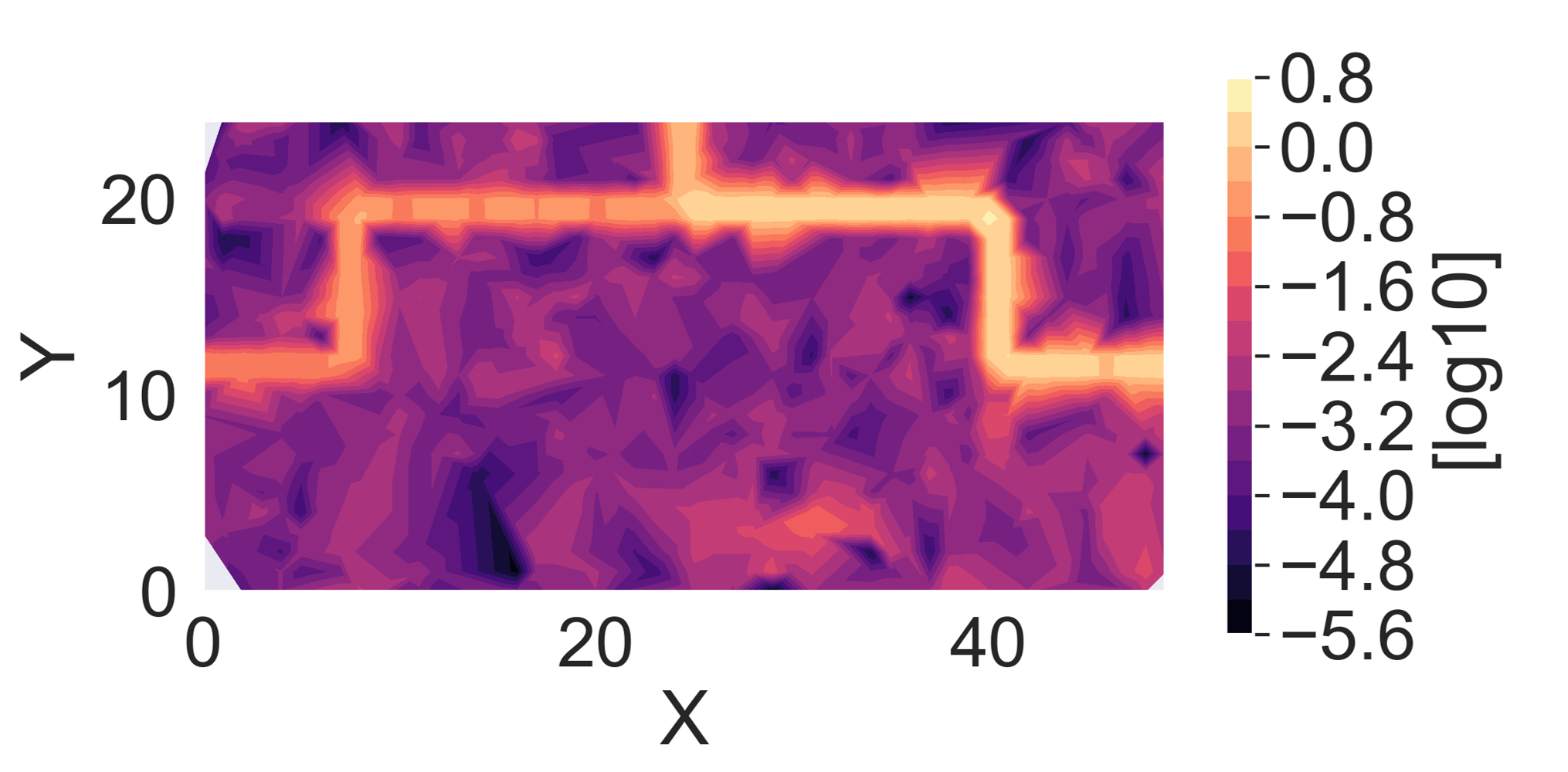}
    }
\end{minipage}

\caption{Local mechanical metrics computed on a representative composite indentation map. 
(a--b) Base mechanical properties. 
(c--d) Scalar mechanical ratios. 
(e) Full KAMM field showing boundary sensitivity. 
(f--g) Single-property mismatch components. 
(h--i) Alternative mismatch formulations. 
These features enhance clustering and interphase detection by capturing spatial and mechanical contrast.}
\label{fig:five_regions_reference}
\end{figure}

The choice of neighborhood impacts sensitivity: $\mathcal{O}_1$ captures sharp phase boundaries, while $\mathcal{O}_2$ offers smoother transitions and broader gradient detection.

To illustrate the behavior of each feature, Figure~\ref{fig:five_regions_reference} shows the computed maps for hardness ($H$), modulus ($E$), scalar ratios ($H/E$, $H^3/E^2$), and the various KAM-based mismatch metrics using the five-region synthetic specimen. The ratio maps ($H/E$ and $H^3/E^2$) highlight differences in mechanical response across regions but often fail to capture precise boundaries or subtle interfacial transitions. In contrast, spatial mismatch metrics such as KAMM, KAEM, and KAPM produce stronger signals at phase boundaries and are particularly effective in highlighting interfaces when mechanical contrast is high. KAMM\textsubscript{RATIO} offers additional sensitivity within relatively homogeneous regions, revealing substructure or subtle measurement variability.

Notably, KAMM\textsubscript{NORMPROD} displays enhanced responses near triple junctions or corners, suggesting it captures second-order variations and complex local gradients that are less apparent in the base KAMM field. However, both KAMM\textsubscript{RATIO} and KAMM\textsubscript{NORMPROD} may show missing values (i.e., blank pixels), especially in highly homogeneous regions or near specimen corners. These arise from undefined operations, such as division by zero or the absence of significant local variation, where neighboring values are too similar to yield a stable contrast signal.

\paragraph{Clustering pipeline and neighborhood-based integration}
To classify mechanically distinct regions in nanomechanical maps, we implemented a modular and reproducible clustering pipeline. After importing raw indentation data (CSV or Excel), the pipeline performs initial preprocessing, including header normalization, numerical cleaning, and optional computation of dimensionless mechanical ratios such as $H/E$ and $H^3/E^2$.

When neighborhood-based awareness is required, the pipeline computes kernel-based descriptors-specifically KAMM, KAEM, KAPM, KAMM\textsubscript{RATIO}, and KAMM\textsubscript{NORMPROD}-using either first- or second-order neighborhoods. These spatial features capture local mechanical mismatches and gradients, enriching the dataset beyond global $(E, H)$ values. By appending KAMM-based features to the $(E, H)$ feature space, the clustering process becomes both mechanically and neighborhood-informed. This enables the identification of interphase regions and ambiguous transition zones that may otherwise be misclassified using $(E, H)$ alone. As shown in Section~\ref{sec:results}, incorporating KAMM variants significantly improves clustering robustness-particularly in systems with low contrast or complex interfaces.

Following feature construction, all selected variables are standardized using z-score normalization implemented via the \textit{scikit-learn} library~\cite{scikit-learn, sklearn_api}. Dimensionality reduction and visualization are optionally performed using Principal Component Analysis (PCA), also from \textit{scikit-learn}. Clustering is carried out with four unsupervised algorithms provided by the same library: agglomerative clustering (hierarchical), DBSCAN (density-based), Gaussian Mixture Models (probabilistic), and KMeans (distance-based). The algorithms are applied in a consistent order to facilitate comparison across figures and datasets.

The number of clusters can be defined manually or estimated automatically (e.g., using the elbow method based on the KMeans inertia). In this study, the number of clusters is fixed to 2 for specimens a, d, and f (no interphase), and to 3 for all specimens containing interphases, ensuring a consistent comparison across the four algorithms. For DBSCAN, although the algorithm does not inherently require a predefined number of clusters, its hyperparameters ($\varepsilon$ and \textit{min\_samples}) are systematically tuned such that the number of detected non-noise clusters matches the prescribed target ($k = 2$ or $k = 3$), thereby enforcing methodological consistency across all clustering approaches.

\paragraph{Clustering evaluation and error analysis}

Cluster assignments are first mapped back to the spatial domain and projected into feature space to facilitate interpretation. Each cluster is associated with a reference phase according to ascending mean hardness, ensuring a physically consistent ordering under the assumption of monotonic mechanical contrast. For each identified phase, statistical summaries—such as volume fraction and the mean and standard deviation of $E$ and $H$—are computed and exported together with the corresponding visualizations.

For synthetic datasets, where ground-truth phase labels are available, clustering performance is primarily evaluated using supervised metrics. After mapping predicted clusters to reference phases via mean hardness ranking, we compute:

\begin{itemize}
    \item \textbf{Pixel-level accuracy:} the fraction of grid points for which the predicted cluster label matches the ground-truth phase label.
    \item \textbf{Phase-level property error:} the absolute deviation between predicted and true mean mechanical properties (hardness and elastic modulus) for each phase, normalized to the interval $[0,1]$.
\end{itemize}

These metrics are evaluated over a range of mechanical contrast ratios $(H_\text{ratio}, E_\text{ratio})$, clustering algorithms, and configurations with or without KAMM-based spatial features. When analyzing accuracy as a function of hardness contrast, the reported curves correspond to an average over all sampled elastic modulus contrast ratios. Specifically, for each value of $H_\text{ratio}$, pixel-level accuracy is computed as the mean accuracy across the full set of $E_\text{ratio}$ values. This aggregation isolates the influence of hardness contrast while marginalizing variations in elastic stiffness contrast.

In a second step, intrinsic clustering quality is assessed using unsupervised metrics, including the Silhouette score, Calinski--Harabasz index, and Davies--Bouldin index. These scores quantify cluster compactness and separation independently of ground truth, enabling systematic comparison across datasets and clustering strategies \cite{Rubinos_2024}.

\section{Results and Discussion}
\label{sec:results}

To evaluate the impact of KAMM features, we assessed the clustering accuracy of four standard algorithms (Agglomerative Clustering, DBSCAN, GMM, and KMeans) using synthetic composite specimens with progressively increasing interphase widths ($W = 0$, $0.03$, $0.05$). The input features were normalized prior to clustering, and no dimensionality reduction (e.g., PCA) was applied. Accuracy is reported as a function of the hardness contrast $H_\mathrm{ratio}$, with solid curves representing KAMM (1st-order)–enhanced clustering and dashed curves indicating the baseline $(E,H)$-only approach. Different colors denote the various interphase widths, enabling direct comparison of algorithm performance under increasing boundary complexity. This procedure is used consistently throughout the next three sections, with synthetic specimens.

\subsection{Case of rectangular composite with sharp interface - Effect of interphase presence on clustering accuracy}
\label{sec:results_interphasewidth}

\begin{figure}[htbp]
\centering
\includegraphics[width=0.7\textwidth]{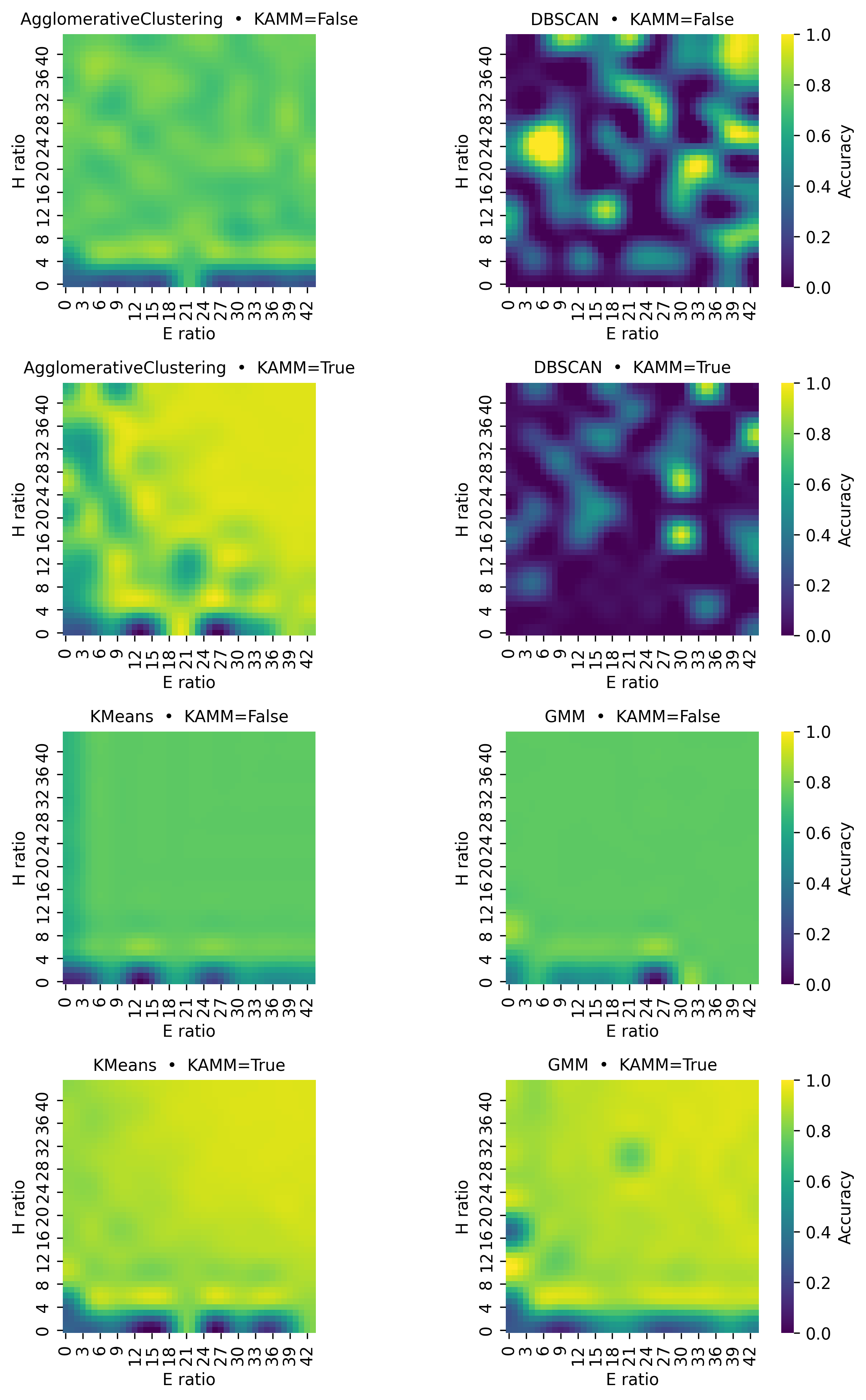}
\caption{Three-dimensional accuracy maps showing pixel-level clustering accuracy as a function of both hardness contrast ($H_\mathrm{ratio}$) and elastic modulus contrast ($E_\mathrm{ratio}$) for the different clustering algorithms. These maps provide the full contrast-dependent behavior, from which the averaged 2D profiles in Fig.~\ref{fig:kamm_accuracy_noise} are derived.}
\label{fig:accuracy_matrix}
\end{figure}

Figure~\ref{fig:accuracy_matrix} presents the full three-dimensional (3D) accuracy landscapes, where pixel-level clustering accuracy is mapped as a function of both hardness contrast ($H_\mathrm{ratio}$) and elastic modulus contrast ($E_\mathrm{ratio}$). These maps provide the complete contrast-dependent behavior for each algorithm. From these surfaces, two-dimensional (2D) accuracy profiles are obtained by averaging over all sampled $E_\mathrm{ratio}$ values. The resulting curves, shown in Fig.~\ref{fig:kamm_accuracy_noise}, therefore represent hardness-dependent projections of the full 3D accuracy maps and summarize the global trend with respect to mechanical contrast. Overall, these results indicate that incorporating KAMM-based spatial descriptors generally improves clustering performance for most algorithms, particularly centroid- and distribution-based methods such as KMeans and GMM.

Quantitatively, the mean pixel-level accuracy of KMeans increases from 0.689 to 0.823 with KAMM, while GMM improves from 0.737 to 0.833. This behavior is consistent with the underlying assumptions of these methods: both rely on global structure in feature space (cluster centroids or Gaussian components), and the addition of spatial descriptors enhances inter-phase separability and cluster coherence. In contrast, DBSCAN exhibits reduced performance when KAMM features are included. As a density-based method, its clustering behavior is sensitive to changes in feature-space geometry, and the inclusion of neighborhood-derived descriptors can weaken density contrasts between phases. Accordingly, DBSCAN generally performs better without KAMM, particularly at higher $H$ ratios and larger interphase widths. Agglomerative clustering shows moderate improvement with KAMM, suggesting that hierarchical linkage criteria benefit from enhanced spatial coherence while remaining less sensitive to density distortions than DBSCAN.

From the intrinsic clustering metrics averaged over all contrast ratios, KMeans without KAMM provides the highest Silhouette score (0.047) and Calinski-Harabasz index (529.8), while Agglomerative clustering yields the lowest Davies-Bouldin index (1.85), indicating the most compact and well-separated clusters in feature space. When KAMM is included, these internal metrics decrease for all centroid- and linkage-based methods: for KMeans, the Silhouette score drops from 0.047 to -0.013, the Calinski-Harabasz index decreases from 529.8 to 34.7, and the Davies-Bouldin index increases from 5.58 to 24.48. GMM shows a similar reduction in Calinski-Harabasz (371.3 to 85.1) and increase in Davies-Bouldin (5.04 to 26.36). DBSCAN exhibits consistently negative Silhouette values (-0.339 without KAMM and -0.503 with KAMM), reflecting limited cluster compactness under density-based assumptions. Overall, although KAMM improves pixel-level classification accuracy, it reduces standard unsupervised compactness and separation indices across methods.

Finally, no systematic influence of interphase width on clustering accuracy is observed within the investigated range, indicating that diffuse interfaces do not significantly compromise phase separability in the adopted feature representations.

\begin{figure}[htbp]
\centering

\begin{minipage}{0.45\textwidth}
    \subfloat[Agglomerative Clustering]{
        \includegraphics[width=\linewidth]{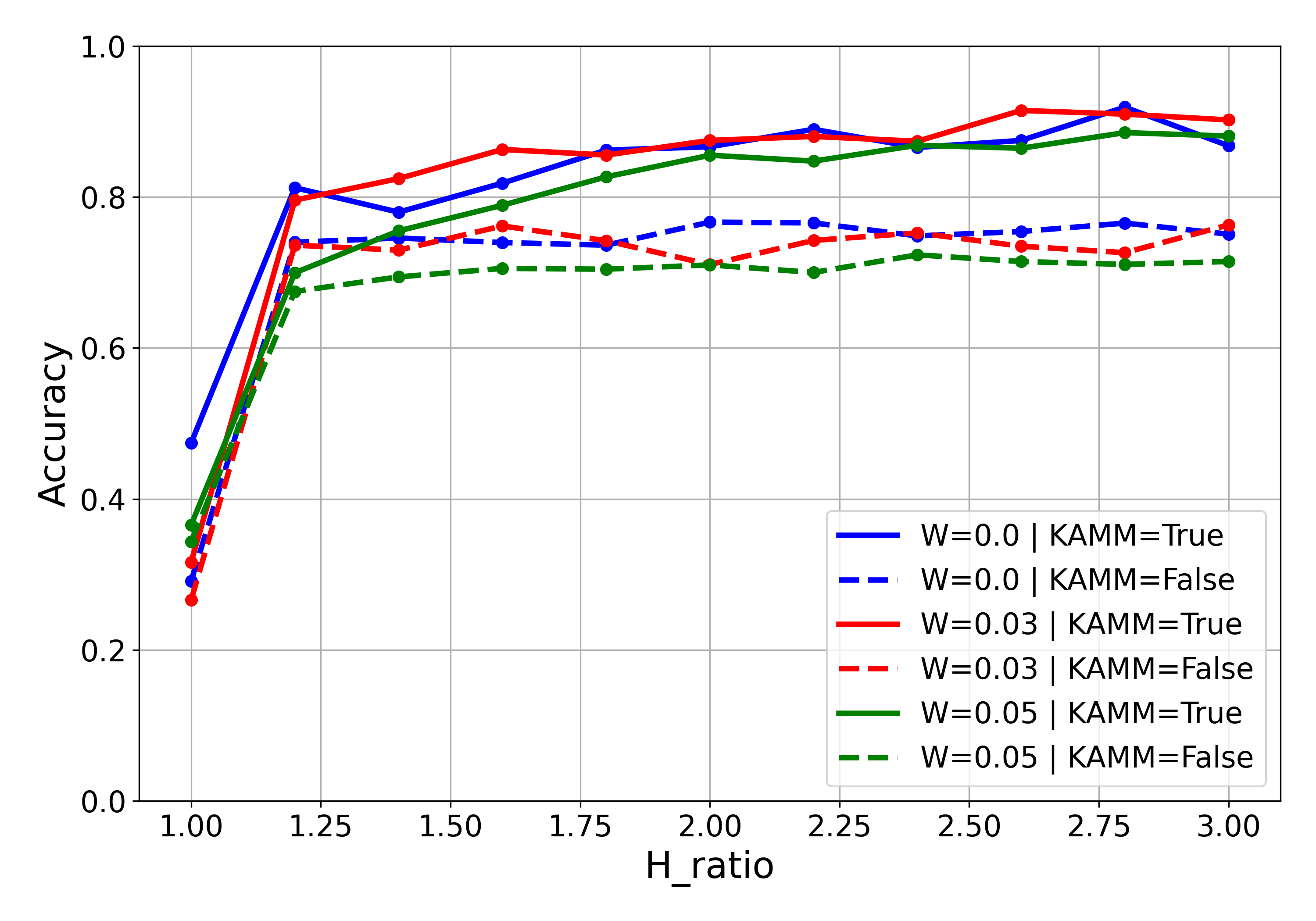}
    }
\end{minipage}
\hfill
\begin{minipage}{0.45\textwidth}
    \subfloat[DBSCAN]{
        \includegraphics[width=\linewidth]{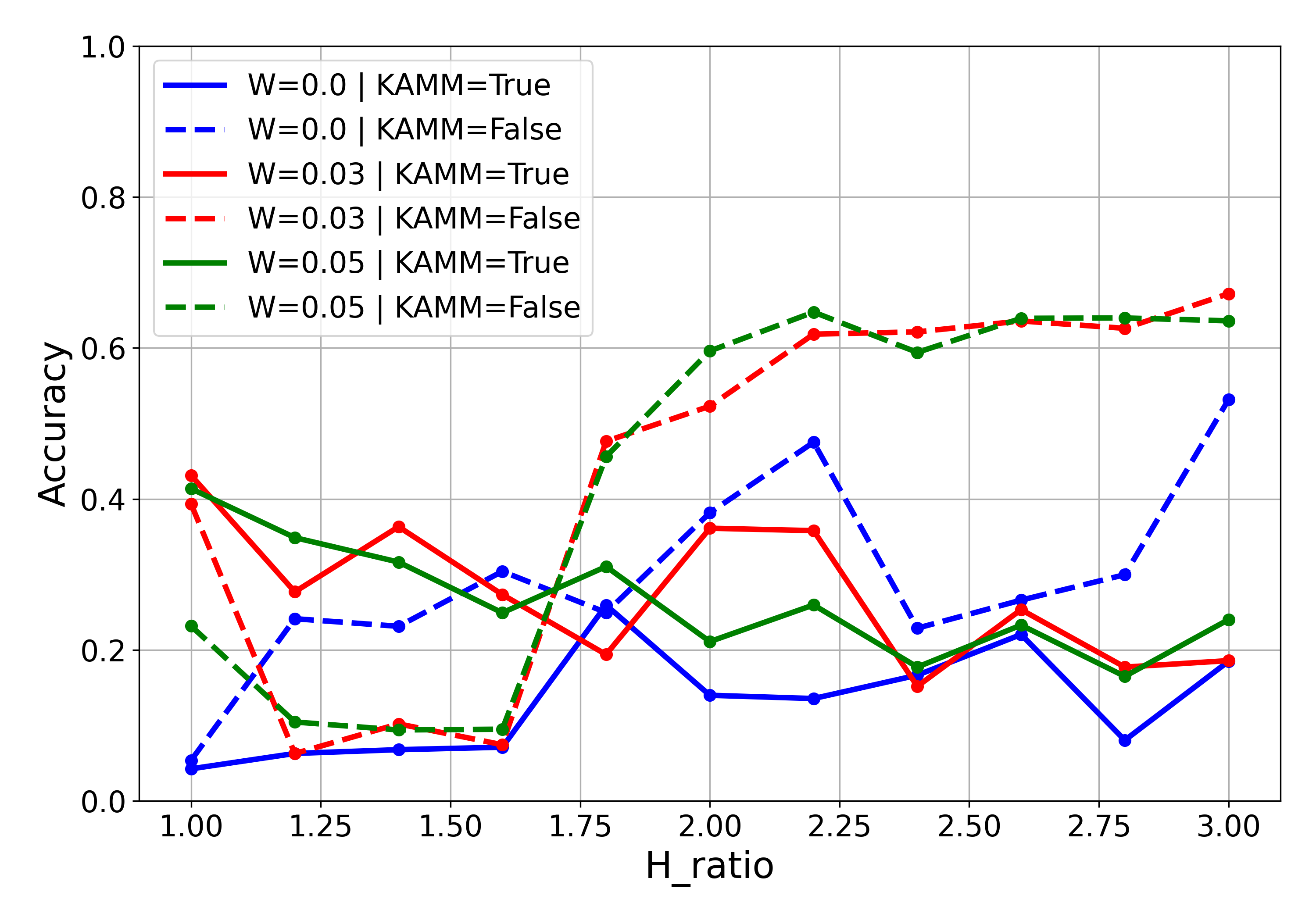}
    }
\end{minipage}

\vspace{0.3cm}

\begin{minipage}{0.45\textwidth}
    \subfloat[GMM]{
        \includegraphics[width=\linewidth]{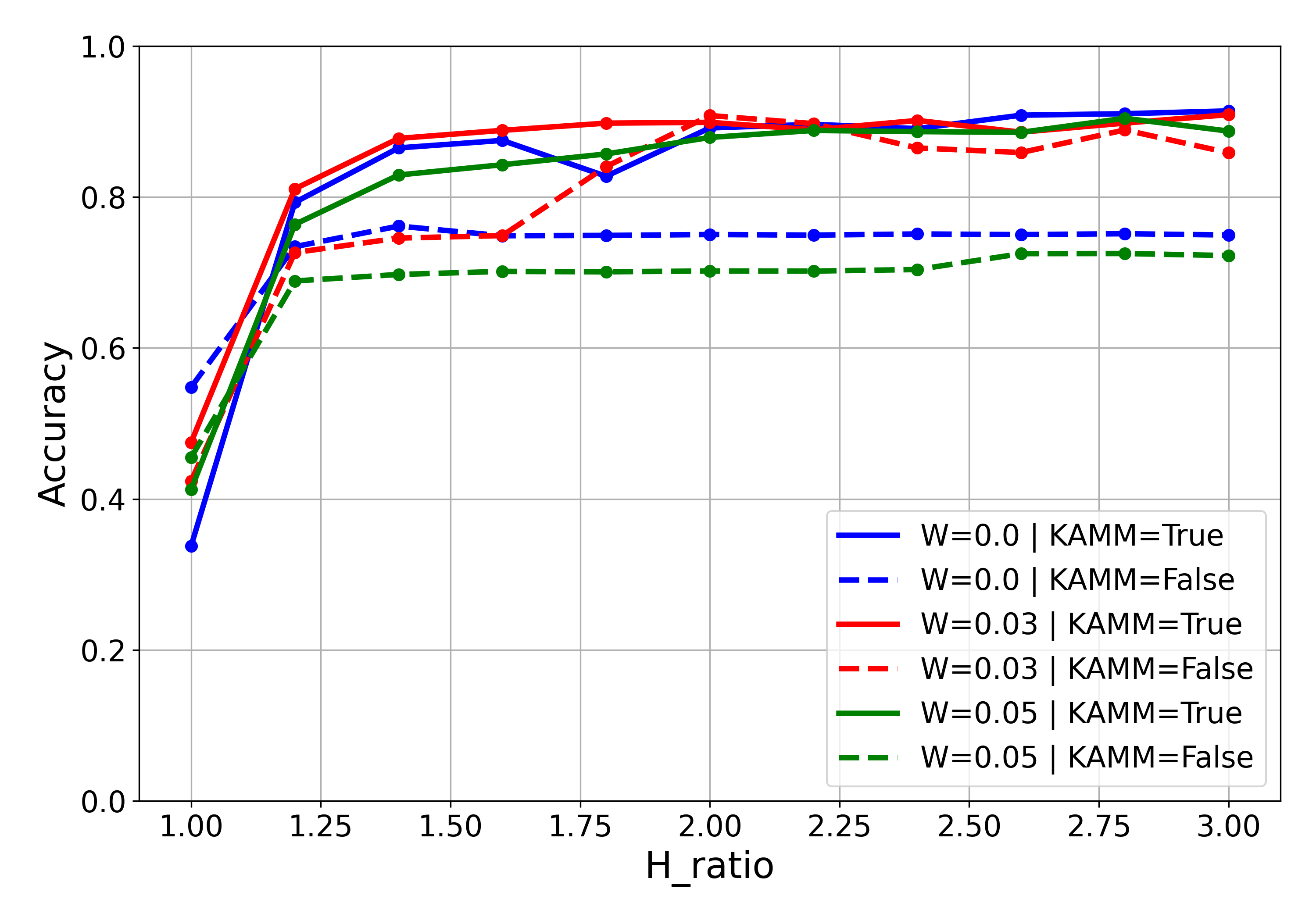}
    }
\end{minipage}
\hfill
\begin{minipage}{0.45\textwidth}
    \subfloat[KMeans]{
        \includegraphics[width=\linewidth]{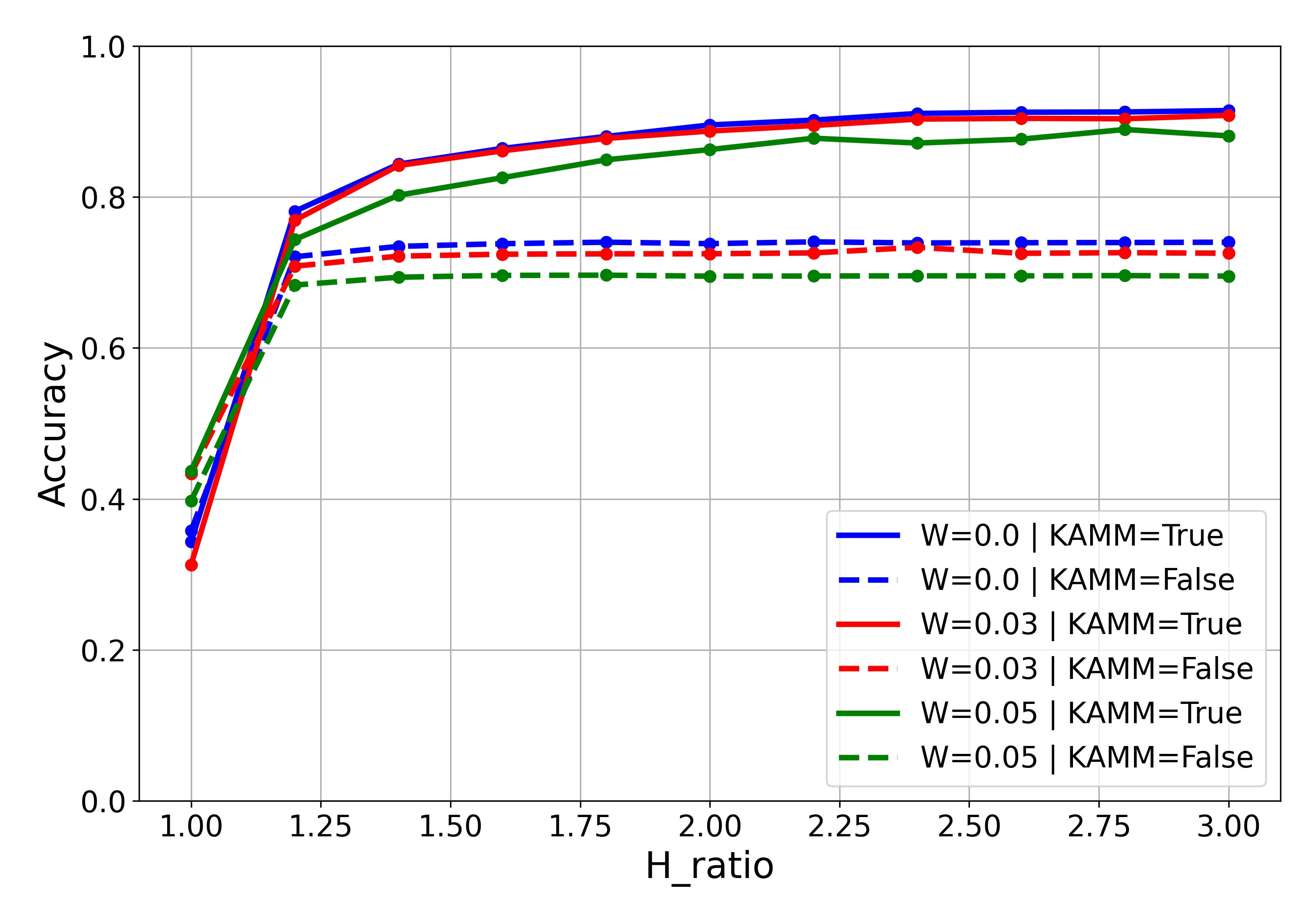}
    }
\end{minipage}

\vspace{0.4cm}

\caption{Pixel-level clustering accuracy as a function of hardness contrast ($H_\mathrm{ratio}$) for the rectangular composite specimen. Solid lines denote clustering with KAMM-based spatial descriptors, whereas dashed lines correspond to baseline clustering using only $(E,H)$ features.}
\label{fig:kamm_accuracy_comparison}
\end{figure}

\subsection{Case of rectangular composite with sharp interface - Effect of noise on clustering accuracy}
\label{sec:results_noise}

\begin{figure}[htbp]
\centering

\begin{minipage}{0.45\textwidth}
    \subfloat[Agglomerative Clustering]{
        \includegraphics[width=\linewidth]{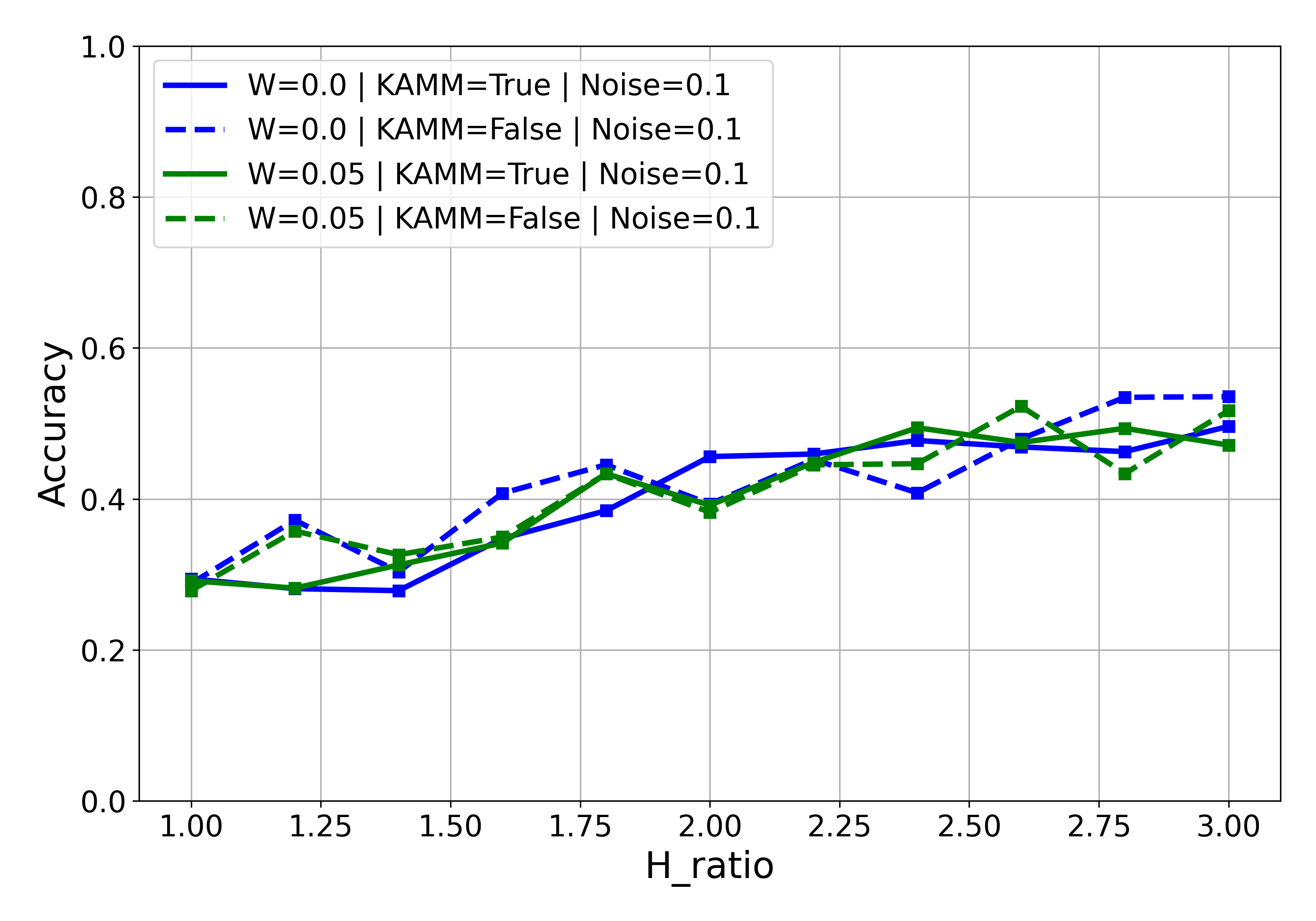}
    }
\end{minipage}
\hfill
\begin{minipage}{0.45\textwidth}
    \subfloat[DBSCAN]{
        \includegraphics[width=\linewidth]{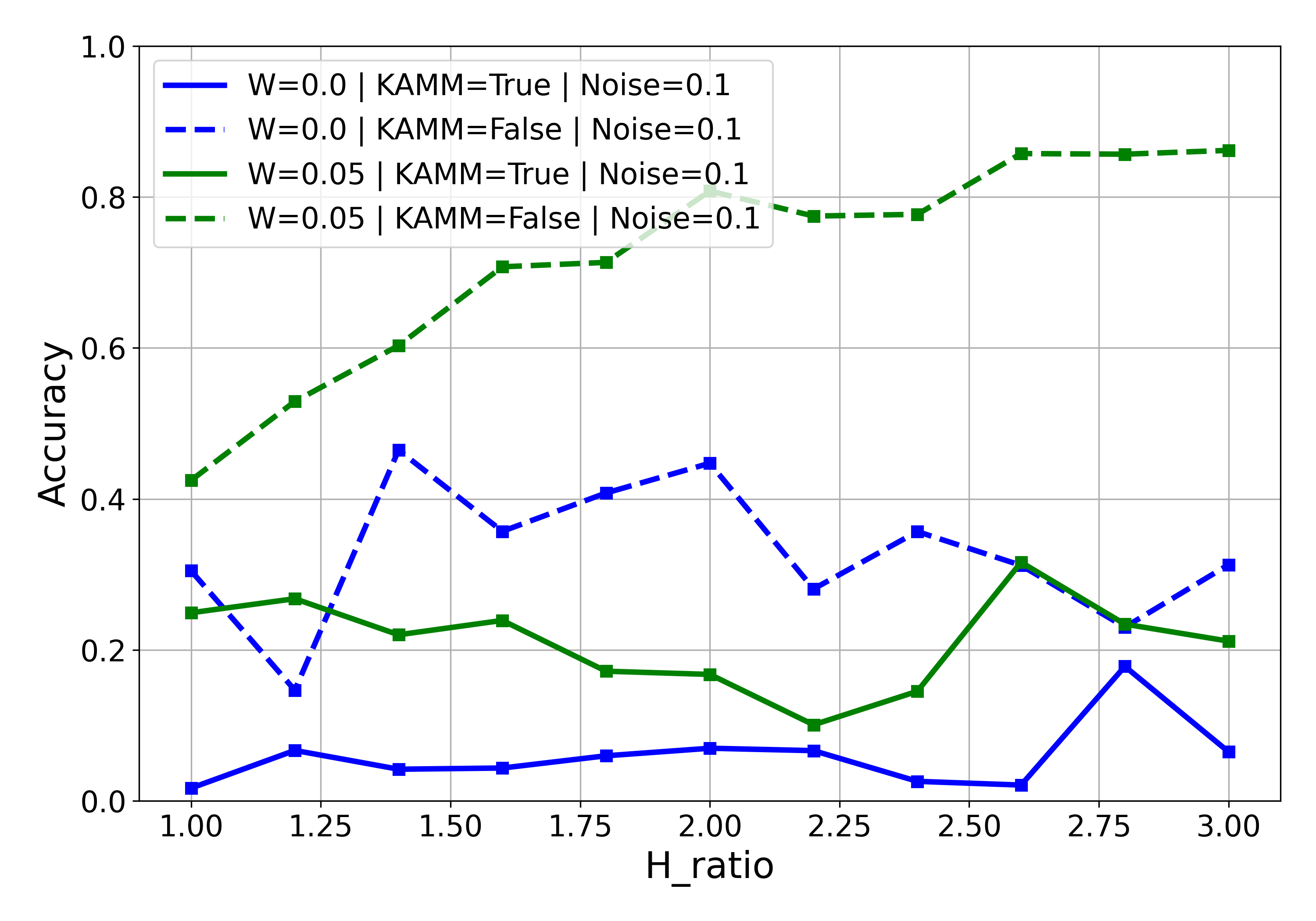}
   }
\end{minipage}

\vspace{0.3cm}

\begin{minipage}{0.45\textwidth}
    \subfloat[GMM]{
        \includegraphics[width=\linewidth]{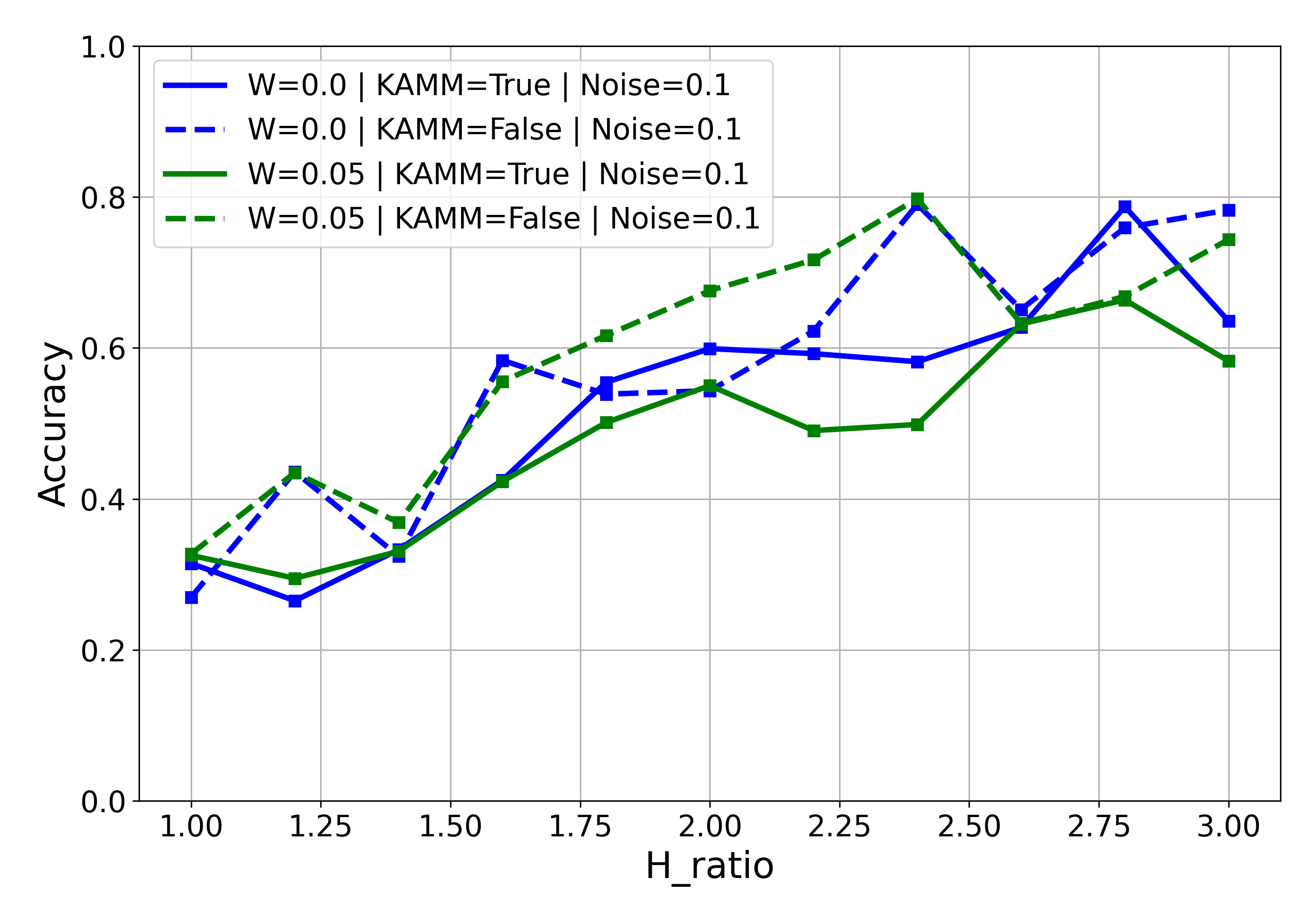}
    }
\end{minipage}
\hfill
\begin{minipage}{0.45\textwidth}
    \subfloat[KMeans]{
        \includegraphics[width=\linewidth]{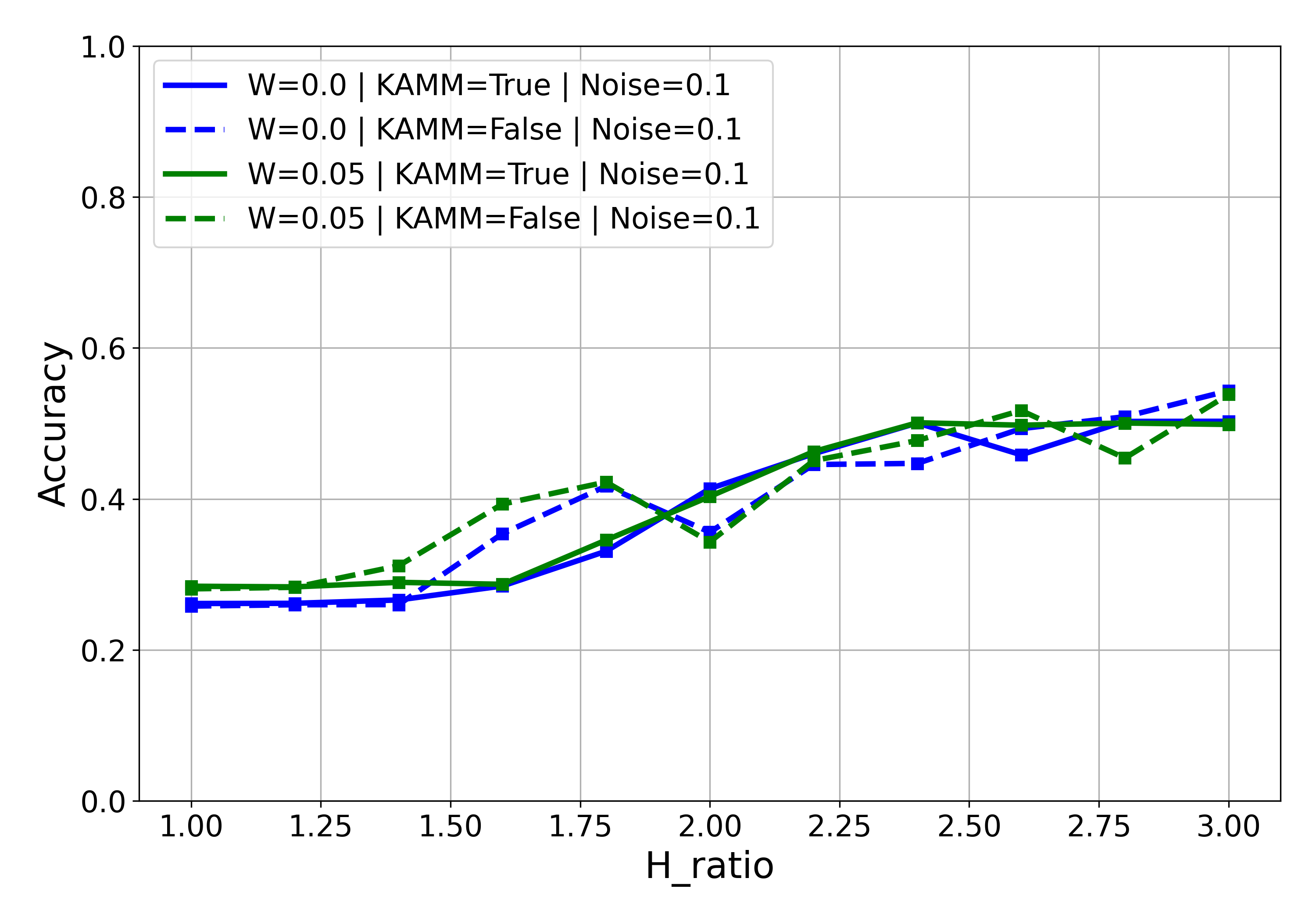}
    }
\end{minipage}

\vspace{0.4cm}

\caption{Pixel-level clustering accuracy vs.\ hardness contrast ($H_\mathrm{ratio}$) for the rectangular composite specimen with 10\% synthetic noise added. Solid lines: clustering with KAMM; dashed lines: baseline $(E,H)$-only clustering.}
\label{fig:kamm_accuracy_noise}
\end{figure}

Figure~\ref{fig:kamm_accuracy_noise} shows the accuracy curves obtained under the noisy condition for the four clustering algorithms considered in this study. Each plot compares the performance with and without the KAMM descriptor for $W=0$ and $W=0.05$.

For the Agglomerative clustering method, the introduction of noise leads to a drastic loss of accuracy across all $H$ ratios. Moreover, the curves obtained with and without KAMM are nearly indistinguishable, indicating that KAMM provides no measurable benefit under noisy conditions. The method appears highly sensitive to perturbations and rapidly loses meaningful clustering capability when noise is present. For DBSCAN, the behavior differs markedly. In the presence of noise, DBSCAN generally performs better without KAMM across most contrast ratios, particularly when the interphase width approaches zero. The inclusion of the KAMM descriptor appears to degrade density-based clustering under noisy conditions, likely due to the alteration of local density structure in feature space. For both GMM and KMeans, noise substantially reduces accuracy, and the curves with and without KAMM remain very similar. This suggests that these centroid- and model-based methods become dominated by noise, rendering the contribution of the KAMM descriptor largely ineffective. Their overall accuracy remains significantly lower compared to the noise-free case. Overall, under noisy conditions, the four methods exhibit distinct sensitivities to the KAMM descriptor: Agglomerative clustering, GMM, and KMeans experience strong performance degradation regardless of KAMM, whereas DBSCAN performs more robustly when KAMM is excluded, particularly at low $H$ ratios.

\subsection{Case of rectangular composite with graded interface}
\label{sec:results_rectComposite}

\begin{figure}[htbp]
\centering

\begin{minipage}{0.45\textwidth}
    \subfloat[Agglomerative Clustering]{
        \includegraphics[width=\linewidth]{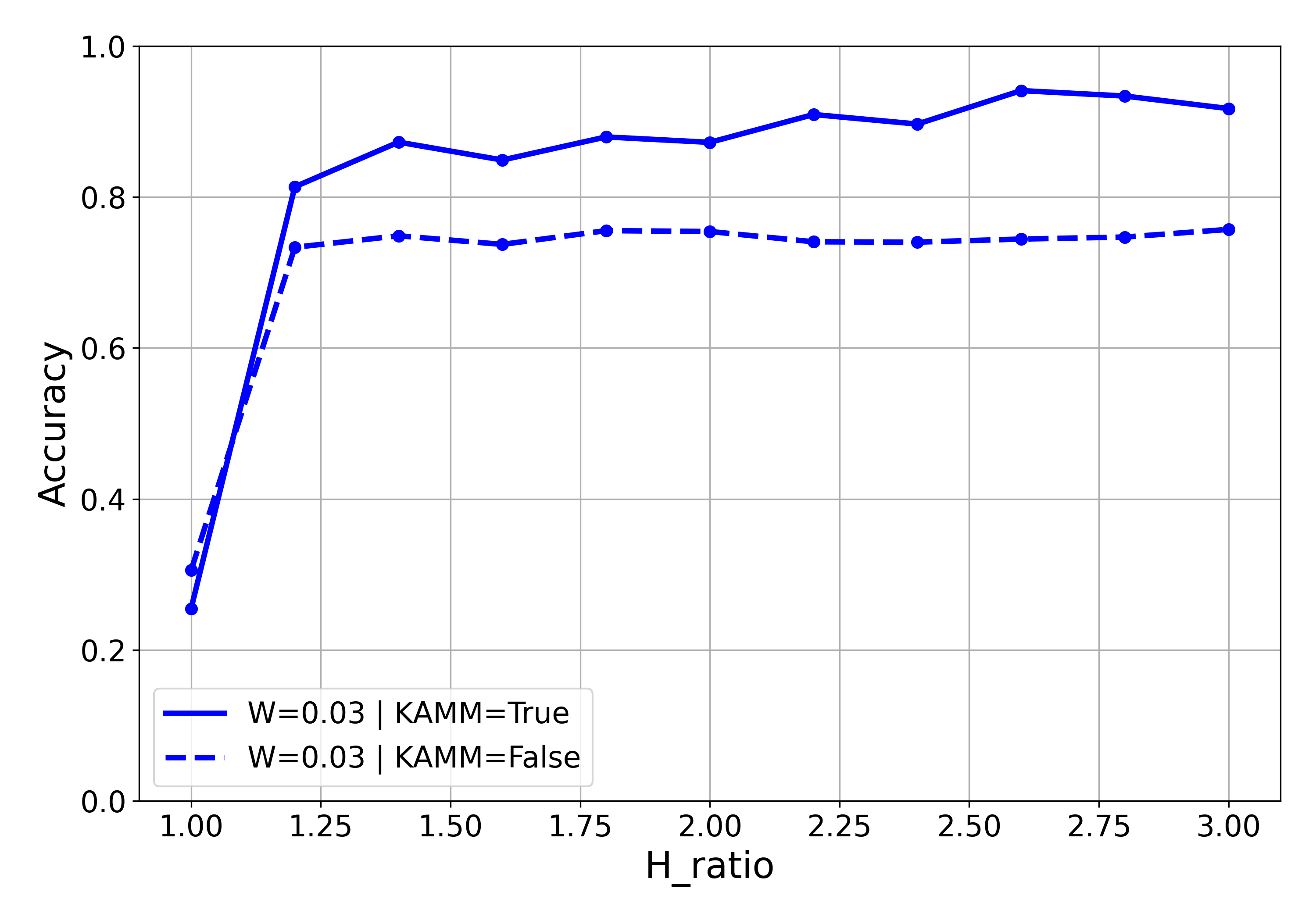}
    }
\end{minipage}
\hfill
\begin{minipage}{0.45\textwidth}
    \subfloat[DBSCAN]{
        \includegraphics[width=\linewidth]{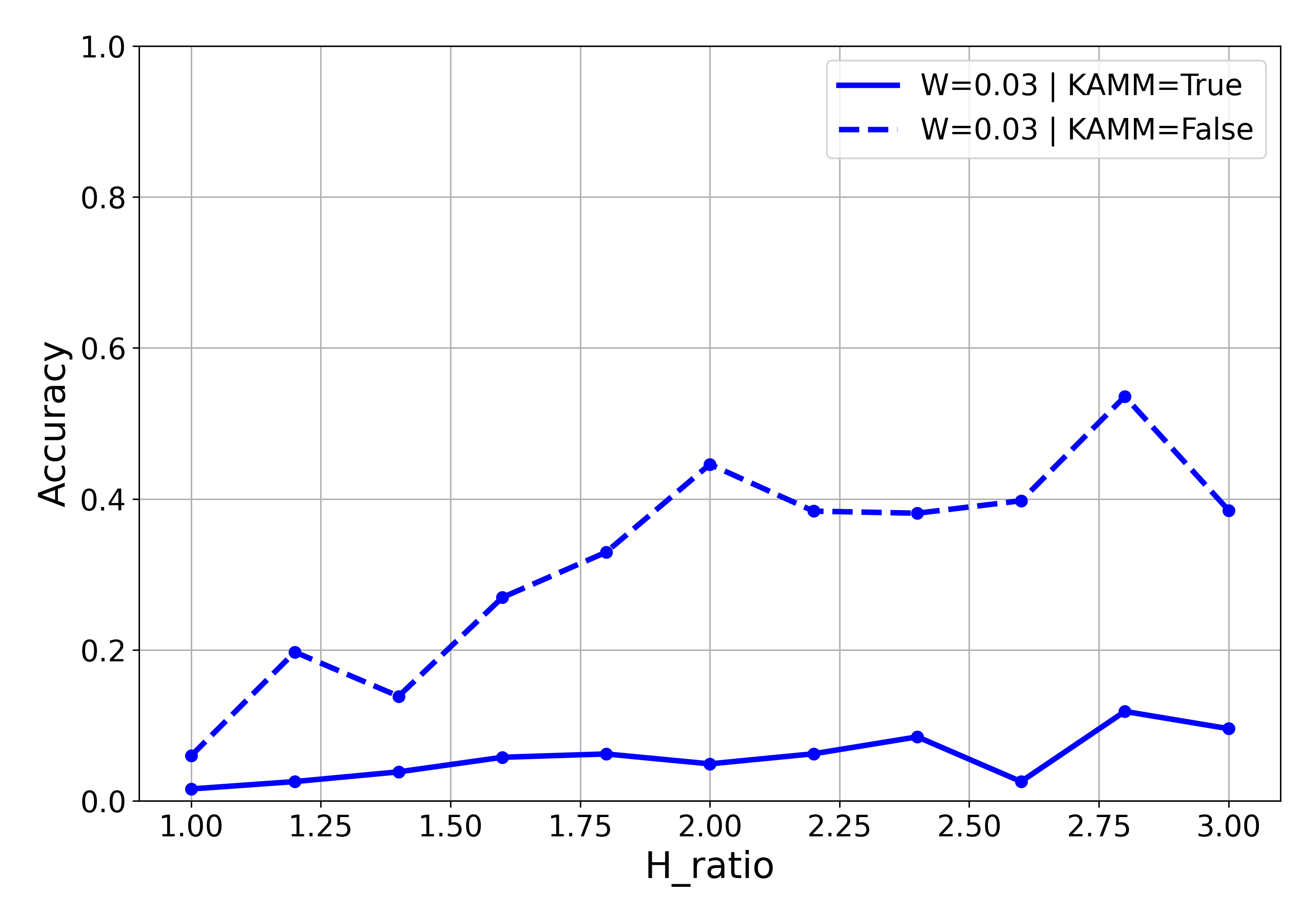}
    }
\end{minipage}

\vspace{0.3cm}

\begin{minipage}{0.45\textwidth}
    \subfloat[GMM]{
        \includegraphics[width=\linewidth]{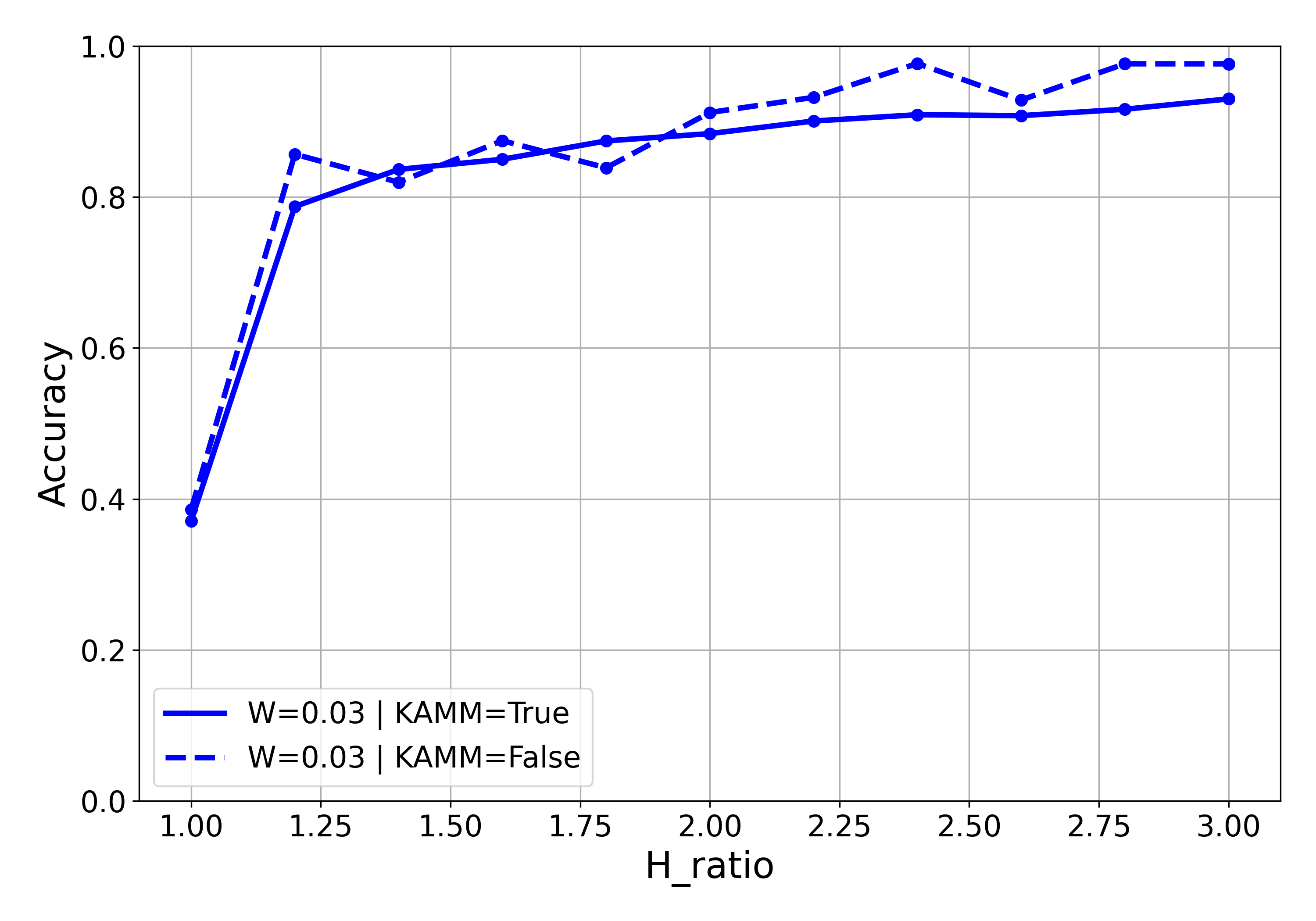}
    }
\end{minipage}
\hfill
\begin{minipage}{0.45\textwidth}
    \subfloat[KMeans]{
        \includegraphics[width=\linewidth]{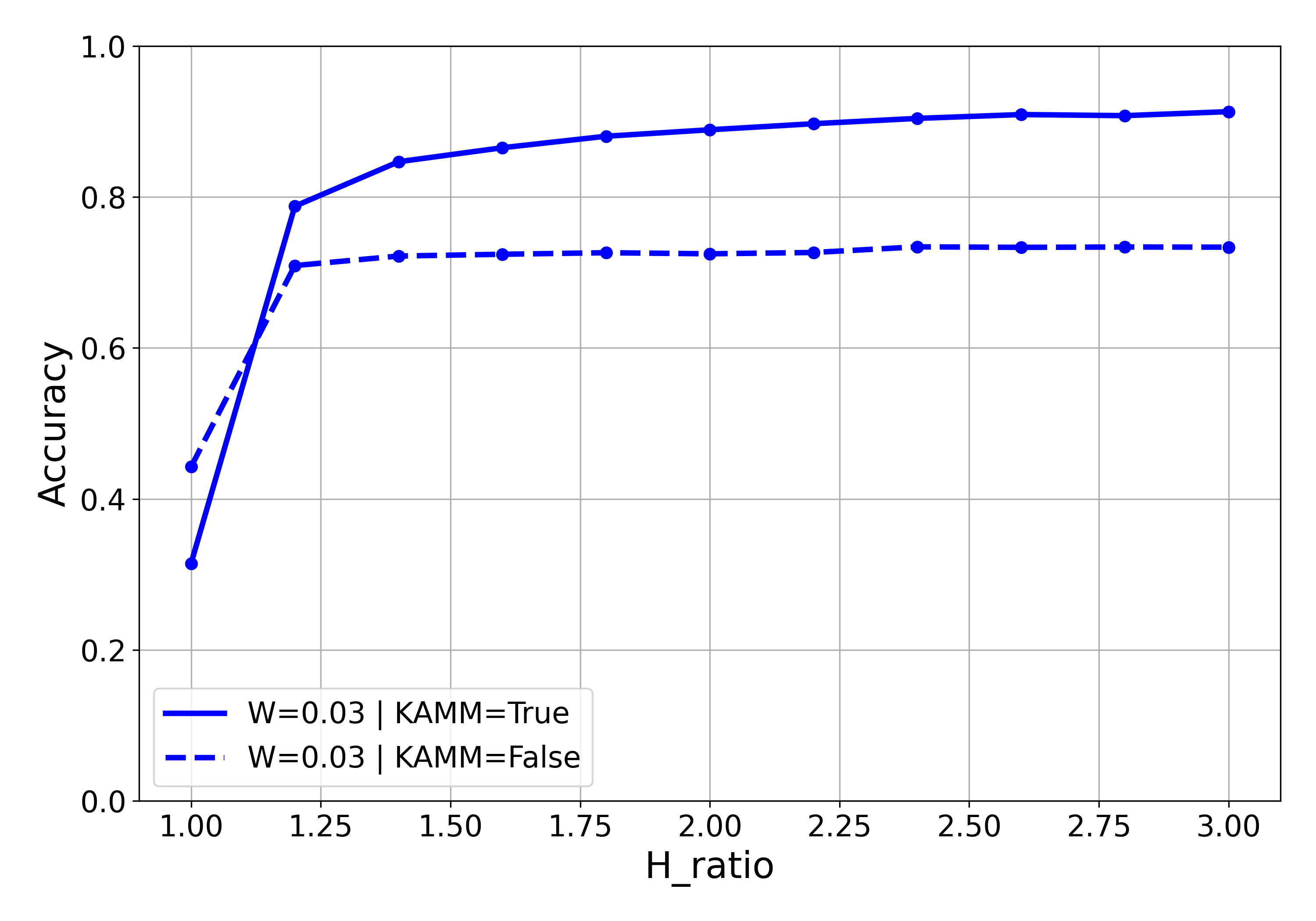}
    }
\end{minipage}

\vspace{0.4cm}

\caption{
Clustering accuracy at pixel level vs.\ hardness contrast ($H_\mathrm{ratio}$) for graded interphase ($W=0.03$) in the case of rectangular composite. Solid lines: clustering with KAMM; dashed lines: $(E,H)$-only baseline.
}
\label{fig:Clustering_GradedInterphase}
\end{figure}

Figure~\ref{fig:Clustering_GradedInterphase} presents the pixel-level clustering accuracy as a function of hardness contrast ($H_\mathrm{ratio}$) for the rectangular composite with a graded interphase ($W = 0.03$). The impact of the KAMM descriptor varies significantly across the different clustering algorithms. Agglomerative clustering exhibits a clear improvement when KAMM is included, with consistently higher accuracy across the investigated hardness contrasts. This suggests that hierarchical linkage criteria benefit from the enhanced spatial coherence introduced by the KAMM descriptor in the presence of a diffuse interface. In contrast, DBSCAN shows its lowest performance when KAMM features are incorporated. The density-based formulation appears to be negatively affected by the additional neighborhood-derived information, leading to reduced phase separability compared to the $(E,H)$-only baseline. GMM displays nearly identical performance with and without KAMM, indicating that the probabilistic modeling of the feature distribution is largely insensitive to the added spatial descriptor in this graded configuration. KMeans, on the other hand, demonstrates a noticeable improvement with KAMM, particularly at intermediate and high $H_\mathrm{ratio}$ values. The inclusion of spatial information enhances cluster compactness and phase discrimination for this centroid-based method. Overall, for the graded interphase case, KAMM clearly benefits centroid- and linkage-based approaches (KMeans and Agglomerative), has negligible influence on GMM, and degrades the performance of the density-based DBSCAN algorithm.

\subsection{Case of matrix-fibers specimens}
\label{sec:results_matrix-particles}

\begin{figure}[htbp]
\centering

\begin{minipage}{0.45\textwidth}
    \subfloat[Agglomerative Clustering]{
        \includegraphics[width=\linewidth]{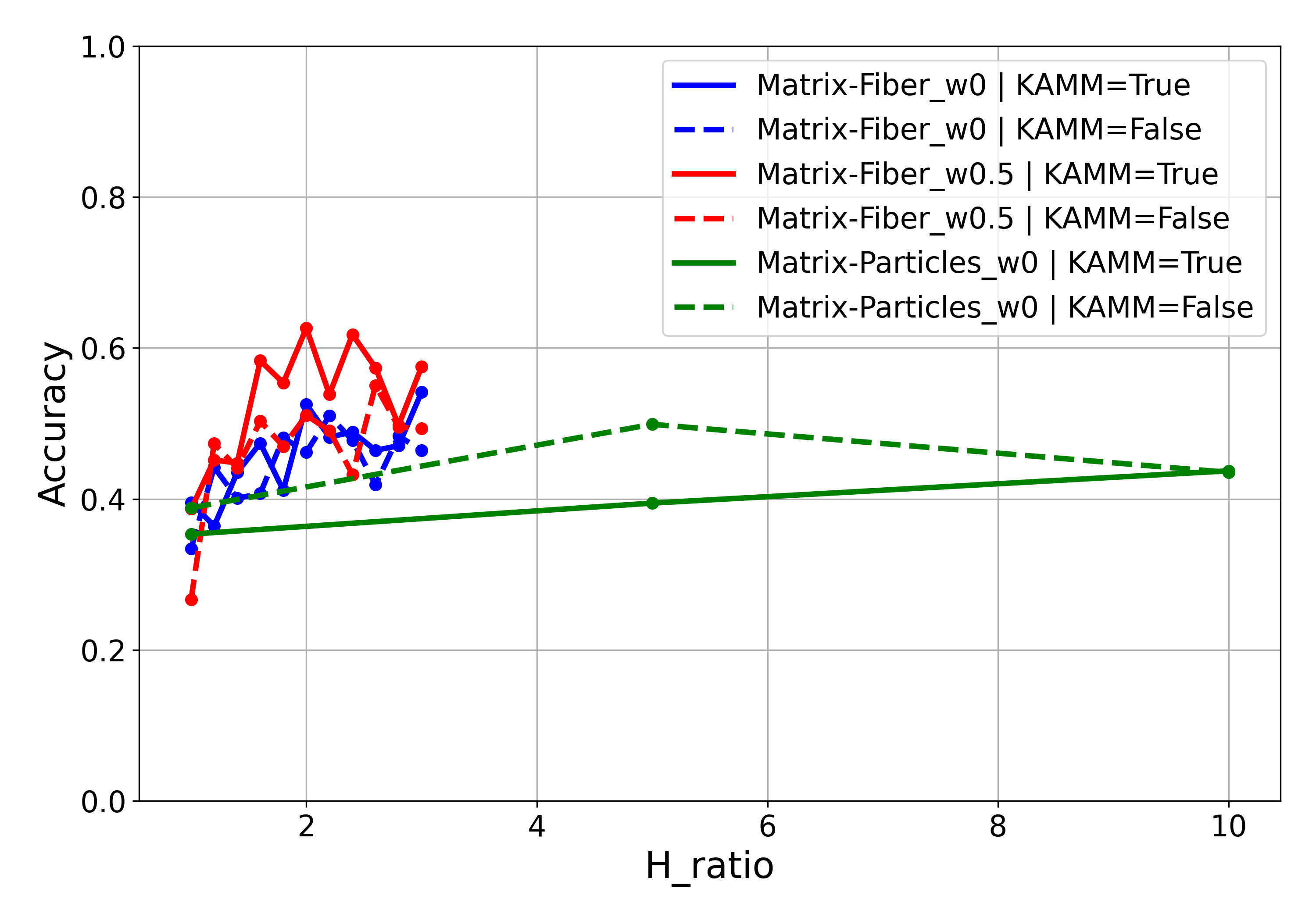}
    }
\end{minipage}
\hfill
\begin{minipage}{0.45\textwidth}
    \subfloat[DBSCAN]{
        \includegraphics[width=\linewidth]{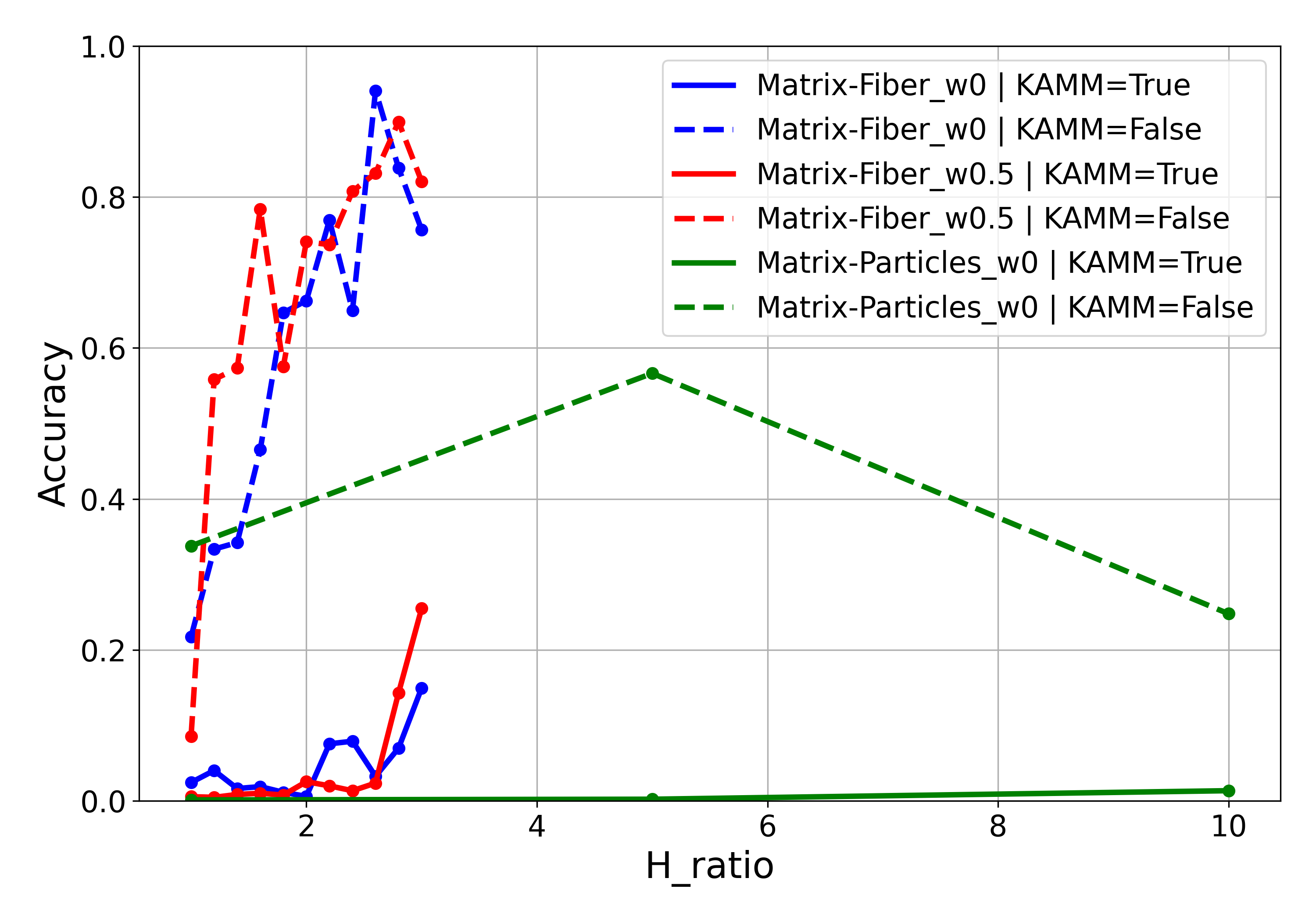}
    }
\end{minipage}

\vspace{0.3cm}

\begin{minipage}{0.45\textwidth}
    \subfloat[GMM]{
        \includegraphics[width=\linewidth]{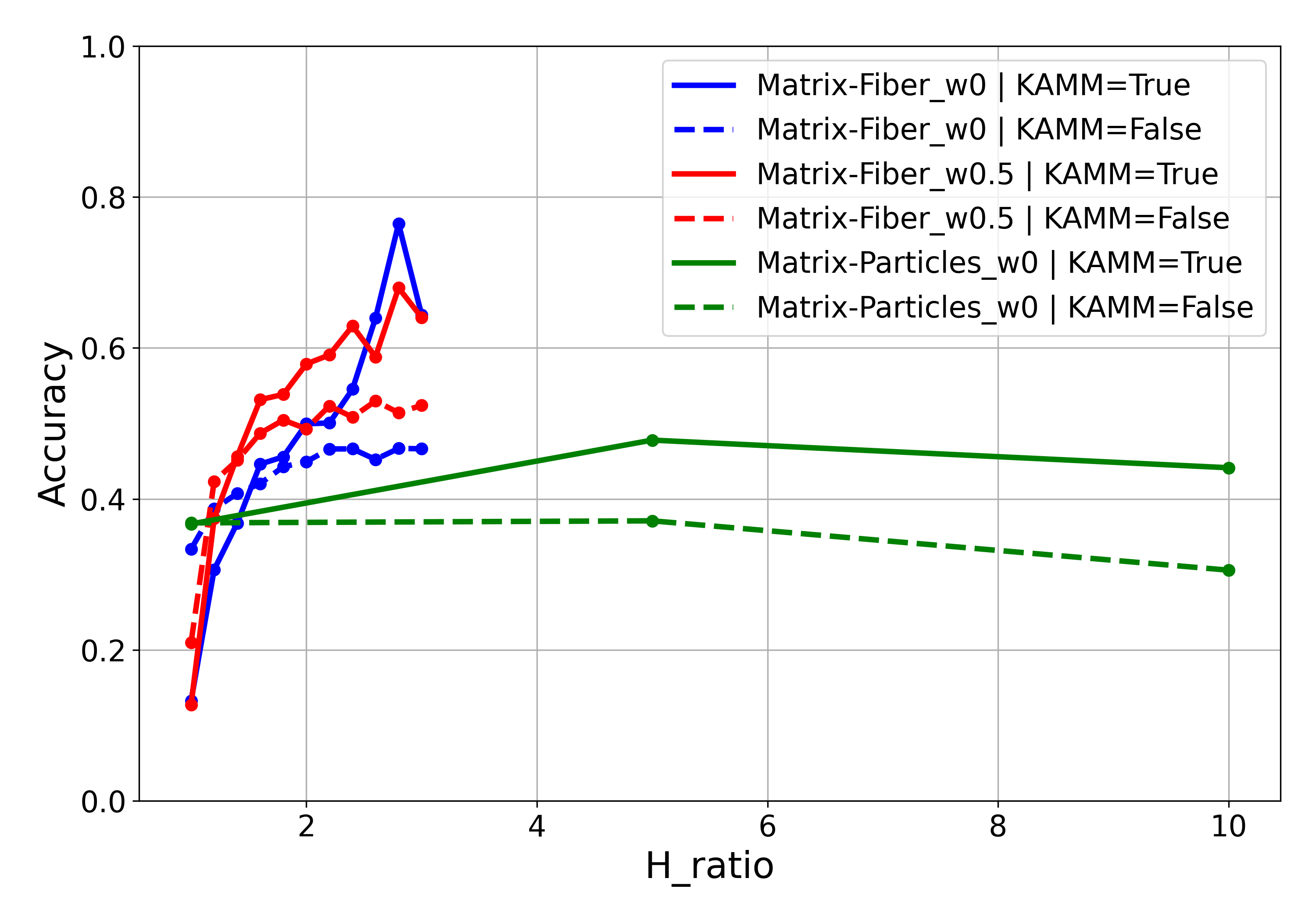}
    }
\end{minipage}
\hfill
\begin{minipage}{0.45\textwidth}
    \subfloat[KMeans]{
        \includegraphics[width=\linewidth]{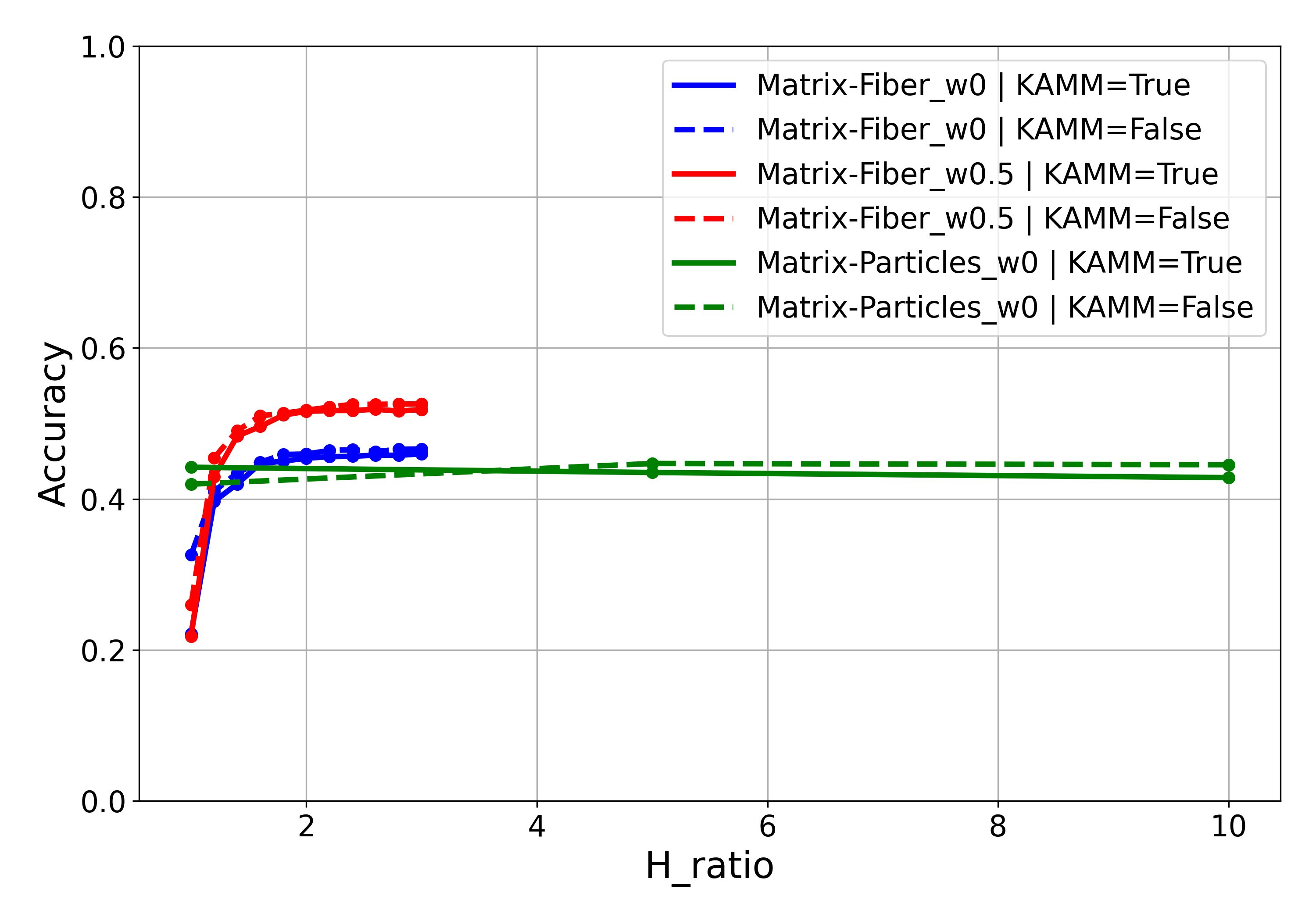}
    }
\end{minipage}

\vspace{0.4cm}

\caption{
Clustering accuracy at pixel level vs.\ hardness contrast ($H_\mathrm{ratio}$) for matrix–particle composites:
sharp interphase, graded interphase ($W=0.05$), and random particle distributions.
Solid lines: clustering with KAMM; dashed lines: $(E,H)$-only baseline.
}
\label{fig:Clustering_MatrixParticles}
\end{figure}

Figure~\ref{fig:Clustering_MatrixParticles} presents the clustering accuracy obtained on the three specimen types considered in this work: matrix–fibers composite respectively with a sharp fiber–matrix interface ($W = 0$), a graded interphase ($W = 0.05$), and a random distribution of particles. For each algorithm, the curves compare the performance with and without the KAMM descriptor as a function of the hardness contrast $H_\mathrm{ratio}$.

For Agglomerative Clustering, the influence of KAMM strongly depends on the microstructure. In the graded interphase case ($W = 0.05$), including KAMM clearly improves clustering accuracy across most hardness ratios. In contrast, little to no difference is observed for the sharp interface and matrix-particles specimens, indicating that the hierarchical scheme benefits from KAMM primarily when the underlying gradients are spatially diffuse. For DBSCAN, the trends differ markedly. Across all three specimen types, accuracy is consistently higher when KAMM is not included. The mismatch-based descriptor appears to deteriorate density-based segmentation, likely because it amplifies local variability that DBSCAN interprets as noise, thereby fragmenting clusters and lowering accuracy. The behavior of GMM is opposite: for all specimen types, introducing KAMM increases accuracy, particularly at intermediate hardness ratios. These results indicate that Gaussian mixture models effectively exploit the additional variability encoded in the KAMM descriptor, leading to a more informative probabilistic partition of the data. Finally, for KMeans, the curves with and without KAMM nearly overlap for all three specimens. This suggests that centroid-based clustering is largely insensitive to KAMM, converging to similar partitions regardless of whether the mismatch descriptor is provided. Overall, the utility of KAMM is strongly algorithm-dependent. It provides clear benefits for GMM, selective improvement for Agglomerative Clustering, minimal influence for KMeans, and systematic degradation for DBSCAN.

\subsection{Real experimental datasets}
\label{sec:results_experimental}

To validate the trends and performance improvements observed on synthetic datasets, the proposed clustering framework is next applied to real experimental nanoindentation data. This step assesses the robustness of the methodology under realistic measurement conditions, where phase boundaries are unknown a priori and experimental noise, microstructural heterogeneity, and indentation size effects may influence feature separability.

\subsubsection{Specimen description and first analysis}
\label{sec:nisic}

The specimen originates from the work of Mercier \textit{et al.}~\cite{mercier_microstructural_2019}. 
It consists of a thick nickel coating reinforced with micrometric silicon carbide (SiC) particles, electro-deposited onto a steel substrate. Microscopic observations (SEM, OM, EBSD) reveal a two-phase surface microstructure composed of approximately 80\% Ni and 20\% SiC. Nanoindentation mapping was performed using a Berkovich indenter on a \(25 \times 25\) grid, with a spacing \(d = 2\,\mu\text{m}\) in both the \(X\) and \(Y\) directions and a maximum indentation depth of \(50\,\text{nm}\). This experimental setup satisfies the recommended length-scale criteria~\cite{randall_nanoindentation_2009, mercier_tridimap_2018}:

\begin{itemize}
    \item \(h_{\max} \leq 0.1\,D\), where \(D\) is the characteristic microstructural length scale  
    (here \(D = 1.3\,\mu\text{m}\), the mean SiC particle diameter);
    \item \(h_{\max} > 3\,R_q\), where \(R_q\) is the average surface roughness;
    \item \(d > (10.5\text{–}21)\,h_{\max}\).
\end{itemize}

Reference mechanical properties indicate a substantial contrast between the Ni matrix and the SiC particles, with reported hardnesses of \(3.9 \pm 0.3\)\,GPa and \(11.7 \pm 3.0\)\,GPa, respectively. Previous deconvolution analyses of nanoindentation data have shown that both hardness and elastic modulus measurements in this composite exhibit multimodal distributions, reflecting contributions from the matrix, the reinforcement, and interfacial regions. This real microstructure therefore provides a representative and challenging benchmark for evaluating the clustering performance of the proposed methodology.

The nanoindentation dataset used in this study is provided as a \texttt{.csv} file containing the \(x\)–\(y\) coordinates of each indent, together with the corresponding measured elastic modulus and hardness values. To enhance the spatial resolution of the resulting mechanical maps, the raw data were subjected to linear interpolation between pixels (by a factor of two) followed by Gaussian smoothing. This pre-processing procedure yields a final dataset represented on a \(97 \times 97\) pixel grid \cite{mercier_tridimap_2018, rossi_revealing_2025}.

\begin{figure}[htbp]
    \centering
    \includegraphics[width=0.95\textwidth]{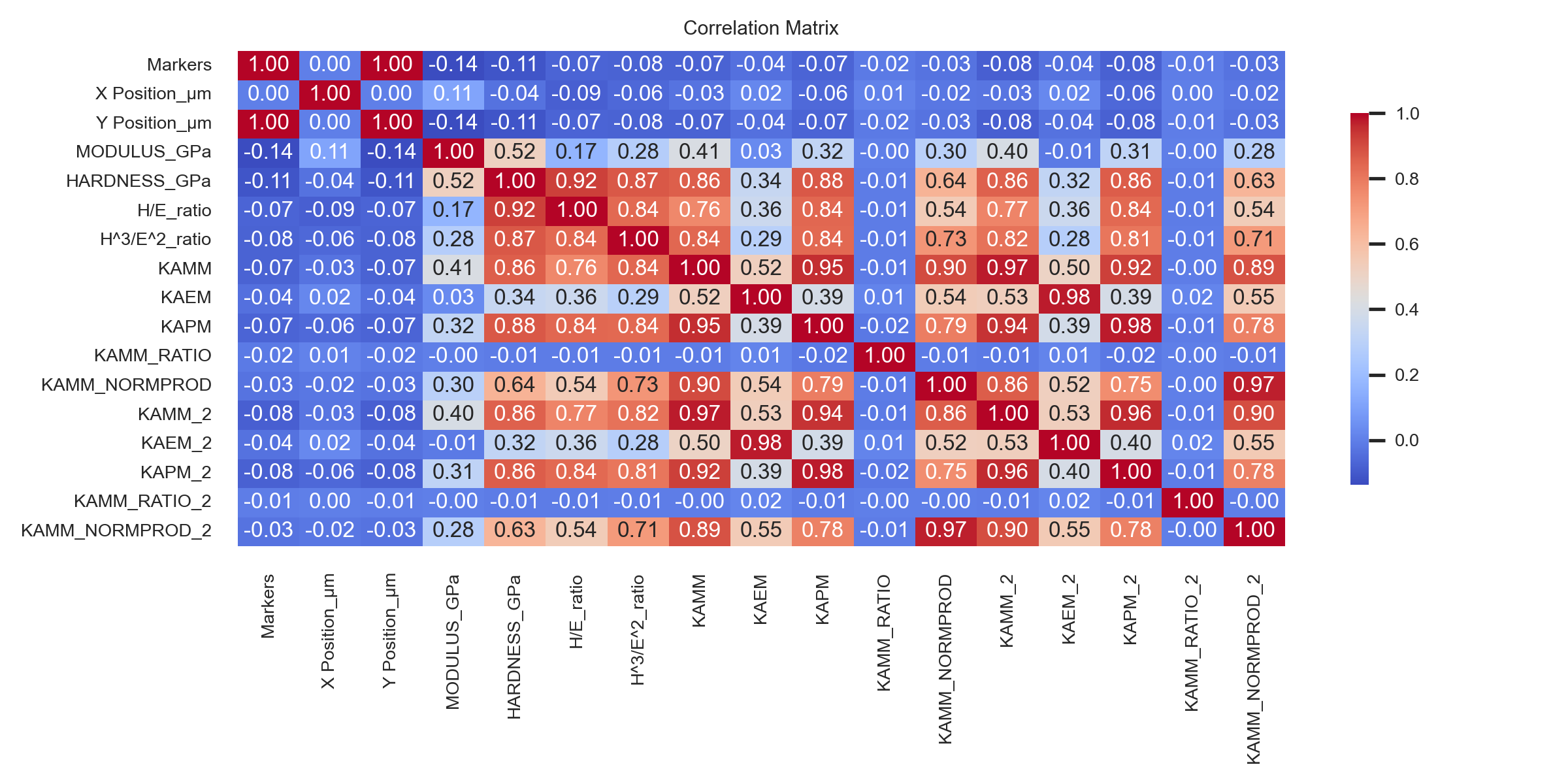}
    \caption{
        Correlation matrix for all mechanical features extracted from the Ni–SiC specimen~\cite{mercier_microstructural_2019}.
    }
    \label{fig:CorrMatrix}
\end{figure}

From these primary measurements, all derived mechanical features (ratios, KAMM descriptors, etc.) were computed, and the full correlation matrix of the resulting feature space was obtained (Fig.~\ref{fig:CorrMatrix}).

For the clustering analysis, four algorithms (DBSCAN, Agglomerative Clustering, GMM, and KMeans) were applied under identical preprocessing conditions (z-score normalization and a fixed number of three clusters, following \cite{mercier_microstructural_2019}).  
Three feature spaces of increasing complexity were considered:

\begin{itemize}
    \item \textbf{Case~1 :} \(H\), \(E\), \(H/E\), \(H^3/E^2\);
    \item \textbf{Case~2 :} Case~1 features augmented with first-order KAMM descriptors;
    \item \textbf{Case~3 :} Case~2 features augmented with first and second order KAMM variants.
\end{itemize}

\begin{table}[htbp]
    \centering
    \caption{Maximum feature--feature correlation for each feature set (identical across all clustering algorithms).}
    \label{tab:max_corr_cases}
    \begin{tabular}{lcc}
        \hline
        \textbf{Case} & \textbf{Feature Set Description} & \textbf{Maximum Correlated Pair} \\
        \hline
        Case 1 & Mechanical features (\(H, E, H/E, H^3/E^2\)) 
               & \(H\) -- \(H/E\) \\[2pt]
        Case 2 & All features (\(H, E, H/E, H^3/E^2\) + KAMM descriptors, 1st order only) 
               & KAMM -- KAPM \\[2pt]
        Case 3 & All features (\(H, E, H/E, H^3/E^2\) + KAMM descriptors, incl. higher-order terms) 
               & KAPM -- KAPM\(_2\) \\
        \hline
    \end{tabular}
\end{table}

Because several descriptors are algebraically or physically related, increasing feature-space complexity may introduce redundancy. In supervised learning, strong inter-feature correlation is often undesirable due to multicollinearity and overfitting risks. However, in unsupervised clustering, correlated features are not inherently detrimental. If the correlation reflects meaningful physical coupling (e.g., spatial coherence or phase continuity), redundant descriptors may reinforce cluster compactness and separability. Nevertheless, high correlation can still influence distance metrics, density estimation, or covariance conditioning depending on the algorithm. For this reason, the maximum feature--feature correlation within each feature set is reported in Table~\ref{tab:max_corr_cases} to document potential redundancy and interpret algorithm-dependent behavior.

\subsubsection{Influence of feature sets on clustering}
\label{sec:nisiC_feature_sets}

\begin{figure}[htbp]
    \centering
    \includegraphics[width=0.95\textwidth]{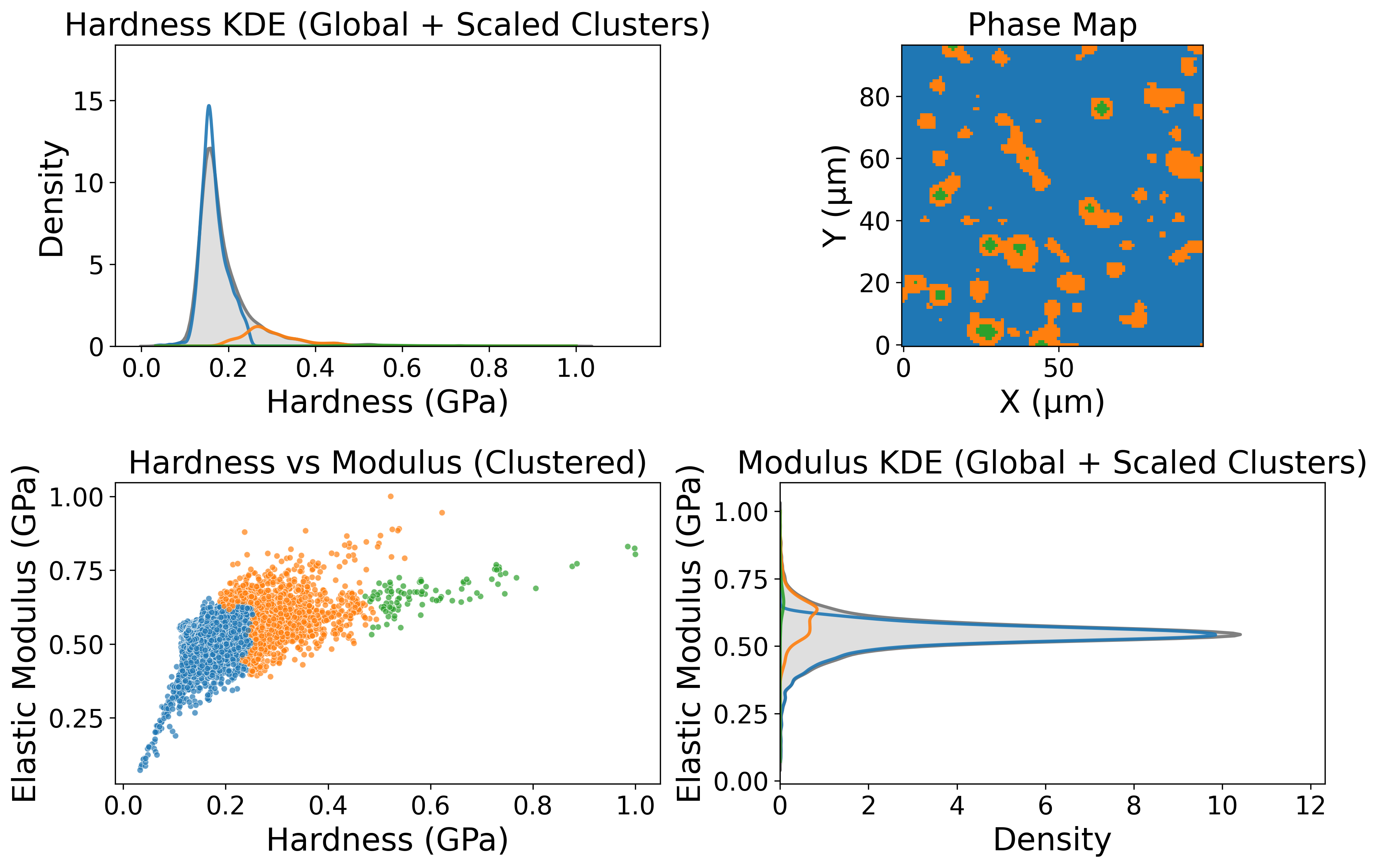}
    \caption{Agglomerative results using mechanical features only.}
    \label{fig:1_Agglo_1}
\end{figure}

\begin{figure}[htbp]
    \centering
    \includegraphics[width=0.95\textwidth]{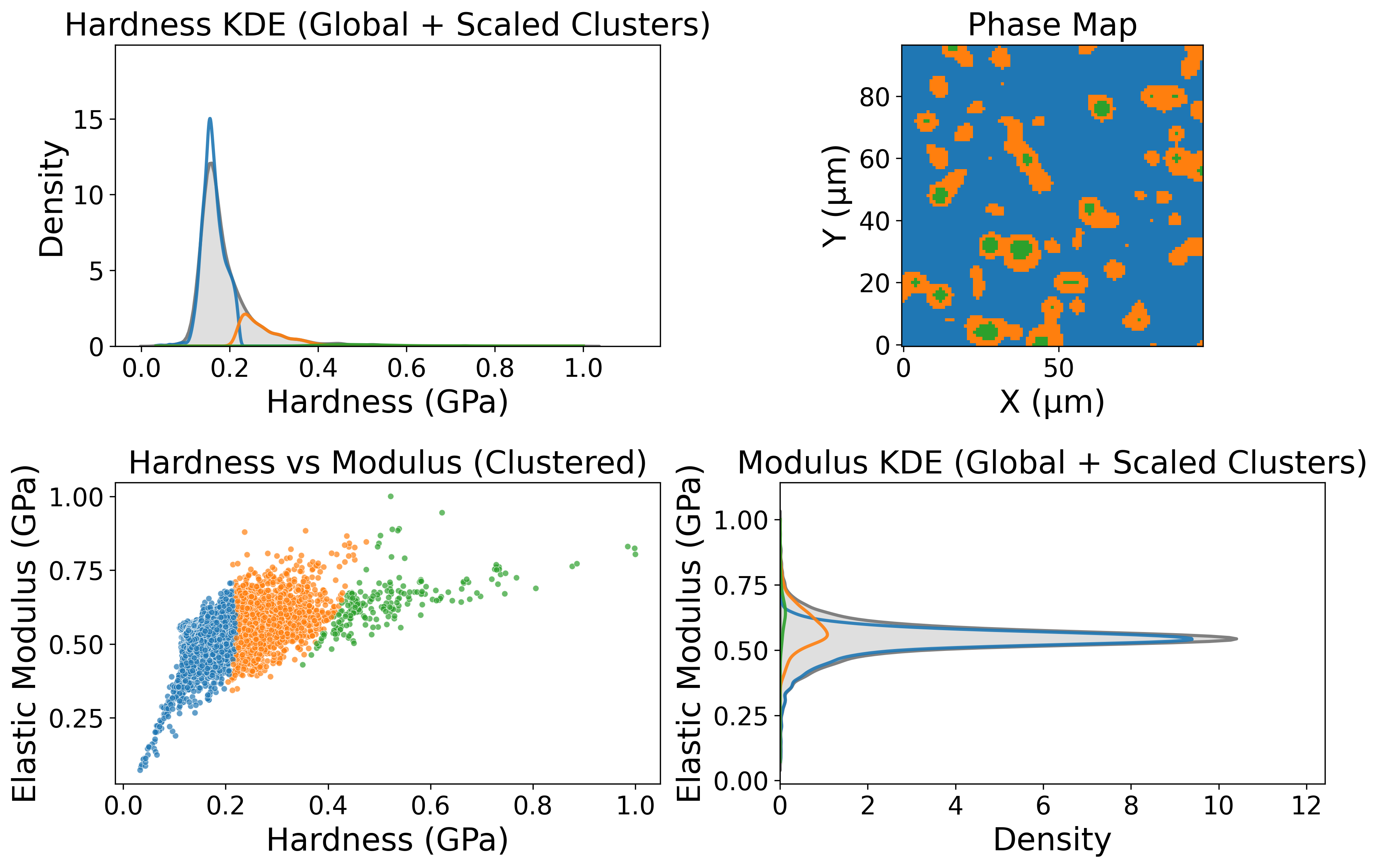}
    \caption{KMeans results using mechanical features only.}
    \label{fig:1_KMEANS_2}
\end{figure}

\begin{figure}[htbp]
    \centering
    \includegraphics[width=0.95\textwidth]{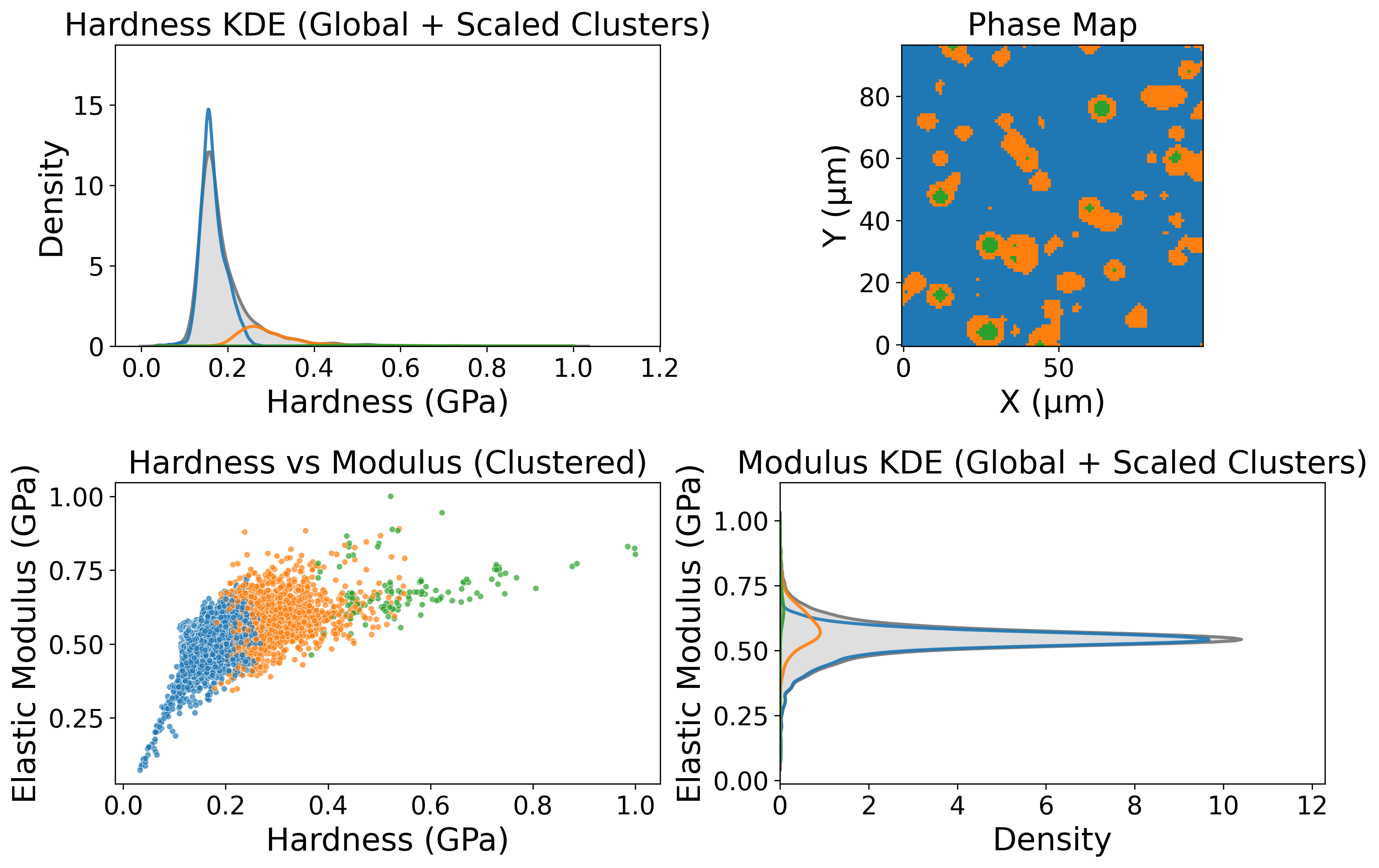}
    \caption{KMeans results using mechanical features and first-order KAMM.}
    \label{fig:2_KMEANS_1}
\end{figure}

\begin{figure}[htbp]
    \centering
    \includegraphics[width=0.95\textwidth]{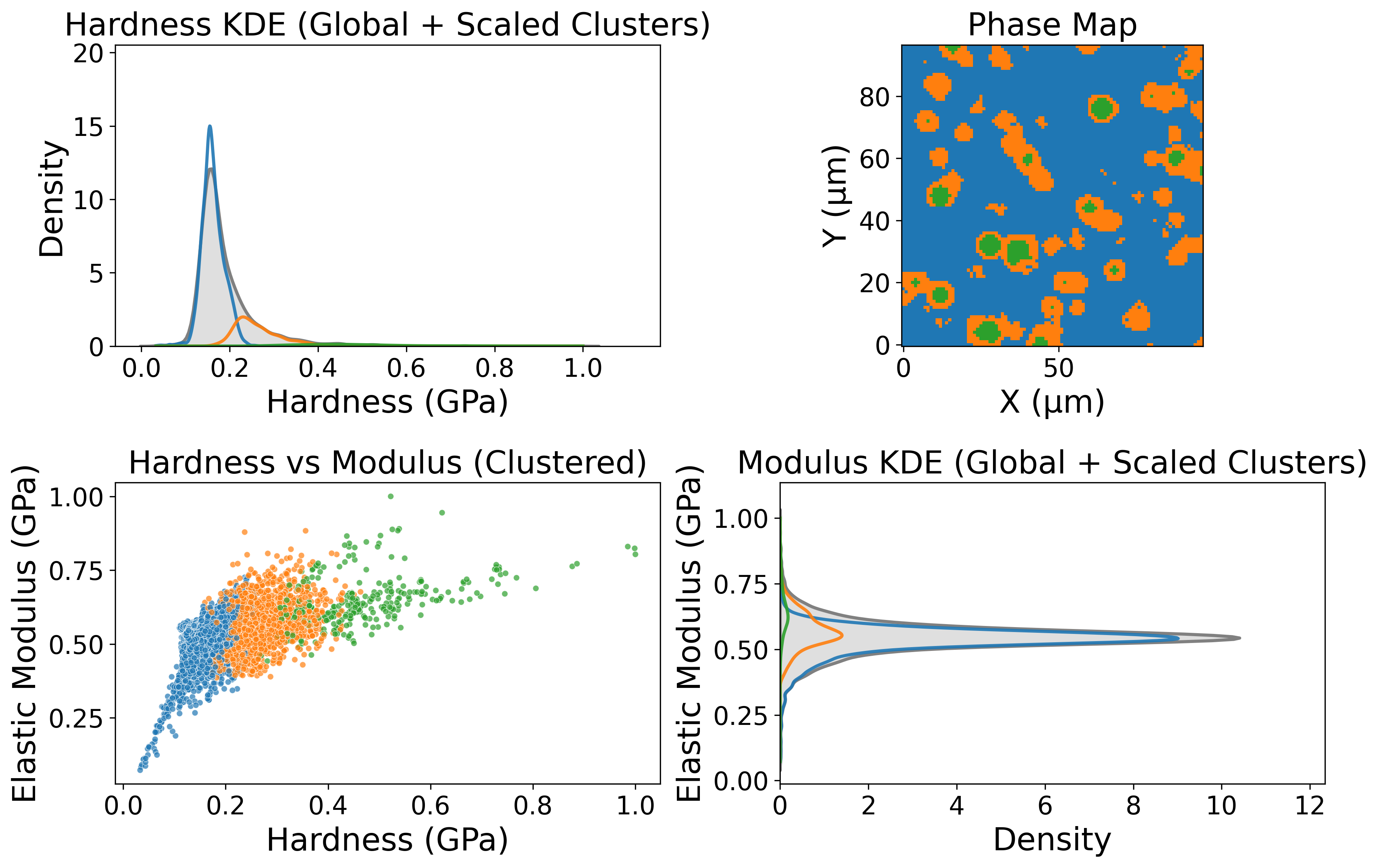}
    \caption{Agglomerative Clustering results using mechanical features and first-order KAMM.}
    \label{fig:2_Agglo_2}
\end{figure}

\begin{figure}[htbp]
    \centering
    \includegraphics[width=0.95\textwidth]{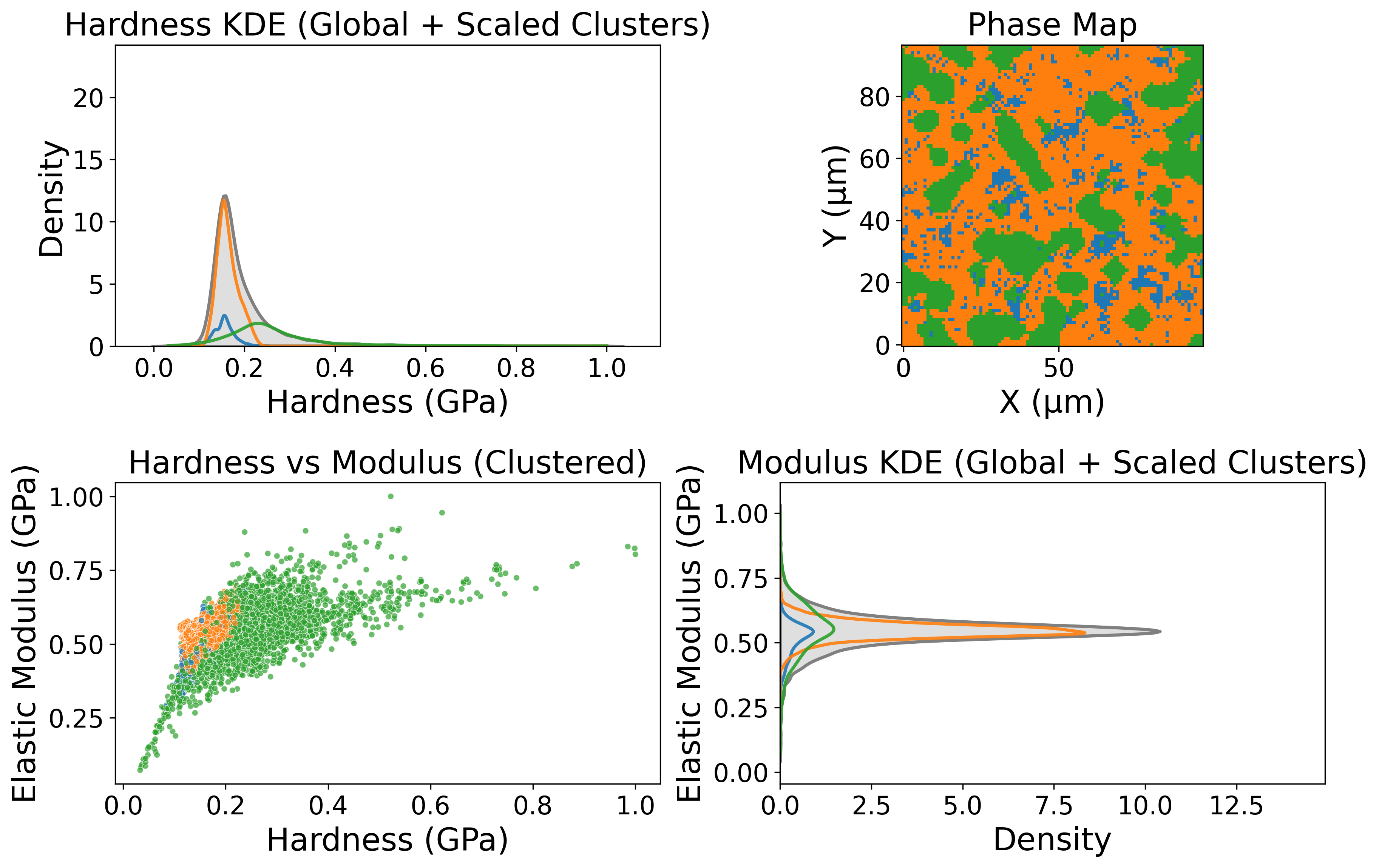}
    \caption{GMM results using the full feature set including higher-order KAMM descriptors.}
    \label{fig:3_GMM_1}
\end{figure}

\begin{figure}[htbp]
    \centering
    \includegraphics[width=0.95\textwidth]{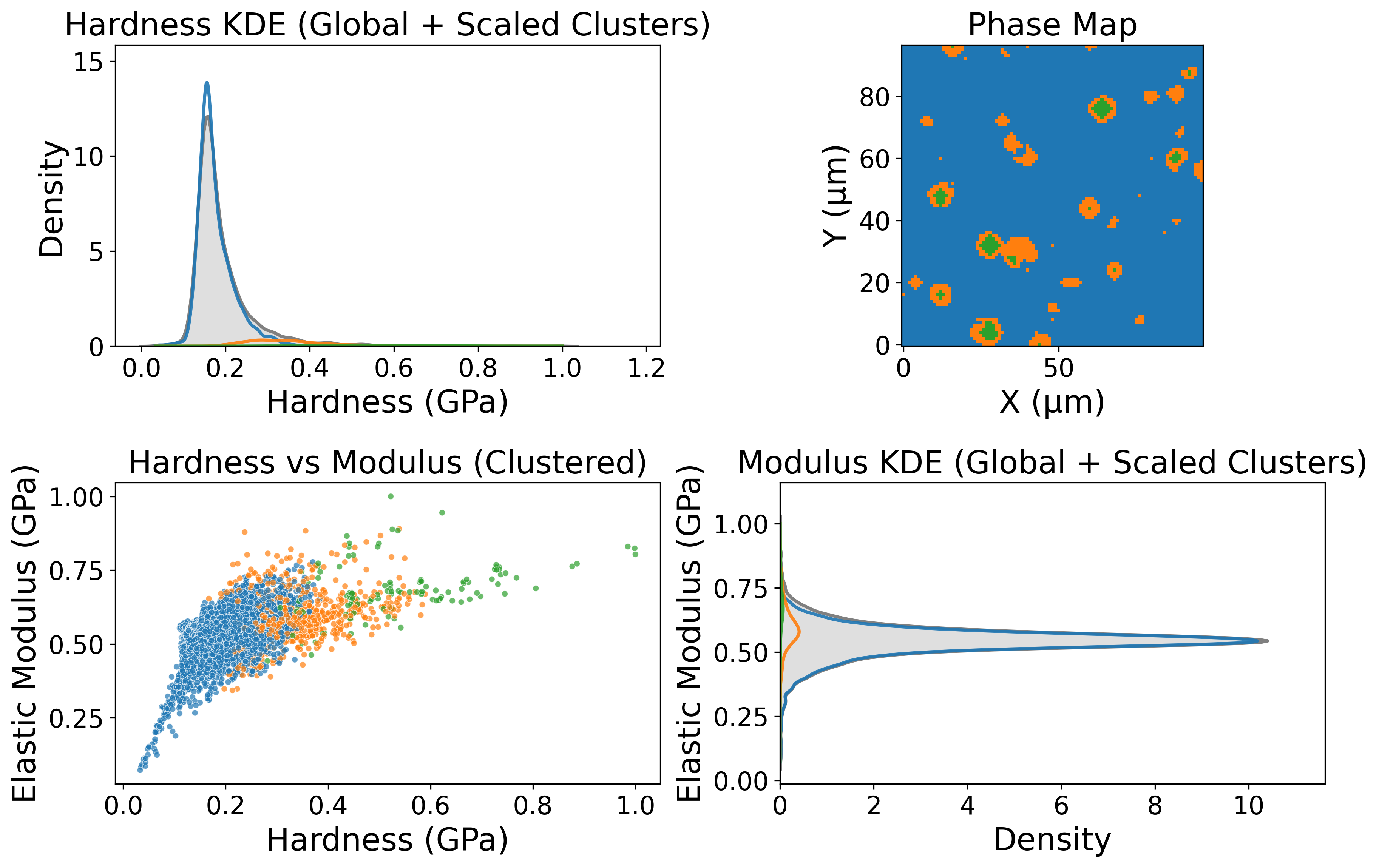}
    \caption{Agglomerative Clustering results using the full feature set including higher-order KAMM descriptors.}
    \label{fig:3_Agglo_2}
\end{figure}

Using the three feature sets defined above (Cases~1--3), clustering results are compared in terms of internal statistical validity, spatial coherence, and phase-wise recovery of mechanical properties. Representative results are shown in Figs.~\ref{fig:1_Agglo_1}, \ref{fig:1_KMEANS_2}, \ref{fig:2_KMEANS_1}, \ref{fig:2_Agglo_2}, \ref{fig:3_GMM_1}, and~\ref{fig:3_Agglo_2}.  For consistency, all visualizations follow a standardized quadrant layout: (top left) global hardness kernel density estimation (KDE) with cluster-wise contributions; (top right) spatial phase map; (bottom left) clustered $H$--$E$ scatter plot; and (bottom right) modulus KDE with cluster-wise contributions. Figures~1 correspond to Case~1 (mechanical features only), Figures~2 to Case~2 (mechanical features augmented with first-order KAMM descriptors), and Figures~3 to Case~3 (full feature set including higher-order KAMM descriptors). Colors are used solely to distinguish clusters within each figure; they do not imply a fixed labeling or direct correspondence to a specific physical phase across different algorithms or feature sets.

\paragraph{Case~1: Mechanical features only (\(H, E, H/E, H^3/E^2\))}

Table~\ref{tab:case1_scores} reports the internal clustering validity metrics obtained using only intrinsic mechanical descriptors. KMeans and Agglomerative Clustering yield clearly positive Silhouette values (0.21 and 0.31, respectively), indicating meaningful separation in feature space. Agglomerative Clustering achieves the best overall performance, combining the highest Silhouette score (0.31), the largest Calinski--Harabasz index (1134.53), and the lowest Davies--Bouldin value (1.43), thus reflecting well-separated and compact clusters. KMeans also performs satisfactorily, though with slightly reduced compactness (Davies--Bouldin = 1.86). In contrast, GMM produces weakly separated clusters (Silhouette = 0.05) and a high Davies--Bouldin index (6.74), indicating strong overlap. DBSCAN yields a negative Silhouette score (-0.38), suggesting poor partitioning despite a moderate Calinski value.

Importantly, higher internal validity scores correspond here to improved physical interpretability of the segmentation. As illustrated in Figs.~\ref{fig:1_Agglo_1} and~\ref{fig:1_KMEANS_2}, Agglomerative Clustering produces a spatially coherent three-phase partition that closely matches the expected Ni--SiC microstructure, clearly distinguishing a Ni-rich matrix, SiC-rich particles, and an intermediate interphase. KMeans identifies the same phases but with reduced spatial coherence, while GMM and DBSCAN generate fragmented or poorly localized regions.

These results indicate that mechanical contrast alone can enable phase discrimination; however, successful recovery of physically meaningful regions strongly depends on the algorithm’s ability to generate compact and well-separated structures in feature space.

\begin{table}[htbp]
    \centering
    \caption{Clustering metrics for Case~1 (mechanical feature set \(H, E, H/E, H^3/E^2\))}
    \label{tab:case1_scores}
    \begin{tabular}{lccc}
        \hline
        \textbf{Method} & \textbf{Silhouette} & \textbf{Calinski} & \textbf{Davies} \\
        \hline
        KMeans                   & 0.21 & 802.90  & 1.86 \\
        GMM                      & 0.05 & 102.99  & 6.74 \\
        Agglomerative Clustering & 0.31 & 1134.53 & 1.43 \\
        DBSCAN                  & -0.38 & 76.86 & 2.02 \\
        \hline
    \end{tabular}
\end{table}

\paragraph{Case~2: Mechanical features with first-order KAMM}
When first-order KAMM descriptors are added, cluster separability decreases compared to Case~1 (Table~\ref{tab:case2_scores}). The silhouette score drops to 0.17 for KMeans and 0.08 for Agglomerative Clustering, while GMM yields a very low value (0.02) and DBSCAN becomes strongly negative (-0.35). Calinski–Harabasz indices also decrease substantially relative to Case~1, and Davies–Bouldin values increase for most methods, indicating reduced compactness and stronger cluster overlap in the augmented feature space. This behavior reflects the strong correlations introduced by the spatial gradient fields, which blur Euclidean distances when KAMM-based descriptors are combined with mechanical ratios.

Despite the reduction in classical clustering metrics, spatial segmentation improves. As shown in Figs.~\ref{fig:2_KMEANS_1} and~\ref{fig:2_Agglo_2}, both KMeans and Agglomerative Clustering reveal a clearer spatial separation between the Ni-rich matrix and the SiC-rich particles. The phase-averaged properties in Table~\ref{tab:ref_cluster_feature_comparison} further show that the mean hardness and modulus of the Ni-like and SiC-like phases are recovered more accurately than in Case~1, with values closer to the corresponding reference data. Thus, first-order KAMM enhances physical interpretability by incorporating spatial gradient information that reflects local mechanical heterogeneity and interfacial transitions around inclusions. Although this additional structure reduces classical clustering metrics based on Euclidean compactness, it improves phase-wise property recovery and spatial coherence of the segmentation.

\begin{table}[htbp]
    \centering
    \caption{Clustering metrics for Case~2 (all mechanical + KAMM features)}
    \label{tab:case2_scores}
    \begin{tabular}{lccc}
        \hline
        \textbf{Method} & \textbf{Silhouette} & \textbf{Calinski} & \textbf{Davies} \\
        \hline
        KMeans                   & 0.17  & 16.10 & 1.76 \\
        GMM                      & 0.02  & 9.27  & 3.91 \\
        Agglomerative Clustering & 0.08  & 15.21 & 2.21 \\
        DBSCAN                   & -0.35 & 15.89 & 1.69 \\
        \hline
    \end{tabular}
\end{table}

\paragraph{Case~3: Full feature set with higher-order KAMM}

Including higher-order KAMM descriptors further modifies the structure of the feature space (Table~\ref{tab:case3_scores}). In contrast to Case~2, the centroid-based methods retain positive Silhouette values, with GMM achieving the highest value (0.19), followed by Agglomerative Clustering (0.15) and KMeans (0.07). However, the Calinski--Harabasz and Davies--Bouldin indices reveal mixed behavior: while GMM shows the largest between-cluster dispersion (Calinski = 58.44), it also exhibits a very high Davies--Bouldin value (10.47), indicating strong cluster overlap. Agglomerative Clustering yields the lowest Davies--Bouldin index (1.99), suggesting comparatively better compactness, though with limited overall separation (Calinski = 6.22). DBSCAN remains unstable, with a negative Silhouette value (-0.33), consistent with fragmented or noisy partitions despite a relatively moderate Davies--Bouldin score.

The phase maps (Figs.~\ref{fig:3_GMM_1} and~\ref{fig:3_Agglo_2}) remain spatially structured; however, the phase-averaged mechanical properties (Table~\ref{tab:ref_cluster_feature_comparison}) tend to shift toward intermediate values. This behavior indicates a form of over-parameterization: the higher-order KAMM descriptors introduce additional variance that reduces the effective contrast between Ni-rich and SiC-rich regions. In practice, these higher-order terms emphasize fine-scale mechanical heterogeneity rather than clear phase separation.

\begin{table}[htbp]
    \centering
    \caption{Clustering metrics for Case~3 (full feature set with higher-order KAMM)}
    \label{tab:case3_scores}
    \begin{tabular}{lccc}
        \hline
        \textbf{Method} & \textbf{Silhouette} & \textbf{Calinski} & \textbf{Davies} \\
        \hline
        KMeans                   & 0.07  & 9.51  & 2.11 \\
        GMM                      & 0.19  & 58.44 & 10.47 \\
        Agglomerative Clustering & 0.15  & 6.22  & 1.99 \\
        DBSCAN                   & -0.33 & 4.14  & 1.75 \\
        \hline
    \end{tabular}
\end{table}

Overall, a consistent hierarchy emerges across the three cases when comparing the phase-averaged properties in Table~\ref{tab:ref_cluster_feature_comparison}. Mechanical features alone (Case~1) already allow a clear separation between the Ni-like (Phase~1) and SiC-like (Phase~3) clusters, with Phase~1 moduli close to the Ni reference value (219.7~GPa) and Phase~3 exhibiting significantly higher stiffness. The addition of first-order KAMM (Case~2) maintains this contrast while slightly improving spatial coherence, without drastically altering the mean phase properties.

In contrast, the inclusion of higher-order KAMM descriptors (Case~3) leads to a redistribution of mechanical contrast. As shown in Table~\ref{tab:ref_cluster_feature_comparison}, the GMM solution produces intermediate hardness and modulus values for Phase~3 (7.99~GPa, 223.00~GPa), substantially closer to the Ni-like phase than to the $\alpha$-SiC reference (343.2~GPa). This indicates that higher-order spatial descriptors partially blur the mechanical distinction between phases. Although Agglomerative Clustering in Case~3 preserves a larger hardness contrast (15.30~GPa for Phase~3), the associated modulus remains far below the $\alpha$-SiC reference value, confirming that the additional descriptors primarily capture local heterogeneity rather than global phase contrast.

These trends confirm that mechanical features are sufficient to recover the dominant phases, while first-order KAMM improves interphase delineation through spatial gradient information. Higher-order KAMM descriptors, however, mainly encode localized deformation patterns and should therefore be used selectively when the objective extends beyond phase identification toward fine-scale heterogeneity analysis.

Finally, the elastic modulus values—particularly within the SiC-like clusters—remain influenced by outliers associated with unusually high measured stiffness on certain particles. As reflected by the large standard deviations in Table~\ref{tab:ref_cluster_feature_comparison} (e.g., 43.42~GPa for Phase~3 in Case~3 using GMM), these extreme values broaden the cluster distributions and may artificially increase or destabilize the apparent modulus of the hard phase. This sensitivity to noise is consistent with the synthetic benchmark analysis, where added variability similarly broadened property distributions and reduced cluster compactness.

\begin{table}[htbp]
\centering
\caption{Comparison between reference phases and clustered phases (absolute values). Phase~1 corresponds to the lowest mechanical properties (Ni-like), and Phase~3 to the highest mechanical properties ($\alpha$-SiC-like). Values are reported as mean $\pm$ standard deviation.}
\label{tab:ref_cluster_feature_comparison}
\begin{tabular}{l l l c c c c}
\hline
\textbf{Case / Features} & \textbf{Phase} & \textbf{Method} &
\textbf{Hardness (GPa)} & \textbf{Std (GPa)} &
\textbf{Modulus (GPa)} & \textbf{Std (GPa)} \\
\hline

\multicolumn{7}{l}{\textbf{References}} \\
Ni \cite{mercier_microstructural_2019}             & -- & -- & 3.90 & 0.30 & 219.70 & 7.20 \\
$\alpha$-SiC \cite{mercier_microstructural_2019}   & -- & -- & 11.70 & 3.00 & 343.20 & 50.00 \\
\hline

\multicolumn{7}{l}{\textbf{Case 1: $H, E, H/E, H^3/E^2$}} \\
& Phase 1 & Agglo  & 5.30 & 0.95 & 216.88 & 23.64 \\
& Phase 3 & Agglo  & 18.95 & 3.64 & 274.82 & 22.77 \\
& Phase 1 & KMeans & 5.17 & 0.81 & 217.05 & 24.17 \\
& Phase 3 & KMeans & 16.53 & 3.62 & 264.11 & 35.22 \\
\hline

\multicolumn{7}{l}{\textbf{Case 2: $H, E, H/E, H^3/E^2$ + KAMM (1st order)}} \\
& Phase 1 & KMeans & 5.27 & 0.96 & 217.41 & 23.80 \\
& Phase 3 & KMeans & 16.93 & 4.26 & 278.89 & 35.61 \\
& Phase 1 & Agglomerative & 5.14 & 0.81 & 217.07 & 24.64 \\
& Phase 3 & Agglomerative & 14.92 & 3.92 & 264.96 & 35.30 \\
\hline

\multicolumn{7}{l}{\textbf{Case 3: Full feature set + higher-order KAMM}} \\
& Phase 1 & GMM    & 4.95 & 0.68 & 211.72 & 23.38 \\
& Phase 3 & GMM    & 7.99 & 3.32 & 223.00 & 43.42 \\
& Phase 1 & Agglomerative & 5.29 & 0.83 & 222.19 & 15.28 \\
& Phase 3 & Agglomerative & 15.30 & 4.44 & 266.16 & 35.05 \\
\hline

\end{tabular}
\end{table}

\section{Integration into materials modelling workflow}
\label{sec:discussion}

The phase maps obtained from the clustering of neighborhood-resolved mechanical measurements are subsequently used to construct a three-dimensional multiphase Representative Volume Element (RVE) suitable for finite element simulations. The clustered phase map is provided as a comma-separated values (CSV) file containing the spatial coordinates $(x,y)$ of each measurement point, expressed in micrometers, together with the corresponding phase label. The experimental grid is assumed to be regular. Unique, sorted coordinate sets $\{x_i\}$ and $\{y_j\}$ are extracted to define the in-plane discretization, resulting in a structured mesh of $(N_x-1)\times(N_y-1)$ quadrilateral cells.

To enable three-dimensional simulations while preserving the essentially two-dimensional nature of the experimental data, the phase map is extruded along the out-of-plane direction. A single layer of hexahedral elements is generated, with a thickness $D_z$ chosen as a fraction of the in-plane dimension,
\begin{equation}
D_z = 0.1\,(x_{\max} - x_{\min}),
\end{equation}
leading to a thin three-dimensional RVE representative of the local microstructure. The finite element model is generated programmatically using the Python interface to Ansys Mechanical APDL (PyMAPDL). Eight-node hexahedral elements (SOLID185) are employed. Nodes are created at the experimental $(x,y)$ positions and duplicated at $z=0$ and $z=D_z$, after which elements are formed by connecting adjacent nodes in the structured grid.

\begin{figure}[htbp]
\centering
\includegraphics[width=0.6\linewidth]{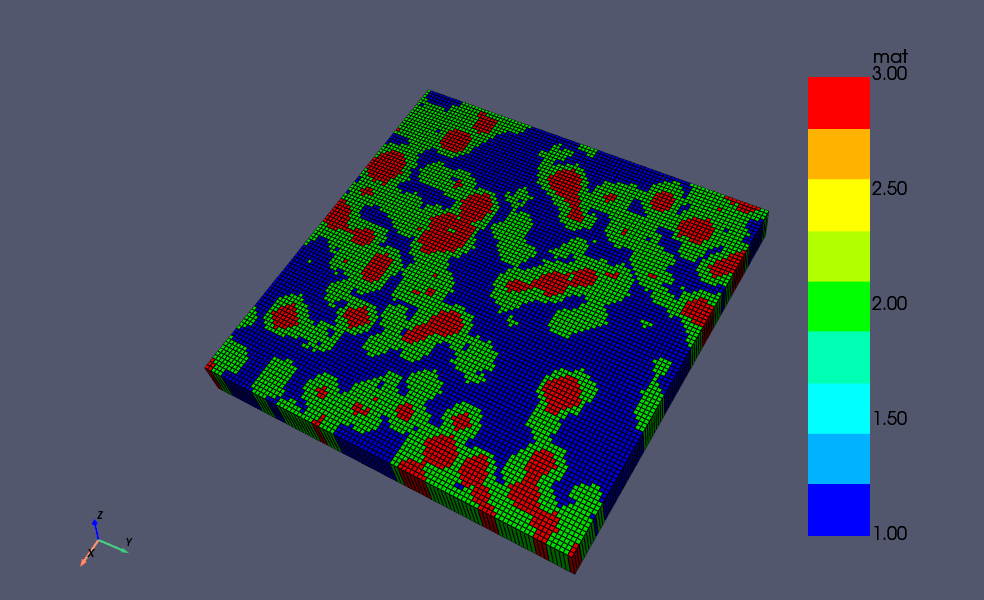}
\caption{Three-dimensional representative volume element (RVE) generated from clustered mechanical phase maps using the KAMM-based feature set. The experimentally identified phases are extruded from the two-dimensional measurement grid into a thin three-dimensional domain and assigned element-wise as distinct material regions, enabling direct use in finite element simulations.}
\label{fig:rve_multiphase}
\end{figure}

An example of such a multiphase RVE generated for the Ni--SiC system is shown in Fig.~\ref{fig:rve_multiphase}, corresponding to Case~2 using the GMM clustering approach. The reconstructed RVE clearly highlights the coexistence of matrix, reinforcement, and an intermediate interphase region, demonstrating the ability of the proposed workflow to capture neighborhood-resolved phase heterogeneity and to translate clustering results into simulation-ready microstructural models.

Each distinct phase identified by the clustering procedure is mapped to a unique material identifier. Material properties are assigned on a phase-wise basis, while the spatial distribution of phases is imposed at the element level. For each element, the material identifier is modified according to the phase label associated with the corresponding $(x,y)$ location in the clustered map, ensuring a one-to-one correspondence between the experimentally identified phase distribution and the numerical discretization. The resulting model is saved as a Mechanical APDL database file, enabling reproducible reloading, post-processing, and extension to subsequent mechanical simulations.

The correctness of the phase assignment is verified by inspecting element material identifiers directly within Mechanical APDL. Visualization of the multiphase RVE is performed using element plots colored by material identifier and complemented by Python-based visualization, in which element-wise material numbers are explicitly extracted and mapped onto a VTK-compatible unstructured grid. Figure~\ref{fig:rve_multiphase} illustrates the resulting multiphase RVE, where different colors correspond to the distinct phases identified from clustering of the mechanical maps. The spatial distribution of phases is preserved from the experimental data, providing a physically consistent numerical representation of the heterogeneous microstructure.

Beyond finite element model generation, the clustering results and derived phase statistics are integrated into a broader ICME workflow. The experimentally measured mechanical maps constitute the entry point of the pipeline and are processed through a clustering-based analysis to identify mechanically distinct phases. For each identified phase, statistical descriptors such as phase volume fraction, as well as the mean and standard deviation of mechanical properties, are computed. Metadata describing the clustering methodology, including the clustering algorithm, number of clusters, and relevant preprocessing steps, are preserved to ensure traceability.

\begin{figure}[htbp]
\centering
\includegraphics[width=1\linewidth]{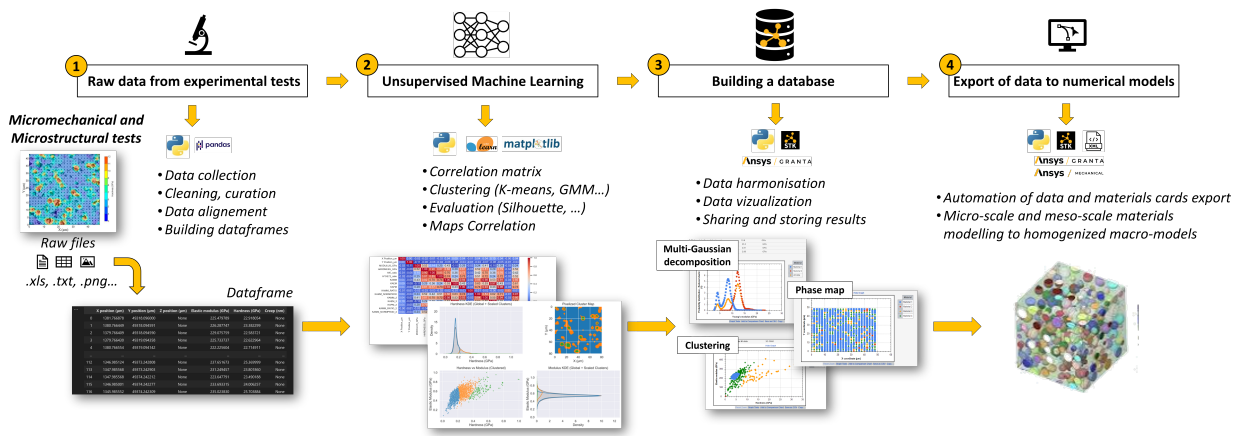}
\caption{Schematic overview of the ICME workflow connecting experimental mechanical mapping, clustering-based phase identification, materials data management (Granta MI), and the generation of simulation-ready material cards for finite element micro- and meso-scale simulations.}
\label{fig:icme_workflow}
\end{figure}

These phase-resolved descriptors are subsequently exported to a materials data management system (Granta MI), where they are stored as structured material records. This enables a consistent linkage between experimental measurements, reduced-order phase descriptions, and simulation-ready material definitions. From this database, material cards can be automatically generated and deployed in finite element simulations at the micro- or meso-scale. At these scales, the identified phase properties can be combined with representative microstructural descriptions to perform homogenization-based analyses, allowing the estimation of additional effective mechanical parameters (e.g. elastic, elastoplastic, or damage-related properties) that are not directly accessible from local indentation measurements. Such homogenization strategies, as implemented for instance in tools like Ansys Material Designer, enable the construction of enriched, physics-consistent material models. This supports a closed-loop workflow from experiments to modeling, where homogenized properties can be fed back into higher-scale simulations within an ICME framework. Figure~\ref{fig:icme_workflow} provides an overview of this integrated workflow.

\section{Conclusion}
\label{sec:conclusion}

This work introduced the \textit{Kernel-Averaged Mechanical Mismatch} (KAMM) as a neighborhood-informed descriptor for nanomechanical property maps of multiphase materials. By quantifying local heterogeneity through neighborhood comparisons in the \((E,H)\) fields, KAMM injects neighborhood-based context that is absent from conventional feature spaces. When appended to the clustering space, first-order KAMM improves the physical interpretability of phase maps by enhancing spatial separation between mechanically distinct domains and supporting more accurate phase-wise property recovery compared to purely mechanical feature sets.

The results further highlight that the benefit of KAMM is algorithm-dependent and that classical clustering metrics do not necessarily reflect physical relevance. In particular, neighborhood-based descriptors introduce correlated dimensions that alter the geometry of the feature space, so improvements in spatial coherence and phase identification may occur even when standard separability scores decrease. This underscores the need to assess clustering quality using both statistical indices and physically meaningful criteria, such as spatial continuity, microstructural plausibility, and the consistency of recovered phase properties.

Across both synthetic benchmarks and experimental data, KAMM generally benefits centroid- and distribution based clustering methods (notably KMeans and Gaussian mixture models), whereas density-based approaches can degrade when mismatch features amplify local variability. These observations emphasize that spatial feature enrichment must be paired with clustering assumptions that remain compatible with the transformed feature space, and that practical implementations should include feature selection strategies (e.g.\ limiting to first-order descriptors or pruning highly correlated variants) to preserve stability and interpretability.

Future work will focus on strengthening the methodology under challenging experimental conditions, including strong noise, non-uniform sampling, and measurement artefacts. Promising directions include (i) evaluating more robust density-based paradigms such as OPTICS and HDBSCAN; (ii) investigating spatially constrained or graph-based clustering methods that enforce neighborhood consistency directly; and (iii) integrating systematic outlier detection and removal into the preprocessing pipeline to further improve robustness and the reliability of phase-wise statistics.

Overall, KAMM provides a simple and computationally inexpensive route to incorporate neighborhood-based awareness into nanomechanical phase mapping. Combined with reproducible workflows and integration into ICME toolchains, it offers a practical basis for automated phase identification and for translating high-throughput mechanical mapping into simulation-ready multiphase microstructural models.

\section*{Author Contributions}

D.~Mercier performed the modelling work and led the manuscript writing.  
Y.~El~Gharoussi developed the synthetic dataset generators and implemented the analysis code.

\section*{Acknowledgements}

The authors gratefully acknowledge Davide Di Stefano and Pascal Salzbrenner for rich and insightful discussions that helped shape and refine this work. This research has received support from the NanomeCommons project (Horizon 2020, Grant Agreement No.~952869). The authors are solely responsible for the content presented herein; this does not necessarily reflect the views of the European Union.

\section*{Declaration of Generative AI and AI-Assisted Technologies in the Manuscript Preparation}

During the preparation of this manuscript, the author(s) used Microsoft Copilot for grammar correction, syntax refinement, and improvements in clarity. After using this tool, the author(s) carefully reviewed and edited the content as needed and take full responsibility for the final version of the published article.

\bibliographystyle{unsrt}
\bibliography{references}

\end{document}